\documentclass[doc]{imsart}

\usepackage{amsthm,amsmath,bm,bbm}
\usepackage{amssymb,mathtools}
\usepackage[numbers]{natbib}
\usepackage{hyperref}
\hypersetup{colorlinks,citecolor=blue,urlcolor=blue}
\usepackage[english]{babel}
\usepackage[utf8x]{inputenc}
\usepackage[T1]{fontenc}
\usepackage{graphicx}
\usepackage{capt-of}
\usepackage{tikz}
\usetikzlibrary{shapes.geometric, arrows}
\usepackage{tabularx}
\usepackage[shortlabels]{enumitem}
\usepackage{cancel}
\usepackage{pdflscape}
\usepackage{afterpage}
\usepackage{multirow}
\usepackage{booktabs}
\usepackage{listings}

\usepackage{pdfpages}

\startlocaldefs
\def\independent{\perp\!\!\!\perp}

\def\E{\text{E}}
\def\P{\text{P}}

\definecolor{codegreen}{rgb}{0,0.6,0}
\definecolor{codegray}{rgb}{0.5,0.5,0.5}
\definecolor{codepurple}{rgb}{0.58,0,0.82}
\definecolor{backcolour}{rgb}{0.95,0.95,0.92}

\lstdefinestyle{mystyle}{
    backgroundcolor=\color{backcolour},   
    commentstyle=\color{codegreen},
    keywordstyle=\color{magenta},
    numberstyle=\tiny\color{codegray},
    stringstyle=\color{codepurple},
    basicstyle=\ttfamily\footnotesize,
    breakatwhitespace=false,         
    breaklines=true,                 
    captionpos=b,                    
    keepspaces=true,                 
    numbers=left,                    
    numbersep=5pt,                  
    showspaces=false,                
    showstringspaces=false,
    showtabs=false,                  
    tabsize=2
}

\endlocaldefs

\begin{document}

\begin{frontmatter}

\title{Causal mediation analysis: From simple to more robust strategies for estimation of marginal natural (in)direct effects\support{This is the final version of the paper (version 4), to be published in \textit{Statistics Surveys} volume 17 (2023), \url{https://doi.org/10.1214/22-SS140}. The main changes from \href{https://arxiv.org/pdf/2102.06048v3.pdf}{version 3} are: (i) the section \textit{Leveraging covariates to improve precision} has been mostly left out; (ii) a brief discussion of model compatibility is added (section 4.5); and (iii) the discussion of how to choose an estimator is expanded (new section 6).}}
\runtitle{Causal mediation from simple to robust}

\begin{aug}

\author{\fnms{Trang Quynh} 
\snm{Nguyen}
\corref{}
\ead[label=e1]{trang.nguyen@jhu.edu}}
\address{Johns Hopkins Bloomberg School of Public Health\\ \printead{e1}}

\author{\fnms{Elizabeth L.} 
\snm{Ogburn}
\ead[label=e2]{eogburn@jhu.edu}}
\address{Johns Hopkins Bloomberg School of Public Health\\ \printead{e2}}

\author{\fnms{Ian}
\snm{Schmid}
\ead[label=e3]{ian\_schmid@jhu.edu}}
\address{Johns Hopkins Bloomberg School of Public Health\\ \printead{e3}}

\author{\fnms{Elizabeth B.}
\snm{Sarker}
\ead[label=e4]{esarker1@jhmi.edu}}
\address{Johns Hopkins Bloomberg School of Public Health\\ \printead{e4}}

\author{\fnms{Noah}
\snm{Greifer}
\ead[label=e5]{ngreifer@iq.harvard.edu}}
\address{Harvard University Institute for Quantitative Social Science\\ \printead{e5}}

\author{\fnms{Ina M.}
\snm{Koning}
\ead[label=e6]{i.koning@uu.nl}}
\address{Utrecht University\\ \printead{e6}}

\author{\fnms{Elizabeth A.} 
\snm{Stuart}
\ead[label=e7]{estuart@jhu.edu}}
\address{Johns Hopkins Bloomberg School of Public Health\\ \printead{e7}}

\runauthor{Nguyen et al.}

\end{aug}

\begin{abstract}
This paper aims to provide practitioners of causal mediation analysis with a better understanding of estimation options. We take as inputs two familiar strategies (weighting and model-based prediction) and a simple way of combining them (weighted models), and show how a range of estimators can be generated, with different modeling requirements and robustness properties. The primary goal is to help build intuitive appreciation for robust estimation that is conducive to sound practice. We do this by visualizing the target estimand and the estimation strategies. A second goal is to provide a ``menu'' of estimators that practitioners can choose from for the estimation of marginal natural (in)direct effects. The estimators generated from this exercise include some that coincide or are similar to existing estimators and others that have not previously appeared in the literature. We note several different ways to estimate the weights for cross-world weighting based on three expressions of the weighting function, including one that is novel; and show how to check the resulting covariate and mediator balance. We use a random continuous weights bootstrap to obtain confidence intervals, and also derive general asymptotic variance formulas for the estimators. The estimators are illustrated using data from an adolescent alcohol use prevention study. R-code is provided.
\end{abstract}


\tableofcontents

\end{frontmatter}

\section{Introduction}

Causal mediation methodology is complex. There are different types of causal contrasts: controlled direct effect,  natural (in)direct effects \citep{Robins1992,Pearl2001}, interventional (in)direct \citep{Didelez2006,VanderWeele2014a} and other interventional effects \citep{Nguyen2020MediationEstimands}, etc., each with their own set of identification assumptions \citep{nguyen2022ClarifyingCausalMediation}. The literature on effect estimation is vast, with a wide variety of estimation methods based on regression \citep[e.g.,][]{VanderWeele2010,Valeri2013,Muthen2015a}, weighting \citep[e.g.,][]{Hong2010,Hong2015,TchetgenTchetgen2013MediationOddsWeighting,Huber2014mediationweighting,VanderWeele2013a}, simulation \citep[e.g.,][]{Imai2010,Imai2010psychmethods,Vansteelandt2012a,VanderWeele2013a}, or some combination of these strategies. Further complicating the picture, some methods estimate marginal effects \citep{Imai2010psychmethods,Hong2010,Hong2015,Lange2012,VanderWeele2013a,TchetgenTchetgen2012} while others estimate effects conditional on covariates \citep{VanderWeele2010,Valeri2013,Steen2017,TchetgenTchetgen2014,TchetgenTchetgen2013MediationOddsWeighting}. Most of these methods are parametric and require all the models used to be correctly specified. Some methods have built in robustness to model misspecification; these are often presented in highly technical papers \citep[e.g.,][]{TchetgenTchetgen2012,Zheng2012}. It can be difficult for researchers to find their way through this literature and identify the estimation approach most appropriate for their application.

To help ease this task, this paper explicates a range of estimation options for causal mediation, focusing on options with some robustness properties. 
Rather than reviewing the complex and constantly growing methodological literature \citep[see e.g.,][]{Huber2020}, 
we take a concrete approach of using as inputs two strategies familiar to practitioners (weighting and regression) and a simple way of combining them, and show how to generate a range of estimators with different modeling requirements and robustness properties. The primary goal is to help build intuitive appreciation for robust estimation that is conducive to sound practice (without requiring prior understanding of these methods). This will benefit from the useful notion of \textit{pseudo samples}, as each weighting procedure can be interpreted as creating a certain meaningful pseudo sample. A secondary goal is to provide a ``menu'' of estimators that practitioners can choose from (depending on which modeling components they feel comfortable with given the specific application).

The paper focuses on \textit{natural (in)direct effects}. These decompose the total causal effect and (when identified) provide insight about effect mechanisms. This is a direct match to researchers' common motivation for conducting mediation analysis -- a wish to understand what part of a causal effect is indirect (operating through a specific intermediate variable) and what part is direct (not through that variable). The kind of reasoning used to build estimators here is not specific to these effects, but can also be applied to other effect types in causal mediation analysis, which we will comment on at the end of the paper. (For readers who require an orientation to different effect types, we refer to \cite{Nguyen2020MediationEstimands,nguyen2022ClarifyingCausalMediation} which discuss \textit{interventional} and natural effects, their relevance in practice, and their identification; and to \cite{robins2022InterventionistApproachMediation} which proposes \textit{separable effects}.)

We consider \textit{marginal} natural (in)direct effects. These effects, when defined on the additive scale, correspond to the total effect being the average treatment effect -- a popular effect in causal inference. Adaptation to average effects on the treated or on the controls is trivial. This paper does not address the estimation of conditional effects as functions of covariates, which entails a different set of estimation strategies that should be tackled separately.

As our construction of estimators is a bottom-up exercise, not all the estimators generated have appeared in the literature. 
We connect to work that employs, or is related to, the strategies and estimators discussed in this paper, and comment on the differences (some quite subtle) between some of these estimators. In addition to giving credit where credit is due, this aims to help the reader be a more informed consumer of the related literature.

To make the paper accessible to a broad audience, all proofs (about robustness properties, large-sample variance, and weight formulas) are placed in the Technical Appendix (Supplement 1). To facilitate application, R-code to implement the estimators is provided in the R-package \texttt{mediationClarity} (available at \url{https://github/trangnguyen74/mediationClarity}.

\section{Preliminaries}

\subsection{Effect definitions}

Consider the setting with a binary exposure $A$, followed in time by a mediator variable $M$ (which may be multivariate), followed in time by an outcome $Y$. We define effects using the potential outcome framework \citep{Rubin1974,Holland1986}.
The target estimands in this paper are \textit{marginal} natural (in)direct effects, which decompose the \textit{marginal} total effect.

On the additive scale, the marginal total effect is formally
$\mathrm{TE}:=\E[Y_1]-\E[Y_0]$,
the difference between the population mean of $Y_1$ (potential outcome if exposed to the active treatment) and that of $Y_0$ (potential outcome if exposed to the comparison condition). 
Definition of \textit{natural (in)direct effects} \citep{Pearl2001} additionally employs a nested potential outcome type, $Y_{aM_{a'}}$ (for a hypothetical condition with exposure set to $a$ and mediator set to its potential value under condition $a'$) where $a$ and $a'$ can be either 0 or 1. We assume that $Y_a=Y_{aM_a}$, thus
$\mathrm{TE}=\E[Y_{1M_1}]-\E[Y_{0M_0}]$. Using a third potential outcome with mismatched $a$ and $a'$, either $Y_{1M_0}$ (exposure set to the active treatment but mediator set to its potential value under control) or $Y_{0M_1}$ (the other way around), TE is decomposed in two ways, giving rise to two pairs of natural (in)direct effects:
$$\text{TE}=\underbrace{\E[Y_{1M_1}]-\E[Y_{1M_0}]}_{\textstyle\text{NIE}_1}~+~\underbrace{\E[Y_{1M_0}]-\E[Y_{0M_0}]}_{\textstyle\text{NDE}_0},$$
$$\text{TE}=\underbrace{\E[Y_{1M_1}]-\E[Y_{0M_1}]}_{\textstyle\text{NDE}_1}~+~\underbrace{\E[Y_{0M_1}]-\E[Y_{0M_0}]}_{\textstyle\text{NIE}_0}.$$
On multiplicative scales, marginal effects are ratios of marginal means or marginal odds of potential outcomes.
For example, the marginal total effect is $\E[Y_{1}]/\E[Y_0]$ on the mean/risk ratio scale and $\frac{\E[Y_{1}]/(1-\E[Y_{1}])}{\E[Y_0]/(1-\E[Y_{0}])}$ on the odds ratio scale; other effects are defined accordingly. On both scales, decomposition is by product instead of sum, $\text{TE}=\text{NDE}_0\times\text{NIE}_1$ and $\text{TE}=\text{NIE}_0\times\text{NDE}_1$.

Marginal effects on the additive scale are also average effects. The marginal additive TE is equal to the mean of the causal effect on the individual, $Y_1−Y_0$, and thus is usually known as the \textit{average treatment effect} in the non-mediation literature. Marginal additive natural (in)direct effects can also be seen as averages of effects on individuals. This interpretation does not apply to effects defined on multiplicative scales.

Note that each TE decomposition mentioned here includes only one indirect effect. In a situation where $M$ is a set of more than one mediator (as in our data example), this is the effect mediated by all the mediators combined. Alternatively, one may be interested in path-specific effects involving different mediators or subsets of mediators; that problem is outside the scope of this paper.

For conciseness, the rest of the paper addresses one of the two effect pairs: the $\text{NDE}_0$ and $\text{NIE}_1$ (also called the \textit{pure direct effect} and \textit{total indirect effect} \citep{Robins1992}). The other effect pair mirrors this one in all content covered here.

\subsection{Assumptions for effect identification}

As the current focus is estimation, we simply assume that the effects of interest are identified, noting that this is a matter for careful judgment in applications. By ``identified'' we mean that the effects, which are functions of \textit{potential} outcomes, can be equated (under certain assumptions) to some functions of the \textit{observed} data distribution. It is the latter that we will attempt to estimate. Below are the assumptions we make for ($\text{NDE}_0$, $\text{NIE}_1$) identification; for more detailed explication, see \cite{VanderWeele2009}, \cite{Imai2010}, \cite{Pearl2012} or our companion paper \cite{nguyen2022ClarifyingCausalMediation}.

\paragraph{\textit{Consistency.}} The first assumption is that there is \textit{consistency} between observed and potential outcomes or mediator values, and between potential outcomes of several types. Specifically,
\begin{align*}
    &Y=Y_a\text{ if }A=a,
    \\
    &Y=Y_{1m}\text{ if }A=1,M=m,
    \\
    &M=M_0\text{ if }A=0,
    \\
    &Y_a=Y_{aM_a},
    \\
    &Y_{1M_0}=Y_{1m}\text{ if }A=0,M=m,
\end{align*}
for $a$ being either 0 or 1, and $m$ being any mediator value. Essentially we have invoked this assumption in defining the effects above.

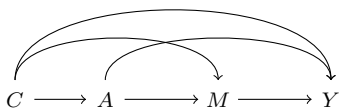
\begin{figure}[t!]
    \caption{Unconfoundedness holds when no mediator-outcome confounder is influenced by exposure and a set of observed pre-exposure covariates $C$ captures all confounding.}
    \label{fig:confounders}
    ~
    \begin{center}
    \begin{tikzpicture}[
        box/.style={rectangle, minimum size=5mm}
    ]
        \node[box]  (C)    {$C$};
        \node[box]  (A)  [right of = C, xshift=2mm] {$A$};
        \node[box]  (M)  [right of = A, xshift=5mm] {$M$};
        \node[box]  (Y)  [right of = M, xshift=5mm] {$Y$};
        
        \draw[->] (C) -- (A);
        \draw[->] (C) .. controls +(up:10mm) and +(up:10mm)  .. (M);
        \draw[->] (C) .. controls +(up:15mm) and +(up:15mm)  .. (Y);
        \draw[->] (A) -- (M);
        \draw[->] (A) .. controls +(up:10mm) and +(up:10mm)  .. (Y);
        \draw[->] (M) -- (Y);
    \end{tikzpicture}
    \end{center}
\end{figure}

\paragraph{\textit{Unconfoundedness.}}
The second assumption may be called \textit{ignorability}, \textit{exchangeability} or \textit{unconfoundedness}. This assumption requires that there is a set of observed pre-exposure covariates $C$ (where ``pre-exposure'' means either preceding exposure in time or simply not being influenced by exposure) that provides several conditional independence relationships. Specifically,
\begin{align*}
    &A\independent Y_a,Y_{1m},M_0\mid C,
    \\
    &M\independent Y_{1m}\mid C,A=1,
    \\
    &M_0\independent Y_{1m}\mid C,
\end{align*}
for $a$ being either 0 or 1, and $m$ being any value in the distribution of the mediator given covariates $C$ in the unexposed. The first two of the three elements above fit with the usual notion of ignorability, where once we condition on some variables the \textit{observed} exposure (or mediator value) does not carry any information about certain \textit{potential} variables. The last element is different in that it involves two potential variables ($M_0$ and $Y_{1m}$) in two different worlds (thus commonly known as the cross-world independence assumption).%
\footnote{This assumption is needed to identify natural effects (which are defined based on a hypothetical situation where exposure is set to \textit{one condition} but mediator is set to the value under \textit{the other condition}) but is not needed to identify interventional effects -- see \cite{nguyen2022ClarifyingCausalMediation}.}

In practice the usual way to deal with the unconfoundedness assumption is to ask (i) whether there is any mediator-outcome confounder (observed or not) that is influenced by exposure (often known as post-treatment confounder);%
\footnote{A post-treatment confounder $L$ results in violation of the cross-world independence assumption $M_0\independent Y_{1m}\mid C$, due to a backdoor path connecting $M_0$ and $Y_{1m}$ that is not blocked by $C$. This path is $M_0\leftarrow L_0\leftarrow U_L\rightarrow L_1\rightarrow Y_{1m}$, where $L_0,L_1$ are potential values of $L$ under exposure and nonexposure, $U_L$ represents the unique causes of $L$ that are not shared with $A,M,Y$ but are shared by $L_0,L_1$. For more thorough treatments of the no post-treatment confounder (or cross-world independence) assumption, see \cite{nguyen2022ClarifyingCausalMediation,VanderWeele2009,Imai2010,Pearl2012}.}
and if not, (ii) whether there is a set of pre-exposure covariates $C$ (all of which observed) that captures all exposure-mediator, exposure-outcome and mediator-outcome confounders. If either the answer to (i) is yes or the answer to (ii) is no, then the unconfoundedness assumption does not hold. Note though that while we can use substantive knowledge to judge the plausibility of these assumptions, these assumptions are not testable using data.

\paragraph{\textit{Positivity.}}
Since identification involves conditioning on covariates $C$, what is also required is that for all covariate levels there are positive chances of observing relevant potential mediator/outcomes. This is the third assumption, termed \textit{positivity}. Specifically,
\begin{align*}
    &\P(A=a\mid C)>0,
    \\
    &\P(M=m\mid C,A=1)>0,
\end{align*}
for $a$ being either 0 or 1, and $m$ being any value in the distribution of the mediator given covariates $C$ in the unexposed. The first element implies positive chances of observing $Y_1,Y_0,M_0$; both combined imply positive chances of observing $Y_{1m}$.

In more practical terms, the positivity assumption means that (i) the covariate range is the same in both the exposed and unexposed groups; and (ii) within each subpopulation homogeneous in covariates $C$, the range of $M$ in the exposed group covers the range of $M$ in the unexposed group.

\smallskip

Two quick notes before we proceed. First, the unconfoundedness (and accompanying positivity) assumptions above with a single covariate set $C$ are a simple version. Alternatively, different (yet overlapping) covariate sets could be used to deconfound the exposure-mediator, exposure-outcome and mediator-outcome relationships -- see details in \citep{nguyen2022ClarifyingCausalMediation}. In that case, the estimation methods discussed here need to be adapted, which is straightforward but involves complicated expressions, and thus is not included to keep the paper manageable. Second, the assumptions above point identify the effects of interest (described shortly). There are cases where we believe or are concerned that an assumption does not hold. For example, the no unobserved mediator-outcome assumption and the cross-world independence assumption are often questioned. In these cases, one strategy is to seek bounds for the effects based on the assumptions one is willing to make (e.g., \cite{miles2017PartialIdentificationNatural}), another is to conduct sensitivity analyses on the assumption that is likely violated (e.g., \cite{hong2018WeightingBasedSensitivityAnalysis,hong2021PostTreatmentConfoundingCausal,qin2021SimulationbasedSensitivityAnalysis,hong2021DidYouConduct,Imai2010,TchetgenTchetgen2012}). We will return to this point at the end of the paper.

\subsection{A heuristic view of identification that clarifies the estimation task}\label{sec:heuristic}

Identification of the $(\text{NDE}_0,\text{NIE}_1)$ pair amounts to identifying the means of the three potential outcomes $Y_1,Y_0$ and $Y_{1M_0}$. Under the assumptions above, the identification results \citep{Pearl2001, VanderWeele2009,Imai2010,Pearl2012} of these three means are
\begin{align*}
    &\E[Y_1]=\E_C\{\E[Y\mid C,A=1]\},
    \\
    &\E[Y_0]=\E_C\{\E[Y\mid C,A=0]\},
    \\
    &\E[Y_{1M_0}]=\E_C(\E_{M\mid C,A=0}\{\E[Y\mid C,M,A=1]\}),
\end{align*}
where the right-hand sides are functions of the \textit{observed} data distribution. 

\begin{figure}
    \caption{Heuristic visualization of the identification result}
    \label{fig:IDResult}
    ~\\
    \includegraphics[width=\linewidth]{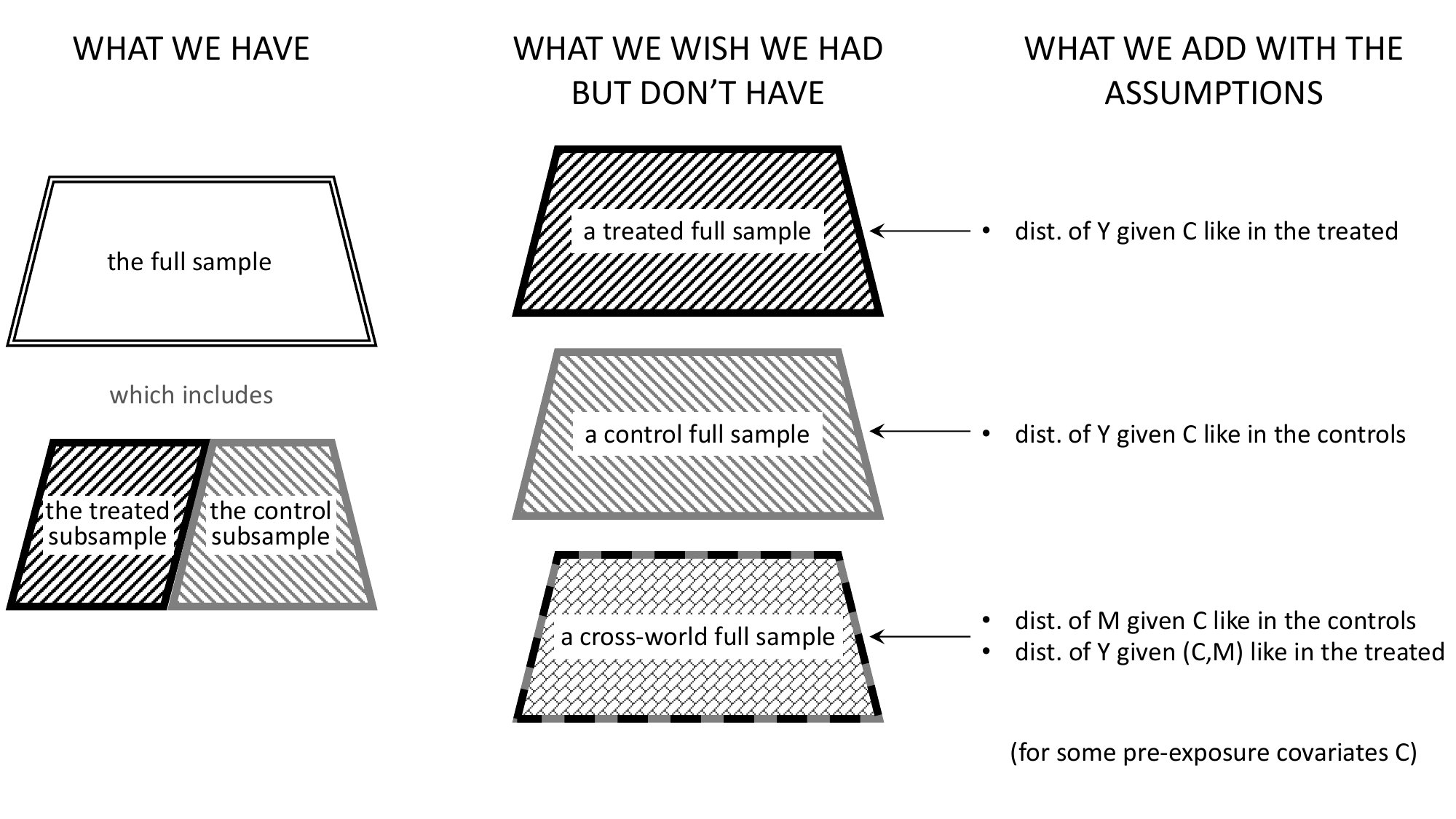}
\end{figure}

To make these results more intuitive to readers who may find them unfamiliar, we offer a heuristic visualization in Figure \ref{fig:IDResult}. 
This figure has three columns. The left column shows the data that we have: the full sample, which is comprised of the treated subsample and the control subsample. As we are interested in the $\text{NDE}_0$ and $\text{NIE}_1$ that together contrast the means of $Y_1,Y_0,Y_{1M_0}$, what we would ideally like to have instead is shown in the middle column: three full samples that all resemble the actual full sample pre-exposure, but are then set to three conditions: the treated (1) condition, the control (0) condition, and the cross-world condition (where exposure is set to 1 but mediator is set to $M_0$); this would allow us to average the outcome in the three samples to estimate the three potential outcome means. Unfortunately, we do not observe these three full samples. To remedy the situation, we invoke the assumptions above, which give us the additional information in the right column: in such a treated (control) full sample, the outcome distribution given $C$ would be the same as that in the observed treated (control) units; and in such a cross-world full sample, the mediator distribution given $C$ would be the same as that in the observed control units, while the outcome distribution given $(C,M)$ would be the same as that in the observed treated units.

This sheds light on the estimation puzzle we need to solve. If we take the obvious approach of estimating the three potential outcome means, the task of estimating $\E[Y_1]$ (or $\E[Y_0]$) would be a puzzle of obtaining the outcome mean for a hypothetical full sample with the distribution of $C$ from the actual full sample and the outcome distribution given $C$ from the actual treated (control) units. The task of estimating $\E[Y_{1M_0}]$ would be another puzzle of obtaining the outcome mean for a hypothetical sample with the distribution of $C$ from the actual full sample, the mediator distribution given $C$ from the control units, and the outcome distribution given $(C,M)$ from the treated units. This is what is conveyed in the identification results stated above.

\subsection{Preview of approaches and strategies for effect estimation}

This paper makes use of two main tools: weighting and model-based prediction. We will first (in Section \ref{sec:weighting}) consider weighting, which can be used as the sole puzzle-solving tool. Here we focus on conceptual clarity of the more complicated cross-world weighting component (including a new simple view based on a novel expression of the weight), methods for weight estimation, and balance checking. Then (in Section \ref{sec:estimatorpairs}) we bring in the model-based prediction tool and examine pairs of estimators of potential outcome means, where each pair includes a nonrobust estimator (requiring all models to be correctly specified) and a more robust estimator (allowing some model misspecification). Here the robustness is due to strategic incorporation of weighting. Section \ref{sec:additivescale} addresses the specific case of effects defined on the additive scale where there is an alternative view of the puzzle, obtaining a pair of estimators of the natural direct effect. 
We do our best throughout to reference work that employs or is related to the strategies mentioned.
Given these various estimators, Section \ref{sec:howtochoose} discusses considerations in choosing an estimator.

With respect to interval estimation (see Section \ref{sec:Bootstrap}), we use a bootstrap procedure to obtain confidence intervals that applies to all the estimators discussed. We also derive general formulas for the asymptotic variance of the estimators.%
\footnote{This is based on parametric specification of components of the model that need to be estimated, e.g., $\P(A\mid C)$, $\P(A\mid C,M)$, $\E[Y\mid C,M,A=1]$, etc. depending on the estimator.}

\subsection{Illustrative example}

We illustrate the estimators using a synthetic dataset generated to mimic real data from The Prevention of Alcohol Use in Students (PAS) trial in the Netherlands. 
In the real trial, middle schools were randomized to one of four conditions: student intervention (promoting healthy attitudes and strengthening refusal skills), parent intervention (encouraging parental rule setting), student and parent combined intervention, and control condition (regular biology curriculum covering effects of alcohol). The combined intervention was effective in reducing drinking onset \citep{Koning2009,Koning2011} and drinking frequency \citep{Koning2009}, and \cite{Koning2010} found that student attitudes towards alcohol, perceived self-control in situations involving alcohol, and student-reported parental rules about alcohol mediated the effect of the combined intervention on onset of weekly drinking. Our analysis of synthetic data considers the effect of the combined intervention relative to control on weekly drinking at 22 months, with the same mediators measured at six months.

The real sample consists of students clustered in schools, and has missing data on covariates, mediators and outcome. As our purpose is to illustrate a range of estimators, not to draw inference on the trial, we ignore the clustering for simplicity, complete the dataset with a single imputation (based on data observed at baseline, six, twelve and 22 months), and use this as the basis to create a synthetic dataset. The imputation and synthesization used the R-packages \texttt{mice} \cite{mice} and \texttt{synthpop} \cite{synthpop}, respectively, and both are nonparametric (using classification and regression trees).
All estimation outputs are specific to the synthetic dataset, and should not be interpreted as results of the original study.

\section{Weighting to create pseudo samples}\label{sec:weighting}

Let us first examine one of the two tools we set out to use, that of weighting. This tool can be used by itself to estimate the effects of interest: we would weight data to create pseudo samples that stand in for the ideal treated, control and cross-world full samples we wish we had (see Figure \ref{fig:IDResult}), average the outcome in those pseudo samples to estimate the potential outcome means, and then contrast those means to estimate the effects. Such an estimator is consistent if the weights are consistently estimated. It tends to have large variance, and may suffer from high influence of observations with large weights. An important value of weighting, though, is that it can also be used in combination with regression-based techniques (as we shall see in Section \ref{sec:estimatorpairs}) for more precise and robust estimation. It is therefore important to clarify how the weighting is done.

\subsection{The pseudo treated and control samples}

These pseudo samples are obtained by weighting treated units and control units to mimic the full sample covariate distribution, using the well-known inverse probability weights, $\omega_1(C)=\frac{1}{\P(A=1\mid C)}$ for treated units and $\omega_0(C)=\frac{1}{\P(A=0\mid C)}$ for control units. 
These weights are commonly estimated via propensity score \citep{Rosenbaum1983} modeling. With such indirect estimation, it is common practice to check covariate balance and possibly adjust the model to achieve good balance. We will use probability models to estimate weights, and it is most familiar.

An alternative approach is to estimate the weights directly, finding weights that reduce the difference between the full and pseudo samples' covariate distributions. For example, several methods (e.g., entropy balancing \cite{Hainmueller2012} and covariate-balancing propensity score \citep{Imai2014}) directly target balance on covariate moments specified by the user, and another method \citep{Huling2020} minimizes a measure of distance between multivariate distributions called \textit{energy distance} \citep{Szekely2013}.

\subsection{The pseudo cross-world sample}

This pseudo sample is obtained by weighting treated units to mimic the $C$ distribution in the full sample and the $M$ given $C$ distribution in control units. It stands in for the hypothetical full cross-world sample that we wish we had: in addition to these two elements, it retains its original $Y$ given $(C,M)$ distribution (which is that of treated units).
Denote the weights that form the pseudo cross-world sample out of treated units by $\omega_\text{x}(C,M)$.
These weights have several equivalent expressions that point to several ways they may be estimated.

\subsubsection{Three expressions (and views) of the cross-world weights}

The first expression of $\omega_\text{x}(C,M)$ builds on the inverse probability weights $\omega_1(C)$, which weight treated units to the full sample with respect to the covariate distribution, in a sense doing half of the job. Such weighting does not change the mediator given covariates distribution (which is the distribution of $M_1$ given $C$). To morph this distribution to mimic the $M_0$ given $C$ distribution, we use density ratio weighting (or probability ratio if the mediator is discrete) with the weighting function $\frac{\P(M\mid C,A=0)}{\P(M\mid C,A=1)}$, where the numerator and denominator are the densities (or probabilities) of the observed mediator value $M$ conditional on $C$ and on $A=0$ and $A=1$, respectively. This weighting scheme was proposed by Hong (2020) \cite{Hong2010} \citep[see also][]{Hong2015}. 
Thus we have
\begin{align}
    \omega_\text{x}(C,M)=\frac{1}{\P(A=1\mid C)}\frac{\P(M\mid C,A=0)}{\P(M\mid C,A=1)}.
\end{align}

A second expression is due to the fact that by Bayes' rule the ratio of mediator densities above is equal to the ratio of two odds of exposure, $\frac{\P(A=0\mid C,M)}{\P(A=1\mid C,M)}\Big/\frac{\P(A=0\mid C)}{\P(A=1\mid C)}$ (noted by Zheng et al., 2012 \cite{Zheng2012}). The resulting expression,
\begin{align}
    \omega_\text{x}(C,M)=\frac{\P(A=0\mid C,M)}{\P(A=1\mid C,M)}\frac{1}{\P(A=0\mid C)},\label{wtform2}
\end{align}
(which appears in an identification result in \cite{Huber2014mediationweighting}) is the product of two terms: an odds weight%
\footnote{Side note: This odds weight component also appears as part of the weight formula in Tchetgen Tchetgen et al.'s inverse odds ratio weighting method \cite{TchetgenTchetgen2013MediationOddsWeighting} for estimation of the conditional natural direct effect. Both there and here, the role of weighting by this odds is to construct a weighted outcome distribution that reflects the distribution of the cross-world potential outcome given covariates.}
and an inverse probability weight. 
This formula provides another interpretation of the weighting: it could be thought of as first morphing the treated subsample to mimic the joint distribution of $(C,M)$ in the control subsample (this is what odds weighting does), and then morphing the $C$ distribution (which now reflects the distribution under control) to mimic that in the full sample (this is what inverse probability weighting does).

\begin{figure}[htbp]
    \caption{Different views of cross-world weighting via alternative expressions of the weight function}
    \includegraphics[width=\linewidth]{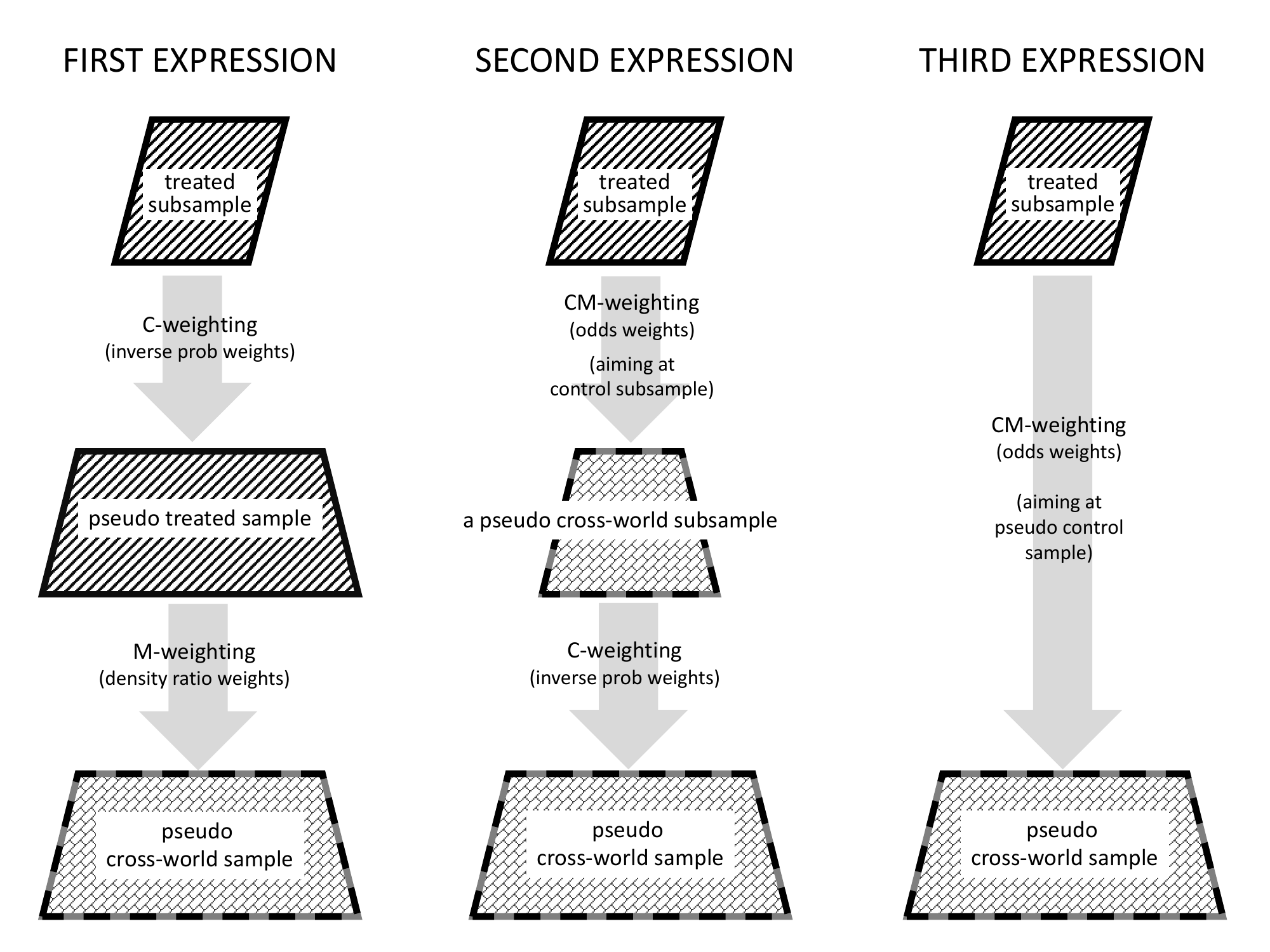}
    \label{fig:crossworldWeighting}
    \\~
    \\
    \captionof{figure}{Desired balance when using weighting to estimate the NDE\textsubscript{0}, NIE\textsubscript{1} pair: covariate balance among all three pseudo samples and full sample, and covariate-and-mediator balance between pseudo cross-world and control samples.}
    \includegraphics[width=.7\linewidth]{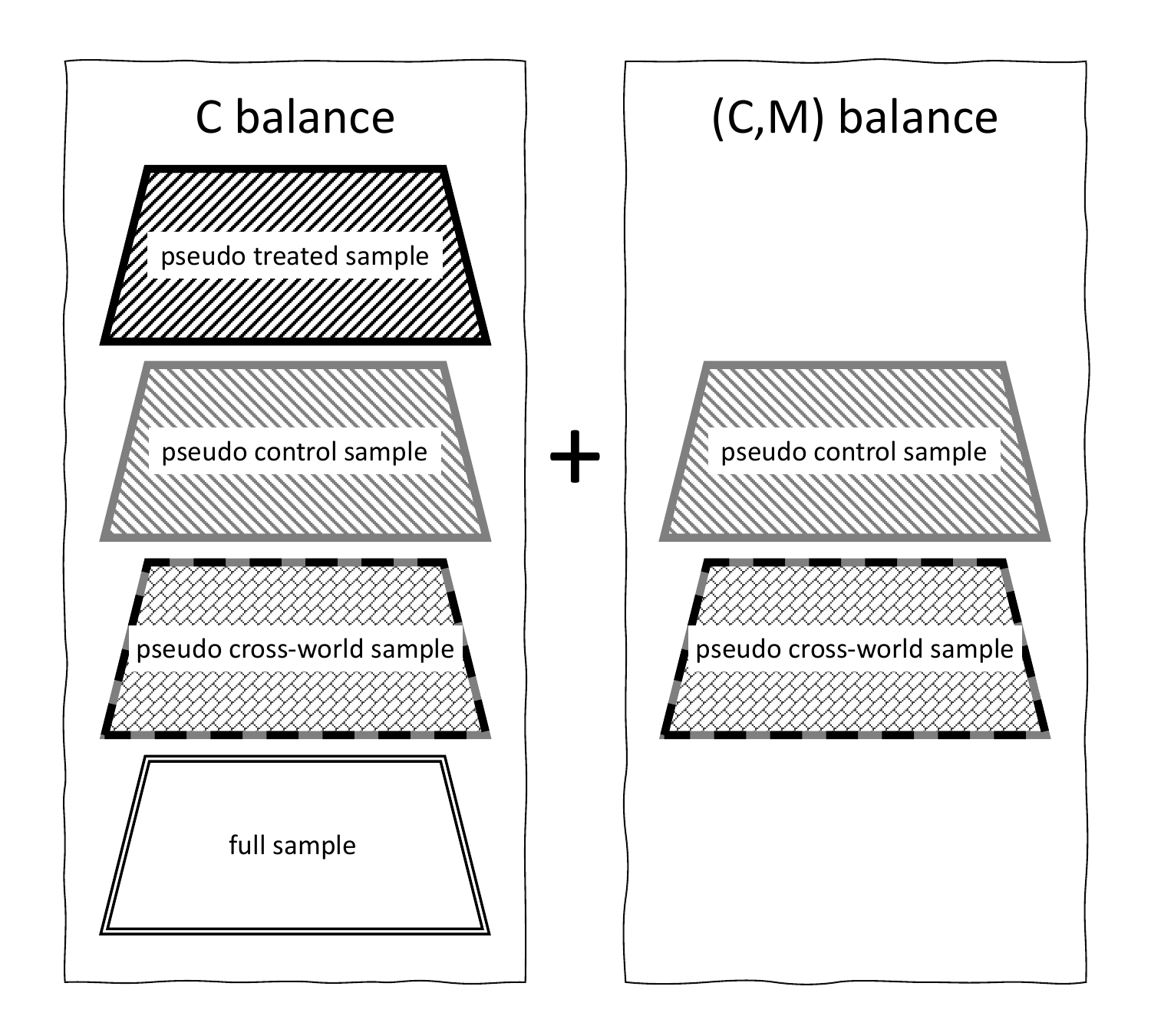}
    \label{fig:desiredbalance}
\end{figure}

In addition, we found a novel third expression (see derivation in the Appendix). This expression is best viewed in its version for stabilized weights. The $\omega_\text{x}(C,M)$ weights in treated units have mean equal to $\frac{1}{\P(A=1)}$; stabilized weights are simply $\omega_\text{x}(C,M)$ scaled down to mean 1 by multiplying with $\P(A=1)$. The third expression is
\begin{align}
    \omega_\text{x}^\text{stabilized}(C,M)=\frac{\P(C,M\mid A=0)\frac{\P(A=0)}{\P(A=0\mid C)}}{\P(C,M\mid A=1)},
\end{align}
which could be seen as the ratio of two densities of $(C,M)$: the denominator is the density in the treated subsample, and the weighted density in the numerator turns out to be the density in the pseudo control sample. That is, the weighting morphs the treated subsample such that it mimics the joint $(C,M)$ distribution in the pseudo control sample. This makes sense, as the pseudo control sample has the $C$ distribution of the full sample and the $M$ given $C$ distribution of control units -- two of the three features desired for the pseudo cross-world sample. 

\subsubsection{Estimation of the cross-world weights}

The first two of the expressions for $\omega_\text{x}(C,M)$ above can be used directly as formulas for estimation purposes. In addition to the propensity score model, we fit either two mediator density models, $\P(M\mid C,A=0)$ and $\P(M\mid C,A=1)$ (if using the first formula), or a model for exposure given covariates and mediators, $\P(A\mid C,M)$ (if using the second formula); and plug the estimated elements into the formula.
Between these two methods, the first one has the appeal that the models are variationally independent,%
\footnote{This weights estimation method uses models for $\P(A\mid C)$, $\P(M\mid C,A=1)$ and $\P(M\mid C,A=0)$, which correspond to a factorization of the likelihood. These model components thus do not put constraints on one another.}
but its disadvantage is that density estimation is generally a harder problem than mean estimation (and especially so for a non-binary or multivariate mediator).
The second method requires fitting fewer models (only two) and they are conditional mean models.

With the third expression of $\omega_\text{x}(C,M)$, rather than treating it as an estimation formula (which would require estimating conditional densities of $(C,M)$), we can use the insight it provides -- to weight the treated subsample to mimic the $(C,M)$ distribution in the pseudo control sample -- and note that this can be achieved by odds weighting. This means stacking the treated subsample with the pseudo control sample, fitting a model for $A$ given $C,M$ to the stacked data, and computing $\omega_\text{x}(C,M)$ as the model-predicted odds of being in the pseudo control sample rather than the treated subsample.  (This is just another instance of the connection between density ratios and odds of group membership.) This method also requires only two conditional mean models. Figure \ref{fig:crossworldWeighting} visualizes these three weights estimation methods.

For readers who wish to use direct weights estimation tools such as moments balancing or distance minimizing (rather than relying on probability models), a couple of notes. First, the third expression of $\omega_\text{x}(C,M)$ provides a simple and elegant way to use such tools: seek weights that morph the treated subsample to mimic the pseudo control sample with respect to the joint $(C,M)$ distribution.%
\footnote{Direct weights estimation seeks to directly mimic a target distribution, thus requires data reflecting that distribution. The pseudo control sample reflects the target $(C,M)$ distribution.}
Second, while the second expression of $\omega_\text{x}(C,M)$ suggests that direct weights estimation can be used for two-step weighting (first mimicking the control subsample's $(C,M)$ distribution, then mimicking the full sample's $C$ distribution), we do not recommend this, as this zigzag weighting may result in unnecessary loss of samples and suboptimal weights.%
\footnote{This weighting scheme is zigzag in the sense that the first step overshoots the target $C$ distribution, as the full sample $C$ distribution is \textit{in between} those in the two subsamples. Therefore the first step may give very small weights to (or even drop) some observations (especially if the treated and control subsamples are dissimilar), which means those observations are essentially lost to the second step.}

\subsection{Balance checking}

With the three pseudo samples, the desired balance includes two components (see Figure \ref{fig:desiredbalance}). The first component is \textit{covariate balance} between the three pseudo samples and the full sample as well as among the three pseudo samples. The second component is the \textit{covariate-and-mediator balance} between the pseudo cross-world sample and the pseudo control sample.

This full balance is important when using weighting as the pure estimation strategy, i.e., the effects are estimated by contrasting the outcome means from the pseudo samples. For some of the other estimators in the next section, certain elements of balance (which we will note) are crucial as they relate directly to the estimator's consistency, while other elements are in a sense of secondary importance as they serve mainly to induce robustness.

\section{Estimating potential outcome means: pairs of nonrobust and more robust estimators}\label{sec:estimatorpairs}

The weighting above gives us one solution to the puzzle described in Section \ref{sec:heuristic}. We can simply average the outcomes in the pseudo samples and contrast the averages to obtain estimates of the total and natural (in)direct effects. We call this the \textit{pure weighting} estimator. This estimator is consistent only if the three weight functions are consistently estimated.

We now explore several other solutions to the puzzle, using our second tool, model-based prediction, either alone or in combination with weighting. These solutions are estimators of the means of $Y_0$ and $Y_1$ (which we refer to as \textit{regular} potential outcomes) and of the \textit{cross-world} potential outcome, which are to be combined to estimate marginal effects on either additive or multiplicative scale.

We present these potential outcome mean estimators in pairs. Each pair consists of a simple estimator that does the minimum needed to solve the puzzle, and a more complex estimator built on the simple one that is more robust as it provides some protection against model misspecification. Our explanations of robustness properties here strive for simple language; proofs for all estimators (here and in the next sections) are provided in the Technical Appendix.

As the paper touches on many estimators, a labeling system is needed. We use labels with two parts separated by ``|'', where the front part signals what is being estimated (e.g., ``reg'' and ``crw'' for regular and cross-world potential outcome means), and the back part signals the estimation method. Within each pair, the more robust estimator is distinguished from the nonrobust one by adding ``MR'' or ``R'' to the back label (the difference between these will be clear shortly). When referring to a pair, we use the base label (without MR or R).

\subsection{Regular potential outcome means}\label{sec:regEstimation}

We start with a single pair of estimators for the regular potential outcome means. While these may be broadly familiar, we will take time in motivating them and clarifying ideas that will later apply in constructing estimators for the cross-world potential outcome mean.

\subsubsection{reg|Ypred: outcome prediction given covariates}

\subsubsection*{The simple version (reg|Ypred)}
Recall from Figure \ref{fig:IDResult} that if we were to observe a full treated/control sample, it would have the same outcome distribution given $C$ as that in the actual treated/control units (under our assumption of unconfoundedness given $C$). This means we can learn this distribution from the treated/control subsample and apply it to the full sample.
We thus fit a model regressing $Y$ on $C$ in the treated subsample, use this model to predict $Y_1$ for every individual in the full sample, and average the predicted values over the full sample to estimate $\E[Y_1]$. For $\E[Y_0]$, we fit the model to the control subsample and use it to predict $Y_0$.%
\footnote{A variant is to combine observed $Y_1$ and $Y_0$ values (in treated and control units, respectively) with predicted $Y_1$ and $Y_0$ values (for control and treated units, respectively).}\label{fn:1}

The two models here are models for $\E[Y∣C,A=1]$ and for $\E[Y∣C,A=0]$. Below we often refer to them collectively as $\E[Y∣C,a]$ (keeping $a=0,1$ implicit); this is just an abbreviation as the key is that these are models that allow predicting two different variables, $Y_1$ and $Y_0$. Instead of fitting the models separately, we can also fit a joint model regressing $Y$ on $C,A$. Separate model fitting has the advantage that it allows tailoring to the subsamples while avoiding the risk of (conscious or unconscious) fishing for a desired treatment effect estimate.

This simple estimator is nonrobust. If the outcome models are misspecified, predictions may be poor, leading to estimation bias. The problem may be exacerbated by extrapolation if the covariate distributions of the subsamples (to which the models are fit) differ substantially from that of the full sample (on which outcomes are predicted).

\begin{figure}[t!]
    \centering
    \caption{A pair of regression-based estimators of regular potential outcome means, depicted as targeting $\textup{E}[Y_1]$. * indicates that the model is required to satisfy the mean recovery condition.}
    \label{fig:regPair}
    ~\\
    \includegraphics[width=.85\linewidth]{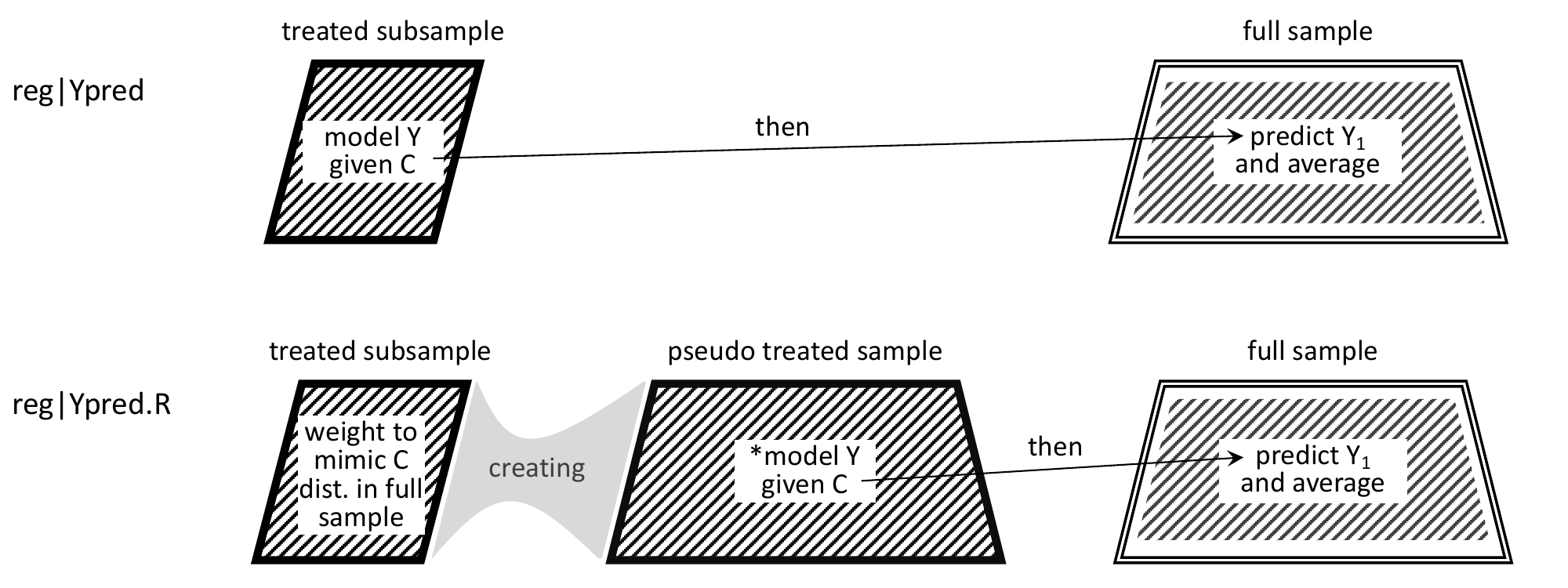}
\end{figure}

\subsubsection*{The doubly robust version (reg|Ypred.R)}

There is a class of estimators that are doubly robust. They combine outcome models and inverse probability weights, and are consistent if one of these two components (but not necessarily both) is correct. Many such estimators exist \citep[see ][]{Kang2007,Robins2007}. We consider one estimator \citep{Robins2007,Vansteelandt2011} based on a strategy that readily extends to the later estimation tasks in the paper.

Like the simple estimator reg|Ypred, this robust estimator reg|Ypred.R relies on predicting $Y_1$ and $Y_0$ for the full sample and averaging predicted values (Figure \ref{fig:regPair}). However, there is a key difference between the two estimators and a mild technical requirement imposed on the robust estimator. The key difference is that the outcome models used for prediction are fit to the \textit{pseudo treated/control samples} instead of the subsamples, i.e., weighted regression models are used.%
\footnote{Related to the nonrobust variant in footnote 5, the corresponding robust variant here would predict $Y_1$ for control units based on an outcome model fit to a weighted treated subsample that mimics the control subsample (using odds weights $\frac{\P(A=0\mid C)}{\P(A=1\mid C)}$), and predict $Y_0$ for treated units based on an outcome model fit to a weighted control subsample that mimics the treated subsample (using odds weights $\frac{\P(A=1\mid C)}{\P(A=0\mid C)}$).}
The technical requirement is that the outcome models satisfy a condition we label \textit{mean recovery}: in the sample to which the model is fit (here a pseudo sample), the average model-predicted outcome equals the average observed outcome \citep[][equation 8]{Robins2007}. Due to these two features combined, the estimator is doubly robust. We offer some intuition about these two points.

First, there is a simple rationale for fitting models to pseudo samples. Generally we do not know the true model that generated the data, so all models we use are just approximations of the true model. One way to improve the approximation (other than using flexible models to reduce misspecification) is to fit the model to the same covariate space on which it will be used for prediction; this is a guard against extrapolation. Compared to the treated/control subsamples, the pseudo samples have covariate distributions that are (at best) the same as or (at least) closer to that of the full sample. Fitting models to the pseudo samples is thus an improvement over the simple prediction estimator.

Second, the technical mean recovery condition serves to make sure that even if the predicted outcome values may be biased, they would be \textit{on average unbiased} (if the weights that form the pseudo samples are correct). This condition is satisfied by generalized linear models with canonical link and an intercept (e.g., the usual linear regression, logistic regression, Poisson regression), which is the option we will use. Note that this is not the only choice. For example, an estimating equations approach may accommodate other link functions while satisfying this condition. Or if the outcome model is fit by machine learning, this condition may be achieved using targeted maximum likelihood estimation (TMLE) \citep{vanderLaan2011targeted}. These topics are outside the scope of the current paper.

\begin{figure}[t!]
    \caption{Four pairs of estimators of the cross-world potential outcome mean $\textup{E}[Y_{1M_0}]$. Outcome models used for MR/R estimators (marked with *) are required to be mean-recovering.}
    \label{fig:crossPairs}
    ~\\
    \includegraphics[width=1.02\linewidth]{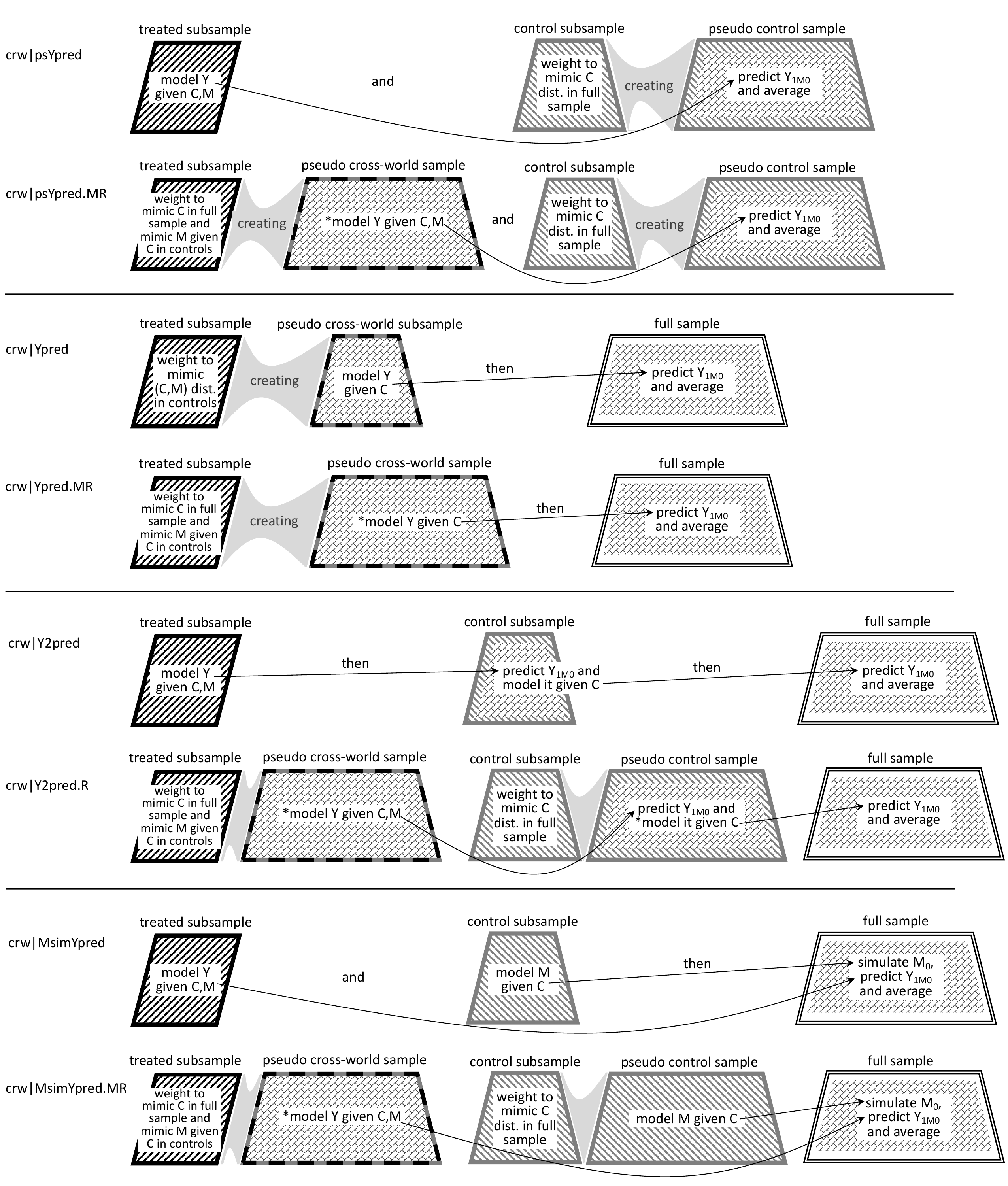}
\end{figure}

\medskip

\subsection{Cross-world potential outcome mean}

Now we turn to the cross-world potential outcome. There are a range of strategies for estimating its mean. This is because the task requires combining several pieces of information from the full sample and from the treated and control conditions -- recall that the cross-world sample in Figure \ref{fig:IDResult} has the $C$ distribution of the full sample, the $M$ given $C$ distribution of the controls, and the $Y$ given $C,M$ distribution of the treated -- and there are different ways those pieces could be obtained and combined. We present four pairs of nonrobust and more robust estimators. The more robust estimators differ from their nonrobust counterparts in that they fit certain models to relevant pseudo samples instead of subsamples, and that outcome models used for them are required to satisfy the mean recovery condition mentioned above. We explain the estimators' robustness properties and point out their nonrobustness if any. To aid explanation and provide a clear view, a visual representation of these four estimator pairs is provided in Figure \ref{fig:crossPairs}. Also, Table \ref{tab:regcrossSteps} lists the steps for implementing them.

\subsubsection{crw|psYpred: outcome prediction given covariates and mediators on pseudo control sample}

These estimators are anchored on the pseudo control sample, where the mediator is $M_0$, and (if the weights are correct) the $C$ distribution mimics that of the full sample. This gives us two of the three required pieces of information. The $Y$ given $C,M$ distribution, however, is off because it is that of control units. To complete the puzzle, we replace the observed outcome with predicted $Y_{1M_0}$ given the units' $(C,M)$ values, where the prediction is based on a model for $\E[Y\mid C,M,A=1]$. We then average these predicted $Y_{1M_0}$ values over the pseudo control sample to estimate $\E[Y_{1M_0}]$.
(These are the only estimators in the paper that average predicted outcome \textit{over a pseudo sample}, hence the `ps' in the label.)

With the \textit{nonrobust} estimator  in this pair, the $\E[Y\mid C,M,A=1]$ model is estimated by regressing $Y$ on $(C,M)$ in the treated subsample.
For this estimator to be consistent, the control weights $\omega_0(C)$ have to be consistently estimated and this outcome model has to be correctly specified.

Relating to the literature, this nonrobust estimator constitutes part of VanderWeele and Vansteelandt's weighting-based estimator for the multiple mediator setting \cite{VanderWeele2013a}. Albert \cite{Albert2012} employs this strategy -- using a model fit in one exposure condition to predict the cross-world outcome on units in the other condition and weighting those units to standardize the covariate distribution -- but for a general target population (a generalization of our current purpose). Also, this strategy of predicting the cross-world outcome on a pseudo sample is one of the methods used in the R-package medflex \cite{Steen2017} as a first step in estimating effects conditional on covariates.

The \textit{more robust} estimator crw|psYpred.MR fits the outcome model (that satisfies the mean recovery condition) to the pseudo cross-world sample instead of the treated subsample. Like its nonrobust sibling, this estimator is not consistent if the $\omega_0(C)$ weights are not consistent. But if they are, then crw|psYpred.MR is consistent if either the outcome model is correctly specified or the cross-world weights $\omega_\text{x}(C,M)$ are consistent. That is, crw|psYpred.MR has two chances to be correct, while crw|psYpred has only one. 
Here we use the MR (more robust) suffix (instead of simply R) to signal that although this estimator is more robust than its nonrobust sibling, it depends on one estimation component (here the control weights) being correct. 

Because these two estimators average predicted outcome over the pseudo control sample, they depend on the covariate distribution of the pseudo sample mimicking the full sample well. Therefore, when using either of these estimators, it is crucial to obtain good covariate balance between the pseudo control sample and the full sample.

\smallskip

While these two estimators rely on outcome prediction, they are also weighted estimators (as predictions are averaged over weighted control units), and weighted estimators may have large variance due to the variability of the estimated weights. An alternative strategy is to find a way to predict $Y_{1M_0}$ on the full sample instead of on the pseudo control sample. The next three pairs of estimators do this in different ways.

\subsubsection{crw|Ypred: outcome prediction given covariates}

In the full sample, we cannot predict $Y_{1M_0}$ based on observed covariates and mediators (because the observed $M$ is a mixture of both $M_0$ and $M_1$). Instead, this pair of estimators relies on $Y_{1M_0}$ prediction based on covariates only. To do this, we need a model that informs of the mean of the cross-world potential outcome given covariates, $\E[Y_{1M_0}\mid C]$. 

The trick employed by the simpler estimator in this pair is to weight the treated subsample to mimic the control subsample with respect to the joint distribution of $(C,M)$ (using odds weights $o_\text{x}(C,M)=\frac{\P(A=0\mid C,M)}{\P(A=1\mid C,M)}$, which can be estimated based on a model for exposure given covariates and mediator).%
\footnote{An alternative is to mimic the $M$ given $C$ distribution using density ratio weights $\frac{\P(M\mid C,A=0)}{\P(M\mid C,A=1)}$; we do not recommend this because there is no simple way to check balance on conditional distributions.}
This weighted subsample (which we call a pseudo cross-world \underline{sub}sample) has two of the three desired features of the ideal cross-world full sample: the $M$ given $C$ distribution (like that in the controls) and the $Y$ given $(C,M)$ distribution (like that in the treated). That means the $Y$ given $C$ distribution is like that in the ideal cross-world full sample -- the distribution of $Y_{1M_0}$ given $C$. We thus fit to this pseudo subsample a model regressing outcome on covariates to estimate $\E[Y_{1M_0}\mid C]$. (To simplify language, we loosely call this model the $\E[Y_{1M_0}\mid C]$ model.) Based on this model, we predict $Y_{1M_0}$ in the full sample and average the predicted values to estimate $\E[Y_{1M_0}]$.
This estimator is \textit{nonrobust}. For it to be consistent, the weights $o_\text{x}(C,M)$ have to be consistently estimated and the outcome model has to be correctly specified.

The \textit{more robust} estimator crw|Ypred.MR, on the other hand, fits an outcome given covariates model (that satisfies the mean recovery condition) to the pseudo cross-world sample (instead of the cross-world subsample above).
This estimator has two chances to be correct: (1) if the weights $\omega_\text{x}(C,M)$ are consistently estimated, crw|Ypred.MR is consistent even if the outcome model is misspecified; and (2) if the outcome model is correctly specified and only \textit{the mediator-related part} of $\omega_\text{x}(C,M)$ is correct, crw|Ypred.MR is consistent. 

To clarify the second case, the mediator-related part of the weights is the term that controls the mediator distribution in the pseudo cross-world sample. It varies by the weights estimation method: in the first and second methods it is the $\frac{\P(M\mid C,A=0)}{\P(M\mid C,A=1)}$ and $\frac{\P(A=0\mid C,M)}{\P(A=1\mid C,M)}$ terms, respectively; in the third method it is the odds of being in the  pseudo control sample rather than the treated subsample (where the pseudo control sample may be incorrectly weighted). When only the mediator-related part of the weights is correct, the weighting gets the $C$ distribution wrong but gets the $M$ given $C$ distribution right, and it is the latter that ensures that the treated units' $Y$ given $C$ distribution is appropriately morphed to resemble the target $Y_{1M_0}$ given $C$ distribution. As the outcome regression model conditions on $C$, it is (if correctly specified) not affected by the incorrectly weighted $C$ distribution.

Intuitively, both of these estimators rely completely on the weighting to obtain data that reflect the distribution of $Y_{1M_0}$ given $C$ (via getting the $M$ given $C$ distribution right). It is thus crucial to achieve good balance, specifically $(C,M)$ balance between the pseudo cross-world subsample and the control subsample (for the nonrobust estimator) or between the pseudo cross-world sample and the pseudo control subsample%
\footnote{Strictly speaking, the construction of the pseudo control subsample is not required to obtain a crw|Ypred estimate of $\E[Y_{1M_0}]$ (see Figure \ref{fig:crossPairs}). However, this is likely not additional work because the pseudo control subsample is already constructed for estimating $\E[Y_0]$.}
(for the more robust estimator).

\subsubsection{crw|Y2pred: outcome prediction based on double model fit}

These estimators also rely on predicting $Y_{1M_0}$ in the full sample based on a model that estimates $\E[Y_{1M_0}\mid C]$. Here this model is estimated in two steps: first fitting a $\E[Y\mid C,M,A=1]$ model and using it to predict $Y_{1M_0}$ in control units, then regressing the predicted $Y_{1M_0}$ on covariates to estimate $\E[Y_{1M_0}\mid C]$. We loosely refer to these two regression models with one building directly on the other as a \textit{double outcome model fit} (or iterated regression).%
\footnote{This iterated regression procedure is a straightforward implementation, by regression, of the double expectation in the identification result.}

With the \textit{nonrobust} estimator crw|Y2pred, these two models are fit to the treated and control subsamples, respectively. For this estimator to be consistent, both models have to be correctly specified.

The \textit{robust} estimator crw|Y2pred.R fits the outcome models to the pseudo cross-world and pseudo control samples instead; and both models are required to satisfy the mean recovery condition.
This estimator has three chances to be consistent: (1) both of the outcome models are correctly specified; (2) both the $\omega_\text{x}(C,M)$ and $\omega_0(C)$ weights (that form pseudo cross-world and pseudo control samples) are consistently estimated; or (3) the $\omega_\text{x}(C,M)$ weights are consistently estimated and the $\E[Y\mid C,A=1,M]$ model is correctly specified.
Notice that we call crw|Y2pred.R \textit{robust} (rather than \textit{more robust}) to signal that this estimator does not depend on any specific estimation component being correct. 

A technical note: crw|Y2pred.R is a multi-step estimator based on the nonparametric influence function. As such, it has similar robustness properties to Tchetgen Tchetgen and Shpitser's estimator which solves the nonparametric influence function estimating equation \cite{TchetgenTchetgen2012}. The appeal of crw|Y2pred.R is that the steps are intuitively meaningful without requiring knowledge of influence function theory.

\begin{table}[t!]
    \caption{Implementation steps of estimators of potential outcome means in Sections \ref{sec:weighting}, \ref{sec:estimatorpairs} and \ref{sec:additiveNDE}. nonR, MR and R stands for nonrobust, more robust and robust, respectively.}
    \label{tab:regcrossSteps}
    \resizebox{\linewidth}{!}{%
    \begin{tabularx}{1.15\linewidth}{ X }
        \toprule
        \textbf{Purely weighting}: nonR estimator
        \begin{enumerate}[itemsep=0pt, topsep=0pt]
            \item Estimate $\omega_1(C)$, $\omega_0(C)$ and $\omega_\text{x}(C,M)$ weights
            \item Average the observed outcome in the pseudo control, pseudo treated and pseudo cross-world samples
        \end{enumerate}
        \\[-1em]\toprule
        \textbf{reg|Ypred pair}: nonR and R estimators
        \begin{enumerate}[itemsep=0pt, topsep=0pt]
            \item For the R estimator, estimate $\omega_1(C)$ and $\omega_0(C)$ weights
            \item Model $Y$ given $C$ in the treated and control subsamples (nonR) or pseudo treated and control samples* (R)
            \item Based on these models, predict $Y_1$ and $Y_0$ given $C$ in full sample
            \item Average predicted $Y_1$ and $Y_0$ in full sample
        \end{enumerate}
        \\[-1em]\toprule
        \textbf{crw|psYpred pair}: nonR and MR estimators
        \begin{enumerate}[itemsep=0pt, topsep=0pt]
            \item Estimate $\omega_0(C)$ weights
            \item For the MR estimator, also estimate $\omega_\text{x}(C,M)$ weights
            \item Model $Y$ given $C,M$ in the treated subsample (nonR) or pseudo cross-world sample* (MR)
            \item Based on model, predict $Y_{1M_0}$ given $C,M$ in control units
            \item Average predicted $Y_{1M_0}$ in pseudo control sample
        \end{enumerate}%
        \\[-1em]\midrule
        \textbf{crw|Ypred pair}: nonR and MR estimators
        \begin{enumerate}[itemsep=0pt, topsep=0pt]
            \item Estimate $o_\text{x}(C,M)$ weights (nonR) or $\omega_\text{x}(C,M)$ weights (MR)
            \item Model $Y$ given $C$ in the pseudo cross-world subsample (nonR) or pseudo cross-world sample* (MR)
            \item Based on model, predict $Y_{1M_0}$ given $C$ in full sample
            \item Average predicted $Y_{1M_0}$ in full sample
        \end{enumerate}
        \\[-1em]\midrule
        \textbf{crw|MsimYpred pair}: nonR and MR estimators
        \begin{enumerate}[itemsep=0pt, topsep=0pt]
            \item For the MR estimator, estimate $\omega_0(C)$ and $\omega_\text{x}(C,M)$ weights
            \item Model the density of $M$ given $C$ in the control subsample (nonR) or pseudo control sample (MR)
            \item Model $Y$ given $C,M$ in the treated subsample (nonR) or pseudo cross-world sample* (MR)
            \item Do many times in full sample:
            \begin{enumerate}[itemsep=0pt, topsep=0pt]
                \item[i.] Based on first model, simulate $M_0$ given $C$
                \item[ii.] Based on second model, predict $Y_{1M_0}$ given the combination of $C$ and predicted $M_0$
            \end{enumerate}
            \item Average all predicted $Y_{1M_0}$ values in full sample
        \end{enumerate}
        \\[-1em]\midrule
        \textbf{crw|Y2pred pair}: nonR and R estimators
        \begin{enumerate}[itemsep=0pt, topsep=0pt]
            \item For the R estimator, estimate $\omega_0(C)$ and $\omega_\text{x}(C,M)$ weights
            \item Model $Y$ given $C,M$ in the treated subsample (nonR) or pseudo cross-world sample* (R)
            \item Based on model, predict $Y_{1M_0}$ given $C,M$ in control units
            \item Model predicted $Y_{1M_0}$ given $C$ in the control subsample (nonR) or pseudo control sample* (R)
            \item Based on model, predict $Y_{1M_0}$ given $C$ in full sample
            \item Average predicted $Y_{1M_0}$ in full sample
        \end{enumerate}
        \\[-1em]\toprule
        \textbf{NDE|YpredEpred pair}: nonR and R estimators
        \begin{enumerate}[itemsep=0pt, topsep=0pt]
            \item For the R estimator, estimate $\omega_0(C)$ and $\omega_\text{x}(C,M)$ weights
            \item Model $Y$ given $C,M$ in the treated subsample (nonR) or pseudo cross-world sample* (R)
            \item Based on model, predict $Y_{1M_0}$ given $C,M$ in control units, and
            
            compute proxy of the individual $N\!D\!E_0$ as predicted $Y_{1M_0}$ minus $Y$
            \item Model the $N\!D\!E_0$ proxy given $C$ in the control subsample (nonR) or pseudo control sample* (R)
            \item Based on model, predict $N\!D\!E_0$ given $C$ in full sample
            \item Average predicted $N\!D\!E_0$ in full sample
        \end{enumerate}
        \\[-1em]\bottomrule
        * This regression model for the MR/R estimator is required to satisfy the mean recovery condition.
    \end{tabularx}%
    }
\end{table}%

\subsubsection{crw|MsimYpred: mediator simulation and outcome prediction} 

These estimators involve fitting models for the conditional mediator density $\P(M\mid C,A=0)$ and the conditional outcome mean $\E[Y\mid C,M,A=1]$. Having learned these models, we put the observed mediator and outcome aside. For all units in the full sample, we simulate $M_0$ based on the first model, and with the simulated $M_0$ and observed $C$, predict $Y_{1M_0}$ based on the second model.%
\footnote{A variant is  to use the observed mediator  in control units and only simulate $M_0$ for treated units.}\label{fn:2}
We do this multiple times, resulting in multiple sets of predicted $Y_{1M_0}$ values, and average these predicted values to estimate $\E[Y_{1M_0}]$.%
\footnote{This is an implementation of the double expectation in the identification result where the outer expectation is evaluated by numerical integration.}

With the \textit{nonrobust} estimator in this pair, the mediator model is fit to the control subsample and the outcome model to the treated subsample.
For this estimator to be consistent, both models have to be correctly specified.
This mediator simulation strategy is used in Imai et al.'s natural (in)direct effects estimation method \cite{Imai2010psychmethods}, implemented in the R package \textit{mediation} \citep{Tingley2014}. This other estimator differs from crw|MsimYpred in that it uses this strategy for all potential outcome means (not only the cross-world one) therefore relies on more models. Also, the implementation in the package uses models for $\P(M\mid C,A)$ and $\E[Y\mid C,A,M]$ fit to the full sample rather than exposure-specific  models.

The \textit{more robust} crw|MsimYpred.MR instead fits the mediator model to the pseudo control sample and the outcome model to the pseudo cross-world sample, with the outcome model safisfying the mean recovery condition. Like its nonrobust sibling, this estimator is inconsistent if the mediator density is misspecified. Assuming correct specification of this model, this estimator has two chances to be correct: either the $\omega_\text{x}(C,M)$ weights are consistently  estimated or the outcome model is correctly specified.%
\footnote{In the \textit{more robust} version of the variant mentioned two foot notes ago, the mediator model is fit to a weighted control subsample that mimics the treated subsample's $C$ distribution -- using odds weights $\frac{\P(A=1\mid C)}{\P(A=0\mid C)}$.}

Relating to the existing literature, a specific version of crw|MsimYpred.MR where the cross-world weights are estimated based on mediator density models is an implementation of the estimator in section 5 in \cite{TchetgenTchetgen2012}. The estimator in \cite{TchetgenTchetgen2012} integrates the conditional outcome mean function $\E[Y\mid C,M,A=1]$ over the conditional mediator density $\P(M\mid C,A=0)$; the simulation-prediction-averaging procedure here is a numerical evaluation of that integral.

An interesting point: with both estimators being inconsistent if the mediator density model is misspecified, is anything gained by fitting this model to the pseudo control sample instead of the control subsample? Yes, what is gained is a partial correction in the sense that with a wrong model, the density fitted with correct weights is closer (in KL-divergence, see section B.4.2.4 of the Technical Appendix) to the true density than the density fitted without weights is.

These two estimators' dependence on correct specification of the mediator density model is an important drawback, as density estimation is a harder problem than mean estimation, an issue raised in \cite{Albert2012}. (Estimators that do not involve simulating $M_0$, or more generally, integrating over an estimated conditional density of $M_0$, avoid this problem.) For example, with a continuous variable, if the conditional mean is of interest, a common model choice is the linear model, which assumes a functional form for the mean but makes no other assumption. If the conditional density is of interest, one might still use the linear model but has to make additional distributional assumptions (e.g., the error is normally distributed or follows some other distribution) which are likely incorrect. In the special case with a single binary mediator, the distribution is fully described by the probability so the model reduces to a conditional mean model.

In a setting with a multivariate mediator (like in our data example), we need to model the joint distribution of the mediators given covariates in control units. To do this, we factor the joint into conditional densities/probabilities. Let $M=(M^a,M^b,M^c)$ where $M^a,M^b,M^c$ are three mediators.
\begin{align*}
    \P(M\mid C&,A=0)=
    \\
    &\P(M^a\mid C,A=0)\,\P(M^b\mid C,A=0,M^a)\,\P(M^c\mid C,A=0,M^a,M^b).
\end{align*}
In the control subsample (or the pseudo control sample if using the more robust version), we fit three models for the three mediators. All three models condition on $C$, the second model conditions additionally on $M^a$, and the third model conditions additionally on both $M^a$ and $M^b$. The order of variables in the factorization can be chosen for modeling convenience (see our data application for an example). Simulation follows the order of the fitted models.

\medskip

As a mini recap, the four pairs of estimators of $\E[Y_{1M_0}]$ above represent different solutions to the puzzle of finding the outcome mean in a target condition where the $C$ distribution is the same as that in the full sample, the $M$ given $C$ distribution is the same as that in the controls, and the $Y$ given $(C,M)$ distribution is the same as that in the treated. In each pair, the second estimator is more robust than the first as it does not require that all estimation components are correct. 
Among the four more robust estimators, crw|Y2pred.R is the most robust as it does not require any specific estimation component to be correctly specified/consistent; in contrast crw|psYpred.MR and crw|Ypred.MR are not robust to inconsistent weights, and crw|MsimYpred.MR is not robust to misspecification of the conditional mediator density model. 

\subsection{A weighting-centric view of the more robust estimators}

The presentation of estimators in pairs above shows that each MR/R estimator is an improvement over a simpler regression-based estimator by incorporating weighting. Several of these estimators can also be seen as a direct improvement on the pure weighting estimator by incorporating regression-based prediction. This is easily seen from the visualization in Figures \ref{fig:regPair} and \ref{fig:crossPairs}.

Consider reg|Ypred.R as an estimator of $\E[Y_1]$. As shown in Figure \ref{fig:regPair}, it is a modification of the pure weighting estimator. The latter solves the puzzle by obtaining the pseudo treated sample and stops there. reg|Ypred.R goes one step further: using regression-based prediction to correct for discrepancy in outcome mean due to the remaining difference between this pseudo treated sample and the \textit{target covariate distribution} (contained in the full sample).

crw|Ypred.MR can also be seen as a direct improvement upon the pure weighting estimator of $\E[Y_{1M_0}]$. The pure weighting estimator solves the puzzle by obtaining the pseudo cross-world sample. crw|Ypred.MR takes an additional step to correct for the remaining difference between the pseudo cross-world sample and the \textit{target covariate distribution} (but not the target conditional mediator distribution).

crw|psYpred.MR also starts with creating the pseudo cross-world sample like the pure weighting estimator. The additional regression-based prediction on the pseudo control sample adjusts for any difference between the pseudo cross-world sample and pseudo control sample. This effectively is a correction for the remaining difference between the pseudo cross-world sample and the \textit{target conditional mediator distribution} (contained in control units).

Like the two previous estimators, crw|Y2pred.R also starts with creating the pseudo cross-world sample by weighting. Then it goes two additional steps to correct for the remaining differences from the \textit{target conditional mediator distribution} and the \textit{target covariate distribution}.

\afterpage{
    \clearpage
    \begin{landscape}
    \captionof{table}{Robustness and nonrobustness properties of estimators from Sections \ref{sec:weighting}, \ref{sec:estimatorpairs} and \ref{sec:additivescale}}
    \label{tab:estimatorproperties}
    \centering
    \resizebox{.95\linewidth}{!}{%
    \begin{tabular}{lllllc}
        \textbf{\begin{tabular}[b]{@{}l@{}}
            Estimator\\label
        \end{tabular}}
        &
        \textbf{\begin{tabular}[b]{@{}l@{}}
            Estimator\\summary
        \end{tabular}}
        &
        \textbf{\begin{tabular}[b]{@{}l@{}}
            Estimation components used\\to estimate $\mathrm{E}[Y_{1M_0}]$ (or $\mathrm{NDE}_0$)
        \end{tabular}}
        &
        \textbf{\begin{tabular}[b]{@{}l@{}}
            Estimation components used\\to estimate $\mathrm{E}[Y_1],\mathrm{E}[Y_0]$ (or $\mathrm{TE}$)
        \end{tabular}}
        &
        \textbf{\begin{tabular}[b]{@{}l@{}}
            Combination of components that need to be\\correct for the estimator to be consistent
        \end{tabular}}
        &
        \textbf{\begin{tabular}[b]{@{}l@{}}
            Components\\not allowed to\\be inconsistent
        \end{tabular}}
        \\
        \toprule\toprule
        \\[-.5em]
        wtd
        &
        pure weighting
        &
        wts: $\omega_\text{x}(C,M)$
        &
        wts: $\omega_1(C),\omega_0(C)$
        &
        all components correct
        &
        all
        \\[.6em]
        \toprule\toprule
        psYpred1
        &
        \begin{tabular}{@{}l}
            crw|psYpred,\\reg|Ypred
        \end{tabular}
        &
        \multirow{3}{*}{%
        \begin{tabular}{@{}l}
            wts: $\omega_0(C)$
            \\
            omod: $\E[Y\mid C,M,A=1]$
        \end{tabular}}
        &
        omods: $\E[Y\mid C,A=a]$~~for $a=1,0$
        &
        all components correct
        &
        all
        \\
        \cmidrule{1-2}\cmidrule{4-6}
        psYpred2
        &
        \begin{tabular}{@{}l}
            $Y_{1M_0}$,$Y_1$|psYpred,\\$Y_0$|wtd
        \end{tabular}
        &
        &
        \begin{tabular}{@{}l}
            wts: $\omega_0(C)$
            \\
            omod: $\E[Y\mid C,A=1]$
        \end{tabular}
        &
        all components correct
        &
        all
        \\
        \midrule
        Ypred
        &
        \begin{tabular}{@{}l}
            crw|Ypred,\\reg|Ypred
        \end{tabular}
        &
        \begin{tabular}{@{}l}
            wts: $o_\text{x}(C,M)$
            \\
            omod: $\E[Y_{1M_0}\mid C]$
        \end{tabular}
        &
        omods: $\E[Y\mid C,A=a]$~~for $a=1,0$
        &
        all components correct
        &
        all
        \\
        \midrule
        MsimYpred1
        &
        \begin{tabular}{@{}l}
            crw|MsimYpred,\\reg|Ypred
        \end{tabular}
        &
        \multirow{3}{*}{%
        \begin{tabular}{@{}l}
            mmod: $\P(M\mid C,A=0)$
            \\
            omod: $\E[Y\mid C,M,A=1]$
        \end{tabular}}
        &
        omods: $\E[Y\mid C,A=a]$~~for $a=1,0$
        &
        all components correct
        &
        all
        \\
        \cmidrule{1-2}\cmidrule{4-6}
        MsimYpred2
        &
        \begin{tabular}{@{}l}
            $Y_{1M_0}$,$Y_0$|MsimYpred,\\$Y_1$|Ypred
        \end{tabular}
        &
        &
        \begin{tabular}{@{}l}
            mmod: $\P(M\mid C,A=0)$
            \\
            omod: $\E[Y\mid C,M,A=0]$, $\E[Y\mid C,A=1]$
        \end{tabular}
        &
        all components correct
        &
        all
        \\
        \midrule
        Y2pred
        &
        \begin{tabular}{@{}l}
            crw|Y2pred,\\reg|Ypred
        \end{tabular}
        &
        omods: $\E[Y\mid C,M,A=1],\E[Y_{1M_0}\mid C]$
        &
        omods: $\E[Y\mid C,A=a]$~~for $a=1,0$
        &
        all components correct
        &
        all
        \\[.3em]
        \midrule
        NDEpred*
        &
        \begin{tabular}{@{}l}
            NDE|YpredEpred,\\TE|Ypred
        \end{tabular}
        &
        \begin{tabular}[c]{@{}l}
            omod: $\E[Y\mid C,M,A=1]$
            \\
            emod: $\E[N\!D\!E_0\mid C]$
        \end{tabular}
        &
        omods: $\E[Y\mid C,A=a]$~~for $a=1,0$
        &
        all components correct
        & 
        all
        \\
        \toprule\toprule
        psYpred1.MR
        &
        \begin{tabular}{@{}l}
            crw|psYpred.MR,\\reg|Ypred.R
        \end{tabular}
        &
        \multirow{3}{*}{%
        \begin{tabular}[c]{@{}l}
            wts: $\omega_0(C),\omega_\text{x}(C,M)$
            \\
            omod: $\E[Y\mid C,M,A=1]$
        \end{tabular}}
        &
        \begin{tabular}[c]{@{}l}
            wts: $\omega_1(C),\omega_0(C)$
            \\
            omods: $\E[Y\mid C,A=a]$~~for $a=1,0$
        \end{tabular}
        &
        \multirow{3}{*}{%
        \begin{tabular}[c]{@{}l} 
            \textbullet~~$\omega_0(C)$ correct, and
            \\
            \textbullet~~either $\omega_1(C)$ or $\E[Y\mid C,A=1]$ correct, and
            \\
            \textbullet~~either $\omega_\text{x}(C,M)$ or $\E[Y\mid C,M,A=1]$ correct
        \end{tabular}}
        &
        \multirow{3}{*}{$\omega_0(C)$}
        \\
        \cmidrule{1-2}\cmidrule{4-4}
        psYpred2.MR
        &
        \begin{tabular}{@{}l}
            $Y_{1M_0}$,$Y_1$|psYpred.MR,\\$Y_0$|Ypred.R
        \end{tabular}
        &
        &
        \begin{tabular}{@{}l}
            wts: $\omega_1(C),\omega_0(C)$
            \\
            omod: $\E[Y\mid C,A=1]$
        \end{tabular}
        \\
        \midrule
        Ypred.MR
        &
        \begin{tabular}{@{}l}
            crw|Ypred.MR,\\reg|Ypred.R
        \end{tabular}
        &
        \begin{tabular}[c]{@{}l}
            wts: $\omega_\text{x}(C,M)$
            \\
            omod: $\E[Y_{1M_0}\mid C]$
        \end{tabular}
        &
        \begin{tabular}[c]{@{}l}
            wts: $\omega_1(C),\omega_0(C)$
            \\
            omods: $\E[Y\mid C,A=a]$~~for $a=1,0$
        \end{tabular}
        &
        \begin{tabular}[c]{@{}l} 
            \textbullet~~either $\omega_1(C)$ or $\E[Y\mid C,A=1]$ correct, and
            \\
            \textbullet~~either $\omega_0(C)$ or $\E[Y\mid C,A=0]$ correct, and
            \\
            \textbullet~~either $\omega_\text{x}(C,M)$ correct, or
            \\ 
            ~~~the $M$-related part of $\omega_\text{x}(C,M)$ and $\E[Y_{1M_0}\!\mid\!C]$ correct
        \end{tabular}
        &
        \begin{tabular}[c]{@{}c}
            the 
            \\
            $M$-related part
            \\
            of $\omega_\text{x}(C,M)$
        \end{tabular}
        \\
        \midrule
        MsimYpred1.MR
        &
        \begin{tabular}{@{}l}
            crw|MsimYpred.MR,\\reg|Ypred.R
        \end{tabular}
        &
        \multirow{5}{*}{%
        \begin{tabular}[c]{@{}l}
            wts: $\omega_\text{x}(C,M),\omega_0(C)$
            \\
            mmod: $\P(M\mid C,A=0)$
            \\
            omod: $\E[Y\mid C,M,A=1]$
        \end{tabular}}
        &
        \begin{tabular}[c]{@{}l}
            wts: $\omega_1(C),\omega_0(C)$
            \\
            omods: $\E[Y\mid C,A=a]$~~for $a=1,0$
        \end{tabular}
        &
        \begin{tabular}[c]{@{}l}
            \textbullet~~either $\omega_1(C)$ or $\E[Y\mid C,A=1]$ correct, and
            \\
            \textbullet~~either $\omega_0(C)$ or $\E[Y\mid C,A=0]$ correct, and
            \\
            \textbullet~~either $\omega_\text{x}(C,M)$ or $\E[Y\mid C,M,A=1]$ correct, and
            \\
            \textbullet~~$\P(M\mid C,A=0)$ correct
        \end{tabular}
        &
        \multirow{5}{*}{$\P(M\mid C,A=0)$}
        \\
        \cmidrule{1-2}\cmidrule{4-5}
        MsimYpred2.MR
        &
        \begin{tabular}{@{}l}
            $Y_{1M_0}$,$Y_0$|MsimYpred.MR,\\$Y_1$|Ypred.R
        \end{tabular}
        &
        &
        \begin{tabular}[c]{@{}l}
            wts: $\omega_1(C),\omega_0(C)$
            \\
            mmod: $\P(M\mid C,A=0)$
            \\
            omods: $\E[Y\mid C,A=1]$, $\E[Y\mid C,M,A=0]$
        \end{tabular}
        &
        \begin{tabular}[c]{@{}l}
            \textbullet~~either $\omega_1(C)$ or $\E[Y\mid C,A=1]$ correct, and
            \\
            \textbullet~~either $\omega_0(C)$ or $\E[Y\mid C,M,A=0]$ correct, and
            \\
            \textbullet~~either $\omega_\text{x}(C,M)$ or $\E[Y\mid C,M,A=1]$ correct, and
            \\
            \textbullet~~$\P(M\mid C,A=0)$ correct
        \end{tabular}
        \\
        \midrule
        Y2pred.R
        &
        \begin{tabular}{@{}l}
            crw|Y2pred.R,\\reg|Ypred.R
        \end{tabular}
        &
        \begin{tabular}[c]{@{}l}
            wts: $\omega_\text{x}(C,M),\omega_0(C)$
            \\
            omods: $\E[Y\mid C,M,A=1],\E[Y_{1M_0}\mid C]$
        \end{tabular}
        &
        \begin{tabular}[c]{@{}l}
            wts: $\omega_1(C),\omega_0(C)$
            \\
            omods: $\E[Y\mid C,A=a]$~~for $a=1,0$
        \end{tabular}
        &
        \begin{tabular}[c]{@{}l}
            \textbullet~~either $\omega_1(C)$ or $\E[Y\mid C,A=1]$ correct, and
            \\
            \textbullet~~either $\omega_0(C)$ correct, or
            \\
            ~~~both $\E[Y\mid C,A=0]$ and $\E[Y_{1M_0}\mid C]$ correct, and
            \\
            \textbullet~~either $\omega_\text{x}(C,M)$ or $\E[Y\mid C,M,A=1]$ correct
        \end{tabular}
        &
        NONE
        \\
        \midrule
        NDEpred.R*
        &
        \begin{tabular}{@{}l}
            NDE|YpredEpred.R,\\TE|Ypred.R
        \end{tabular}
        &
        \begin{tabular}[c]{@{}l}
            wts: $\omega_\text{x}(C,M),\omega_0(C)$
            \\
            omod: $\E[Y\mid C,M,A=1]$
            \\
            emod: $\E[N\!D\!E_0\mid C]$
        \end{tabular}
        &
        \begin{tabular}[c]{@{}l}
            wts: $\omega_1(C),\omega_0(C)$
            \\
            omods: $\E[Y\mid C,A=a]$~~for $a=1,0$
        \end{tabular}
        &
        \begin{tabular}[c]{@{}l}
            \textbullet~~either $\omega_1(C)$ or $\E[Y\mid C,A=1]$ correct, and
            \\
            \textbullet~~either $\omega_0(C)$ correct, or
            \\
            ~~~both $\E[Y\mid C,A=0]$ and $\E[N\!D\!E_0\mid C]$ correct, and
            \\
            \textbullet~~either $\omega_\text{x}(C,M)$ or $\E[Y\mid C,M,A=1]$ correct
        \end{tabular}
        &
        NONE
        \\
        \bottomrule\bottomrule
        \\[-.8em]
        \multicolumn{6}{l}{Notes: ``wts'' = weights. ``omod'' = outcome mean model. ``mmod'' = mediator density model. ``emod'' = effect model. * = only for additive effects.}
    \end{tabular}%
    }
    \end{landscape}
    \clearpage
}

\newpage

\subsection{Combining reg| and crw| estimators to estimate the effects }\label{sec:combinedPOsEstimators}

The marginal natural (in)direct effects are estimated by contrasting the estimated means of the three potential outcomes, using the difference or ratio definition of choice. We combine each of the four nonrobust regression-based crw| estimators with the nonrobust reg|Ypred, and each of the more robust crw| estimators with reg|Ypred.R. We label the resulting effect estimators using simple labels that mostly reflect the crw| method, e.g., Y2pred.R is the combination of crw|Y2pred.R and reg|Ypred.R. 

For two crw| strategies (crw|psYpred and crw|MsimYpred, both nonrobust and more robust versions), we also form a second combination with a modified reg| strategy. Note that crw|psYpred is anchored on the pseudo control sample. The first psYpred combination is with reg|Ypred, which is anchored on the full sample. In the second combination, however, the reg| part is also anchored on the pseudo control sample: $\E[Y_1]$ and $\E[Y_0]$ are estimated by averaging predicted 
$Y_1$ values and observed $Y_0$ values on the pseudo control sample. The other case is crw|MsimYpred. The first combination uses reg|Ypred for both $\E[Y_1]$ and $\E[Y_0]$; the second combination uses mediator simulation to estimate both $\E[Y_{1M_0}]$ and $\E[Y_0]$ (here seen as $\E[Y_{0M_0}]$), and uses reg|Ypred to estimate $\E[Y_1]$ only.

These estimators of natural (in)direct effects inherit the properties of the reg| and crw| estimators they combine. Table \ref{tab:estimatorproperties} summarizes the estimation components involved in, and the (non)robustness properties of, each of these effect estimators. It also covers the pure weighting estimator and a pair of estimators that will be considered in the next section.

\subsection{A quick note on model compatibility}\label{sec:compatibility}

This section is included for the more technically inclined readers and might not be of general interest. In response to helpful comments from the referees, we explored the topic of model compatibility or lack thereof for the estimators covered in this paper. Generally it is undesirable to use incompatible modeling components, because then at least one component in the conflict is mis-specified, regardless of what the actual distribution is. The specific concern here is whether an estimator's use of variationally dependent modeling components means the estimator has a model incompatibility issue. 

Interestingly, we find that practically one needs not worry about model incompatibility for these estimators. Let us consider three relevant cases of variationally dependent models: (i) combination of two conditional outcome mean models $\E[Y\mid C,A=1]$ and $\E[Y\mid C,M,A=1]$; (ii) combination of two conditional exposure models $\P(A\mid C)$ and $\P(A\mid C,M)$; and (iii) combination of the last two models with the mediator density model $\P(M\mid C,A=0)$. In cases (i) and (ii), although the models are variationally dependent, as long as their specification does not restrict the range of the conditional means/probabilities, they are compatible. To make this concrete, an example of restriction-induced incompatibility is that for a certain value $c$ of $C$, one specify $\P(A\mid C=c)=.2$ but specify $\P(A\mid C=c,M)\in(.4,.7)$; these specifications are incompatible because there exists no density $\P(M\mid C=c)$ that satisfies $\P(A\mid C=c)=\E_{M\mid C=c}[\P(A\mid C=c,M]$. It is hard to think of a case where one would specify such weirdly constrained models, though, so this is not really a practical concern. We thus exclude this kind of conflicting specification from consideration.

Case (iii), the combination of the two conditional exposure models with a model for $\P(M\mid C,A=0)$, is only present in a version of MsimYpred.MR that estimates the cross-world weights $\omega_\text{x}(C,M)$ using the second formula. (The other choice for MsimYpred.MR is to estimate $\omega_\text{x}(C,M)$ using the first formula based on mediator densities, where model components are variationally independent so there is no incompatibility.) The case of combining $\P(M\mid C,A=0)$ for mediator simulation with $\P(A\mid C)$ and $\P(A\mid C,M)$ for $\omega_\text{x}(C,M)$ estimation is an interesting case where it turns out that there is also no model incompatibility. Here the explicit specification of $\P(A\mid C)$ and $\P(A\mid C,M)$ implies an implicit specification of the ratio $\frac{\P(M\mid C,A=0)}{\P(M\mid C,A=1)}$ (as this is equal to $\frac{\text{odds}(A=1\mid C)}{\text{odds}(A=1\mid C,M)}$). This implicit specification combined with the explicit specification of $\P(M\mid C,A=0)$ implies an implicit specification of $\P(M\mid C,A=1)$. Since we do not explicitly model $\P(M\mid C,A=1)$, there is no model incompatibility. This estimator essentially ``escapes'' model incompatibility by simulating the mediator only for the cross-world condition, thus relying on estimating the mediator density under only one treatment condition.

We note that the assurance of model compatibility here does not tell us whether the model is correct. Model compatibility is a quality of the estimator; correct or mis-specification is a quality of the correspondence between the model/estimator and the truth. 

\smallskip

At the request of the Editor, we now respond to a specific point raised by a Referee (in the second round review of this paper), which references model compatibility but in our opinion is more about model mis-specification. The point raised is: when modeling the two conditional outcome mean functions $\E[Y\mid C,A=1]$ and $\E[Y\mid C,M,A=1]$ (case (i) above), this implies an implicit specification of the mediator distribution, and the concern is that this implicit specification may be mis-specified. The Referee comments that this would not be an issue with the alternative choice of modeling \mbox{$\E[Y\mid C,M,A=1]$} and $\P(M\mid C,A=1)$. We respond in two parts, one technical and one practical. The technical part is that the latter choice also has the same issue, as it implies an implicit specification for $\E[Y\mid C,A=1]$, and one may also be concerned that this implicit specification is incorrect. In fact, since these three functions (two outcome mean and one mediator density) are tied together by the relationship $\E[Y\mid C,A=1]=\E_{M\mid C,A=1}\{\E[Y\mid C,M,A=1]\}$, explicit specification of any two of the three implies an implicit specification of the third. Also, all specifications, explicit or implicit, may be incorrect. The practical part of our response is that modeling choices should be guided by the specific estimation strategy. For strategies that require an estimate of the function $\E[Y\mid C,A=1]$ (as part of estimating $\E[Y_1]$) and an estimate of the function $\E[Y\mid C,M,A=1]$ (as part of estimating $\E[Y_{1M_0}]$), we choose to estimate both of these target functions directly, so (roughly speaking) both have equal chance of being estimated well. This also means they have equal chance of being estimated poorly. The suggested alternative means estimating $\E[Y\mid C,A=1]$ indirectly by estimating the other two functions and putting them together; this would double this target function's chance of being estimated poorly because either of the component functions could be poorly estimated, or perhaps this chance is more than doubled because density estimation is harder than mean estimation.

In making the point above, the Referee also comments that using logit models for both $\E[Y\mid C,A=1]$ and $\E[Y\mid C,M,A=1]$ implies that $\P(M\mid C,A=1)$ is a bridge distribution \citep{wang2003MatchingConditionalMarginal,wang2004MarginalizedBinaryMixedEffects}. This is not the case, though, because the models bridged by a bridge distribution are fundamentally different from our outcome models. We explain this in section D of the Technical Appendix.

\section{If targeting effects on the additive scale: marginal effect as the mean of individual specific effects}\label{sec:additivescale}

If the marginal effects being targeted are defined on the additive scale, there is an alternative view of the puzzle, where what we wish we had is a single full sample in which all potential outcomes are simultaneously observed, which means for each individual the effects are observed. Then the individual $T\!E$, $N\!D\!E_0$, $N\!I\!E_1$, etc. are variables that could simply be averaged to estimate the average (which are also the marginal additive) effects. While these effect variables are not observed (this is the fundamental problem of causal inference), this view suggests we might learn an average effect if we have a good proxy for the individual effect.
It turns out that this works for natural direct effects.

\subsection{NDE|YpredEpred: effect prediction based on a proxy model}\label{sec:additiveNDE}

The key is to this method is to choose a proxy for the individual effect that has the same mean given covariates as the effect itself. Consider the individual $N\!D\!E_0=Y_{1M_0}-Y_0$. For control units, we observe $Y_0$ but not $Y_{1M_0}$. The idea is to replace the unobserved $Y_{1M_0}$ with its predicted value based on an appropriate model.

This leads to an estimator pair that is a slight modification of crw|Y2pred. Recall that crw|Y2pred involves a double model fit where the second model regresses predicted $Y_{1M_0}$ values (in control units) on covariates. The modification is that to estimate NDE\textsubscript{0}, that second model instead regresses the difference between predicted $Y_{1M_0}$ and observed $Y_0$ (a proxy for the individual $N\!D\!E_0$) on covariates. This model, which we loosely call the $\E[N\!D\!E_0\mid C]$ model, is then used to predict $N\!D\!E_0$ for all units in the full sample, and these predicted individual effects are averaged to estimate the average NDE\textsubscript{0}.

Like the crw|Y2pred pair, the NDE|YpredEpred pair includes a nonrobust and a robust estimator. For the nonrobust one, the $\E[Y\mid C,M,A=1]$ and $\E[N\!D\!E_0\mid C]$ models are fit to the treated and control subsamples, respectively. For the robust estimator NDE|YpredEpred.R, these models (which now are required to satisfy the mean recovery condition) are fit to the pseudo cross-world and pseudo control samples instead.
This robust estimator has three chances to be correct: (i) if both the outcome and effect models are correctly specified; or (ii) if the $\omega_0(C)$ and $\omega_\text{x}(C,M)$ weights that form the two pseudo samples are consistent; or (iii) if the $\omega_\text{x}(C,M)$ weights and the outcome model are consistent.

An aside: Instead of using the observed outcome in the construction of the proxy for $N\!D\!E_0$ in control units, a variant replaces it with a predicted value of this outcome based on a $\E[Y\mid C,M,A=0]$ model. The robust version of this variant (where all models are fit to relevant pseudo samples) is closely related to Zheng and van der Laan's TMLE estimator \cite{Zheng2012}. 
This alternative estimator also has three chances to be correct under similar conditions to those listed above, except that condition (i) additionally requires correct specification of the $\E[Y\mid C,M,A=0]$ model.

\medskip

Note that the current strategy only works for direct effects, as no similar proxy for the individual $N\!I\!E_1=Y_1-Y_{1M_0}$ is available. To estimate the indirect effect, one can subtract an NDE|YpredEpred estimator off of a total effect estimate. We obtain the latter using reg|Ypred estimators.

\section{How to choose an estimator}\label{sec:howtochoose}

In addition to the primary goal of building intuition for (more) robust estimation, a secondary goal of this paper is to provide a menu of estimation options for the specific estimands considered. Table \ref{tab:estimatorproperties} summarizes the estimators of marginal natural (in)direct effects discussed so far, with nonrobust estimators in the top panel and more robust estimators in the bottom panel. Columns 3 and 4 of this table list the components involved in each estimator, under the groupings of outcome models (omod), effect models (emod), mediator models (mmod) and weights (wts). Column 5 lists what is required of these components for the estimator to be consistent. For the (more) robust estimators, each bullet in this column is one requirement, and any bullet of the either-or form indicates a robustness property, while any bullet not in either-or form is a nonrobust component that needs to be consistent for the estimator to be consistent. The nonrobust component is also pointed out specifically in column 6.

Given this menu of estimators, which one should be used for a particular application? We take the pragmatic viewpoint that the choice of methods should partly depend on the user's level of comfort with the different types of methods, because implementation is more error prone if a method is more complex and not well understood by the user. Therefore we do not intend to propose or advocate for a single method but to lay out a range of potential choices. We offer some considerations below.

As the nonrobust estimators are simpler, one approach is to pick one of those estimators. Among the nonrobust options we do not recomend the weighting-based estimators, as they are inefficient. Otherwise, we recommend considering the set of estimation components (weights, mediator density and outcome/effect mean models) required by each estimator and deciding which set is most feasible to implement well.
The disadvantage of the simpler estimators, of course, is the lack of robustness to model mis-specification. Another approach is to consider the more robust estimators, looking at components that must be correctly specified as a way to rule out estimators that depend on hard-to-model components. 

Again, we note that density estimation is generally more challenging than mean estimation, and the difficulty of the task depends on the number and types of mediators. It is easier for certain types of variables (e.g., binary variables) than others (e.g., continuous variables). When there are multiple mediators, there are multiple models to fit and more chances to mis-specify them. With multiple mediators, we need to choose an order of factorization for the mediators, and therefore may prefer an order that makes models slightly easier or more convenient to specify (see our data example). We might consider alternatives for density estimation where possible, e.g., using the second formula of the cross-world weight rather than the first formula. When using the MsimYpred estimators, we want to keep in mind that these methods depend on a correct specification of the $\P(M\mid C,A=0)$ model. 

Several of the estimators depend on certain weights being correct; these include the nonrobust weighting-based estimators, as well as the more robust estimators psYpred.MR (requiring correct control weights) and Ypred.MR (requiring that the $M$-related component of the cross-world weight is correct). While correct specification cannot be determined, balance checking is a useful tool to guard against severe misspecification. If certain weights do not achieve excellent balance, it is advisable not to use the estimators whose validity hangs on those weights.

Lastly, on the menu there are two fully robust estimators, Y2pred.R and NDEpred.R, which do not depend on any specific modeling component being correct. Note that for each of these, we still need enough of the estimation components to be correct.

\section{A comment on a common practice}\label{sec:improveprecision}

Above we have shown how simple estimation strategies can be made more robust. Here we comment on a common practice in applied research that looks similar to robust estimation to point out that it should not be seen as such, but should simply be seen as a method to improve precision.

For simplicity, first consider the non-mediation setting where the estimand is the average treatment effect, $\E[Y_1]-\E[Y_0]$. Here this common practice involves first balancing covariates to justify comparing outcomes between the two groups, and then fitting a simple model regressing outcome on exposure and covariates, usually in the form of main effects. Seen through our pseudo samples lens, when covariate balancing is done by propensity score weighting, we achieve the pseudo treated and pseudo control samples, and the regression model is fit to the combination of these two pseudo samples. With the combination of weighting and an outcome model, this type of analysis looks similar to a doubly robust method, and we have heard practitioners describe it as doubly robust. However, the regression model is likely too simple to have a chance at being correct. As the consistency of the method depends on the pseudo samples, this practice should not be seen as a robust method. It would be appropriate, though, to refer to this use of the simple regression model as \textit{leveraging covariates to improve precision}. Leveraging covariates to improve precision is an approach for analysis of randomized trials, where the effect is identified so no covariate adjustment is needed, but the use of a \textit{working regression model} (not assumed to be correct) helps explain outcome variance and thereby makes the effect estimate more precise \cite{Wang2019,Steingrimsson2017}. Intuitively, the pseudo samples mimic a randomized trial, so the simple regression is just a working model to improve precision.

In the current setting of estimating marginal natural (in)direct effects, methods that first create three pseudo samples representing the conditions being contrasted (by weighting only or weighting combined with prediction/imputation of $Y_{1M_0}$) and then fit a simple model regressing outcome on covariates and conditions (indicated by dummy variables) should be seen as methods to leverage covariates to improve precision, not as robust methods.
For interested readers, a preprint of this paper \cite{nguyen2022CausalMediationAnalysis} (version 3, section 6) includes a translation to the current setting of techniques for using covariates to improve precision that apply to different outcome types. It also comments on similarity with and key distinctions from several methods that estimate conditional effects \cite{Vansteelandt2012a,Steen2017}.

\section{Confidence interval estimation}\label{sec:Bootstrap}

The previous sections cover point estimation of the effects. We now turn to interval estimation.
All the estimators in this paper are M-estimators; they are solutions of generalized estimating equations. Applying the calculus of M-estimators \cite{Stefanski2002}, we derive general formulas for the asymptotic variance of each of the estimators when all estimation components (weights, conditional mediator density, conditional mean outcome/effect) are based on (semi)parametric models. The derivations are placed in the Technical Appendix.

As such variance estimators depend on the specific models used for the different estimation components, they are clunky to use in practice. 
We use bootstrapping as a generic tool to obtain confidence intervals for all the estimators. 
As the data example includes quite a few categorical covariate/mediator variables, a challenge when using the simple resampling bootstrap \citep{Efron1979} is that some bootstrap samples do not cover all values (and combinations of values) of those variables, resulting in predictors being dropped from models and predictions being distorted. To avoid this problem we instead use a continuous weights bootstrap \citep{Xu2020fractionalweightbootstrap}. 
With both bootstrap procedures, the making of a bootstrap sample can be seen as weighting the observations of the original sample by a set of random weights that are identically distributed: the resampling bootstrap uses integer weights drawn from a uniform Multinomial distribution; the continuous weights bootstrap draws weights from a continuous distribution. We use the version proposed by Xu et al. \cite{Xu2020fractionalweightbootstrap} based on the uniform Dirichlet distribution, where the weights sum to sample size $n$, have mean 1 and variance $(n-1)/(n+1)$. (For comparison, resampling weights sum to $n$, have mean 1 and variance $(n-1)/n$.) Bootstrap samples based on continuous bootstrap weights retain all observations, thus they do not lose data patterns.

\section{Data example application}\label{sec:application}

In this example $A$ is a binary variable \texttt{treat} indicating whether a student is in the treatment (i.e., combined intervention) or control condition. $Y$ is binary variable \texttt{drink} indicating whether the student engages in weekly drinking at 22 months. $M$ consists of three mediators measured at six months: attitudes towards alcohol consumption (binary variable \texttt{att} indicating attitudes against consumption), self-control in situations involving alcohol (continuous variable \texttt{sfc}), and parental rules regarding alcohol (binary variable \texttt{rul} indicating strict rules). Baseline covariates $C$ include demographic variables age, sex, religion, education track (academic or vocational); baseline measures of the mediators (\texttt{att0}, \texttt{sfc0}, \texttt{rul0}); and baseline measure of the outcome (\texttt{drink0}). Table \ref{tab:1} summarizes the baseline covariates in the synthetic sample, showing some covariate imbalance between the intervention and control conditions. The unconfoundedness assumption means that the listed baseline covariates are sufficient to remove exposure-mediator, exposure-outcome and mediator-outcome confounding.

\begin{table}[t!]
    \caption{Baseline covariates in the synthetic dataset based on the PAS study}
    \label{tab:1}
    \centering
    \begingroup
    \setlength{\tabcolsep}{3pt} 
    \resizebox{.8\linewidth}{!}{%
    \begin{tabular}{lccrrccrrccrrl}
        &&& \multicolumn{2}{c}{\textbf{Treated}} 
        &&& \multicolumn{2}{c}{\textbf{Control}} 
        &&& \multicolumn{2}{c}{\textbf{Total}}
        \\
        &&& \multicolumn{2}{c}{(n=778)} 
        &&& \multicolumn{2}{c}{(n=907)} 
        &&& \multicolumn{2}{c}{(n=1685)}
        \\
        \toprule
        Age
        \\
        ~~~~11 &&& 35 & (4.5\%) &&& 38 & (4.2\%) &&& 73 & (4.3\%)
        \\
        ~~~~12 &&& 559 & (71.9\%) &&& 682 & (75.2\%) &&& 1241 & (73.6\%)
        \\
        ~~~~13 &&& 184 & (23.7\%) &&& 187 & (20.6\%) &&& 371 & (22.0\%)
        \\
        \midrule
        Sex
        \\
        ~~~~female &&& 305 & (39.2\%) &&& 448 & (49.4\%) &&& 753 & (44.7\%)
        \\
        ~~~~male &&& 473 & (60.8\%) &&& 459 & (50.6\%) &&& 932 & (55.3\%)
        \\
        \midrule
        Religion
        \\
        ~~~~Catholic                   &&&  62 &  (8.0\%) &&& 319 & (35.2\%) &&& 381 & (22.6\%)
        \\
        ~~~~Protestant/other Christian &&&  84 & (10.8\%) &&& 114 & (12.6\%) &&& 198 & (11.8\%)
        \\
        ~~~~Islam                      &&&  45 &  (5.8\%) &&&  34 &  (3.7\%) &&&  79 &  (4.7\%)
        \\
        ~~~~not religiously socialized &&& 552 & (71.0\%) &&& 416 & (45.9\%) &&& 968 & (57.4\%)
        \\
        ~~~~other                      &&&  35 &  (4.5\%) &&&  24 &  (2.6\%) &&&  59 &  (3.5\%)
        \\
        \midrule
        Education tract
        \\
        ~~~~vocational &&& 276 & (35.5\%) &&& 547 & (60.3\%) &&& 823 & (48.8\%)
        \\
        ~~~~academic   &&& 502 & (64.5\%) &&& 360 & (39.7\%) &&& 862 & (51.2\%)
        \\
        \midrule
        Baseline weekly drinking
        \\
        ~~~~yes       &&&  96 & (12.3\%) &&& 166 & (18.3\%) &&&  262 &  (9.0\%)
        \\
        ~~~~no        &&& 625 & (80.3\%) &&& 647 & (71.3\%) &&& 1272 & (75.5\%)
        \\
        ~~~~no answer &&&  57 &  (7.3\%) &&&  94 & (10.4\%) &&&  151 &  (9.0\%)
        \\
        \midrule
        Baseline attitude
        \\
        ~~~~negative re. alcohol use &&& 518 & (66.6\%) &&& 594 & (65.5\%) &&& 1112 & (66.0\%)
        \\
        ~~~~less negative.           &&& 260 & (33.4\%) &&& 313 & (34.5\%) &&&  573 & (34.0\%)
        \\
        \midrule
        Baseline parental rule
        \\
        ~~~~strict     &&& 561 & (72.1\%) &&& 580 & (63.9\%) &&& 1141 & (67.7\%)
        \\
        ~~~~not strict &&& 271 & (27.9\%) &&& 327 & (36.1\%) &&&  544 & (32.3\%)
        \\
        \midrule
        Baseline self control
        \\
        ~~~~mean (SD) &&& 3.59 & (0.55) &&& 3.57 & (0.53) &&& 3.58 & (0.54)
        \\
        ~~~~median [min, max] &&& 3.62 & \multicolumn{2}{l}{[1.69,4.85]} && 3.62 & \multicolumn{2}{l}{[2.00,4.92]} && 3.62 & \multicolumn{2}{l}{[1.69,4.92]}
        \\
        \bottomrule
    \end{tabular}%
    }
    \endgroup
\end{table}

With this example, we target marginal effects on the additive scale. The total effect can be understood as a reduction in weekly drinking prevalence that would occur had all students received the treatment versus no students received the treatment. The natural indirect effect is roughly interpreted as the component of that prevalence reduction that is due to the intervention's impact on the mediators, and the natural direct effect is the remaining component.

\subsection{Weighting}

\subsubsection*{Weights estimation}

The estimators have different requirements in terms of weights -- see Tables \ref{tab:estimatorproperties}. The pure weighting estimator and all the MR/R estimators involve the trio of $\omega_0(C)$, $\omega_1(C)$ and $\omega_\text{x}(C,M)$ weights. Several nonrobust estimators involve some (but not all) weights: psYpred estimators involve $\omega_0(C)$ and Ypred involves $o_\text{x}(C,M)$ weights. The nonrobust Y2pred and MsimYpred estimators do not require weights. Here we focus on the $\omega_0(C)$, $\omega_1(C)$ and $\omega_\text{x}(C,M)$ weights.

We estimate weights via parametric models. $\omega_1(C)$ and $\omega_0(C)$ are estimated via on a propensity score model, i.e., a model for $\P(A\mid C)$. $\omega_\text{x}(C,M)$ is estimated by the second method, using the combination of this propensity score model and a model for $\P(A\mid C,M)$. (We avoid the first method which would require fitting six models for the three mediators.) For both models we use logistic regression with spline terms on continuous predictors and some interaction terms in the second model (the result of several rounds of model fitting and balance checking).%
\footnote{The function \texttt{ns(v,d)} from R-package \texttt{splines} implements cubic splines on variable \texttt{v} with \texttt{d} degrees of freedom.}
The model formulas used for $\P(A\mid C)$ and $\P(A\mid C,M)$ are:

{\footnotesize\noindent
\begin{tabular}{@{}l@{\extracolsep{3pt}}l@{}}
    \texttt{treat $\sim$ sex + age + edu + religion + drink0 + att0 + rul0 + ns(sfc0, 4)},
    \\
    \texttt{treat $\sim$ sex + age + edu + religion + drink0 + att0 * att + rul0 * rul + }
    \\
    \texttt{~~~~~~~~~ns(sf0, 4) * ns(sfc, 4)}.
\end{tabular}%
}

\begin{figure}[t!]
    \centering
    \caption{Distributions of weights for the pseudo control (p00), treated (p11) and cross-world (p10) samples. For comparability, stabilized weights are shown.}
    \label{fig:wtdistributions}
    ~\\
    \includegraphics[width=.6\linewidth]{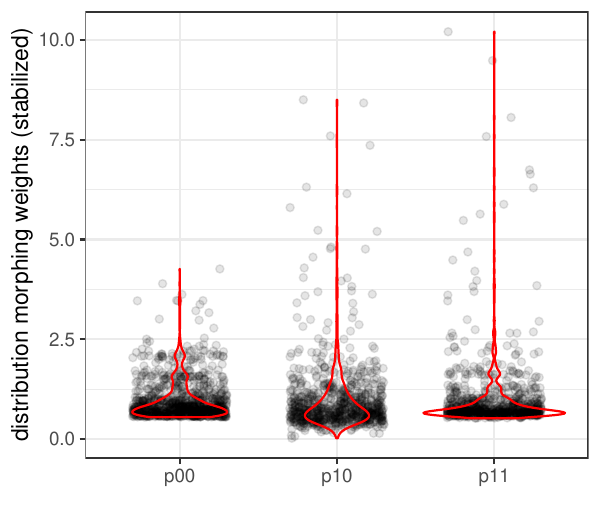}
\end{figure}

The distributions of these weights (in stabilized form, i.e., with mean 1 in each group) are shown in Figure \ref{fig:wtdistributions}. Some of the weights are large, but not extreme.

\subsubsection*{Balance checking}

\begin{figure}[h!]
    \centering
    \caption{Covariate and mediator balance for pseudo treated (p11), pseudo control (p00) and pseudo cross-world (p10) samples. For continuous covariate \texttt{sfc0} and continuous mediator \texttt{sfc} (marked with *), the mean differences are standardized. The parenthesized comments are specific to the wtd and wt-Cadj estimators, indicating that all balance components are important to those estimators.}
    \label{fig:balancecheck}
    ~\\
    \includegraphics[width=\linewidth]{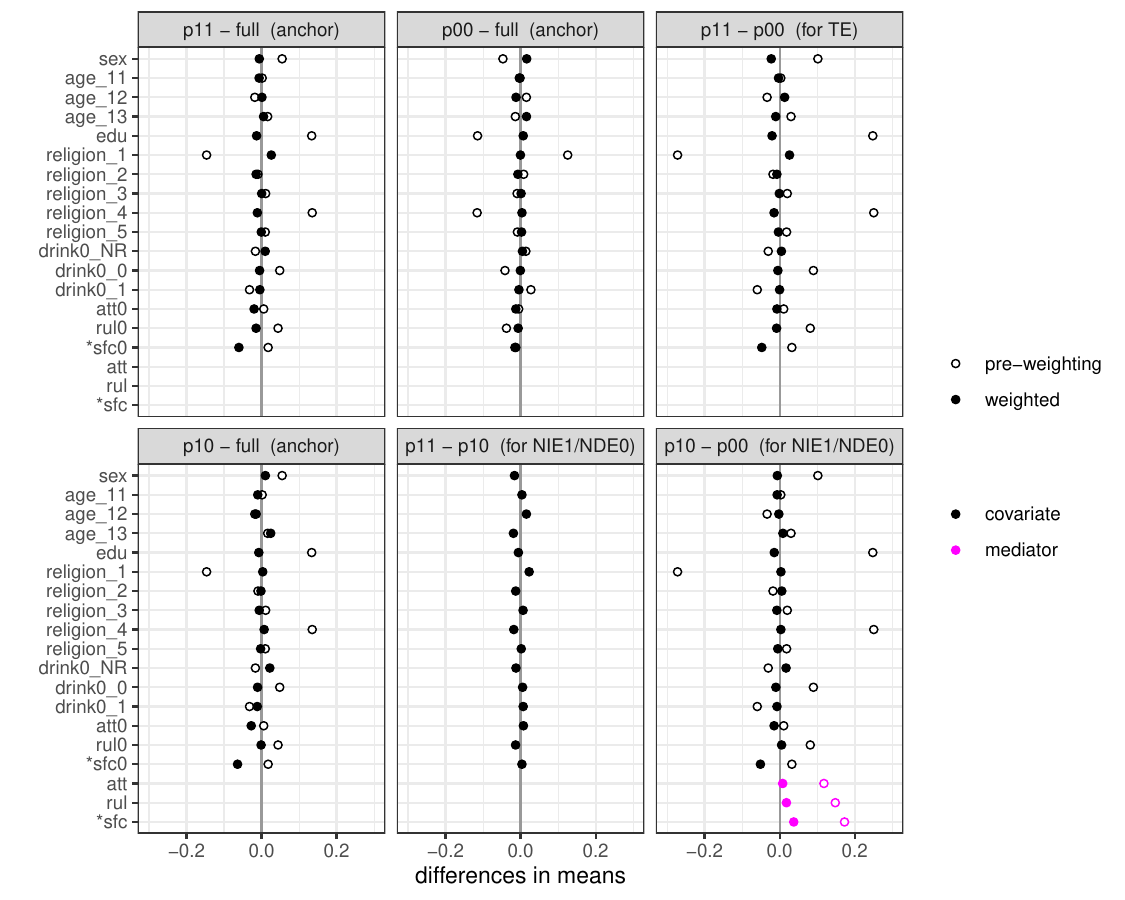}
\end{figure}

Balance on the means of covariates and mediators for the pseudo treated, control and cross-world samples are shown in Figure \ref{fig:balancecheck} (based on the prescription in Figure \ref{fig:desiredbalance}). Overall, balance improves after weighting; this is prominent for covariates sex, education track, religion, and the three mediators. Interestingly, balance on baseline self-control (\verb|sfc0|) is slightly worsened, although the standardized mean difference is still modest. In addition to mean balance, distributional balance on continuous covariates and mediators should also be checked, e.g., using the R-package \verb|cobalt| \cite{cobalt}.

Note that the comments in the plot labels specifically address estimators that depend on all balance components, here the pure weighting estimator (wtd). For estimators with some robustness, we recommend a combination of two plots: a \textit{full balance} plot capturing balance resulting from all the weighting involved in the estimator, and a \textit{key balance} plot capturing the balance component that the estimator absolutely depends on. For Ypred.MR, for example, the full balance plot is the same plot in Figure \ref{fig:balancecheck} without the comments in the plot labels, and the key balance plot picks out the `p10 - p00' component.

For nonrobust estimators that depend on a weighting element, balance checking is specific to the weighting. For example, psYpred1 and psYpred2 require covariate balance between the pseudo control sample and the full sample; Ypred requires covariate-and-mediator balance between the pseudo cross-world subsample and the control subsample. All these variants of balance checking are implemented in the \texttt{mediationClarity} package (see details in package vignette).

\subsection{Other estimation components}

\subsubsection*{Outcome mean models}

The various estimators call for fitting models for the outcome given covariates or given covariates and mediators to subsamples or pseudo samples. We use logistic regression for the binary outcome. 
Models that regress outcome on covariates (estimating $\E[Y\mid C,A=1]$, $\E[Y\mid C,A=0]$ or $\E[Y_{1M_0}\mid C]$) use formula

{\footnotesize\texttt{drink $\sim$ sex + age + edu + religion + drink0 + att0 + rul0 + ns(sfc0, 3)}}.

\noindent Models that regress outcome on covariates and mediators (estimating $\E[Y\mid C,M,A=1]$) use formula

{\footnotesize\texttt{drink $\sim$ sex + age + edu + religion + drink0 + att0 + rul0 + ns(sfc0, 3) +}}

{\footnotesize\texttt{~~~~~~~~~att + rul + ns(sfc, 3)}}.

\subsubsection*{Mediator density model}

The MsimYpred estimators require mediator density modeling in the control subsample or pseudo control sample. We fit logit models for the two binary mediators \texttt{att} ($M^a$) and \texttt{rul} ($M^b$), and a linear model for the continuous mediator \texttt{sfc} ($M^c$), with formulas:

\noindent{\footnotesize\texttt{att $\sim$ age + sex + edu + religion + drink0 + att0 + rul0 + ns(sfc0, 3)}}

\noindent{\footnotesize\texttt{rul $\sim$ age + sex + edu + religion + drink0 + att0 + rul0 + ns(sfc0, 3) + att}}

\noindent{\footnotesize\texttt{sfc $\sim$ age + sex + edu + religion + drink0 + att0 + rul0 + ns(sfc0, 3) + att + rul}},

\noindent and assume the errors in the third model are normally distributed and homoscedastic. These models estimate $\P(M^a\mid C,A\!=\!0)$, $\P(M^b\mid C,M^a,A\!=\!0)$, and $\P(M^c\mid C,M^a,M^b,A\!=\!0)$, respectively.%
\footnote{Of the three mediators, we choose to model \texttt{sfc} last for convenience. This avoids having to specify models with \texttt{sfc} as another continuous predictor, which would be more complicated.}

\subsubsection*{Model for $N\!D\!E_0$ given covariates}

The NDEpred estimators involve regressing the proxy of the individual $N\!D\!E_0$ (predicted $Y_{1M_0}$ minus observed $Y_0$) on covariates in the control subsample or pseudo control sample. As the difference between the two (predicted and observed) binary outcomes is bounded in the $-1$ to 1 interval, we transform it by adding 1 then dividing by 2 to map to the 0 to 1 interval, and fit the regression model using the transformed difference using logit link. Predictions based on this model are back transformed by multiplying by 2 then subtracting 1.%
\footnote{This is equivalent to using the tanh link for a response variable bounded in $[-1,1]$ \citep{Wang2018}; the transformation trick allows fitting the model using standard software with logit link.}
The formula we use for this model is:

{\footnotesize\texttt{trans.diff $\sim$ age + sex + edu + religion + drink0 + att0 + rul0 + ns(sfc0, 3)}}.

\subsection{Results} 

\begin{figure}[h]
    \centering
    \caption{Effect estimates from different estimators, shown as reduction in outcome (weekly drinking) prevalence. The psYpred and MsimYpred estimators shown here are the psYpred2 and MsimYpred1 versions in Table \ref{tab:estimatorproperties}.}
    \includegraphics[width=.9\linewidth]{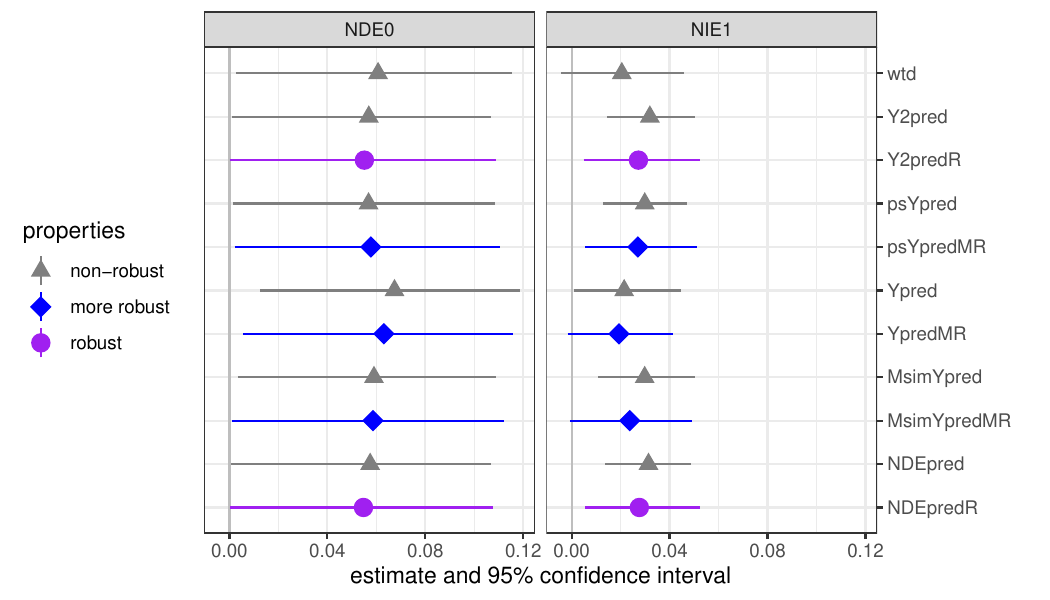}
    \label{fig:results}
\end{figure}

Effect estimates from different estimators are shown in Figure \ref{fig:results}. To avoid clutter, for estimator types that have multiple versions, we show only one version. The estimates are quite similar. Overall, it appears that the effect of the intervention on weekly drinking at follow-up consists of a small part mediated by the mediators being considered (alcohol-related attitudes, parental rules, and self-control), and a larger direct effect.

As these are estimates from one dataset, one should be cautious not to infer characteristics of the estimators. That said, we note that within each pair of nonrobust and (more) robust estimators, the (more) robust one tends to have larger variance than the nonrobust one, with wider confidence intervals. The pure weighting estimator by theory has the largest variance, although for this dataset this is not obvious from the confidence interval widths.

\section{Concluding remarks}

In this paper we have shown how a range of estimators may be constructed based on two strategies that are familiar to many who are involved in statistical analyses (weighting and model-based prediction) and a simple way of combining them (weighted models); this is the paper's primary goal. The key ideas of this exercise, which are not specific to natural effects but apply generally, are (i) to use these tools flexibly to put together the different pieces of the estimation puzzle, where the puzzle is defined by the identification result of the effects of interest; and (ii) to induce robustness on pieces of the puzzle by using weighted models.

Thinking more broadly, this approach to constructing estimators could be applied to other marginal estimands, including interventional effects of various kinds \cite{Nguyen2020MediationEstimands} and causal decomposition of disparities \cite{jackson2021MeaningfulCausalDecompositions}. A key difference is that these other estimands involve setting the mediator (or the target variable of causal decomposition) to one of a range of interventional distributions (depending on the estimand) which may condition on or marginalize over certain pre-exposure and post-exposure covariates. This means the weighting scheme needs to be tailored and may be more complicated, and the density being mimicked may condition (or not) on different types of variables, and may be known or need to be estimated. Whatever the case, the idea of visualizing the identification result to bring clarity to where different types of information (pieces of the puzzle) come from, and a similar exercise of assembling them, will be productive in generating estimators, and importantly will make the estimator transparent to the user (as a sound solution to the puzzle).

On a technical note, the focal estimand in the current case (the identification result of the mean cross-world potential outcome) is an iterated expectation, and there are different ways an iterated expectation can be estimated. For the current case, one way involves fitting repeated conditional mean models (iterated regression), and another involves integrating an inner expectation over an estimated conditional density (here via simulation). Weights can be used to fit the models to the space of predictors where they are used for prediction/simulation, to help correct bias due to model mis-specification. One point we noted is that this provides only a partial correction for misspecified conditional density models. This point is likely relevant to some of the more complicated estimands for which some density estimation cannot be avoided.

One important topic that was not covered in this paper is sensitivity analysis to violation of identifying assumptions. While several sensitivity analysis methods have been proposed, there is room for work that connects each of the many estimation methods that exist and may be used in practice to relevant sensitivity analyses, or at the least point out which estimation methods can (and which cannot) be appropriately paired with which sensitivity analyses. This would be very helpful to the use of methods in practice.

\begin{acks}[Acknowledgments]
This work is supported by the US National Institute of Mental Health through grants R01MH115487 and T32MH122357 (PI Stuart). The authors appreciate Drs. Guanglei Hong and Fan Yang for helpful feedback on an earlier draft; Drs. Ilya Shpitser and Eric Tchetgen Tchetgen for insightful discussions on robust estimation and their seminal article \cite{TchetgenTchetgen2012}; participants of our summer institute mediation course from 2021 and 2022 and participants of the second term 2021 seminar on statistical methods for mental health research at Johns Hopkins Bloomberg School of Public Health for fruitful discussion; and two anonymous Referees, the Associate Editor and the Co-Editor Dr. Richard Lockhart (at \textit{Statistics Surveys}) for their thoughtful and constructive comments. The authors thank the participants, staff and investigators of the PAS trial.
\end{acks}

\begin{supplement}
\stitle{Supplement 1}
\sdescription{This is the technical appendix to the paper, with three parts. Part A derives and connects alternative expressions for the cross-world weight function, including the novel third expression. Part B contains proofs of the robustness (and any nonrobustness) properties of the robust and more robust estimators. Part C derives general asymptotic variance formulas for all the estimators, where estimation components (weights, conditional mediator density functions, conditional outcome/effect mean functions) are based on parametric models. Part D explains our response in section \ref{sec:compatibility} that two logit outcome models do not imply a bridge mediator distribution. This technical appendix can be found at the end of this document, and is also available from the journal website, with doi:10.1214/22-SS140SUPPA.}
\end{supplement}

\begin{supplement}
\stitle{Supplement 2}
\sdescription{This is the vignette of the R package
\texttt{mediationClarity}, which implements the estimators. It can be read after installing the package, available from \url{https://github.com/trangnguyen74/mediationClarity}. This supplement is also available from the journal website, with doi:10.1214/22-SS140SUPPB.}
\end{supplement}

\bibliographystyle{imsart-number}
\bibliography{refs}

\setlength{\voffset}{0cm}
\setlength{\hoffset}{0cm}



\section*{Supplementary material}

# A    Three expressions of the cross-world weights

Of the three expressions of the cross-world weights presented in the main text of the paper (labeled (1), (2) and (3)), the first was proposed by Hong (2010), the second appeared as part of an identification result in Huber (2014), and the equivalence of a density ratio and an odds ratio that connects the two was noted by Zheng and van der Laan (2012). For completeness, however, we will start from a well-known identification result of E[$Y_{1M_0}$], and use it to derive all the three expressions of the cross-world weights. This shows not only where they all come from but also how they connect to one another.

## A.1    Identification result

We use as the basis for deriving the cross-world weights $\omega_{\text{x}}(C, M)$ the identification result

$$\mathrm{E}[Y_{1M_0}] = \mathrm{E}_C(\mathrm{E}_{M|C,A=0}\{\mathrm{E}[Y \mid C, M, A = 1]\}).$$

This is a well-known result under standard assumptions – see, e.g., Pearl (2001); VanderWeele and Vansteelandt (2009); Imai et al. (2010); Pearl (2012). For the sake of completeness, here is how it can be derived.

By iterated expectation,

$$\mathrm{E}[Y_{1M_0}] = \mathrm{E}(\mathrm{E}\{\mathrm{E}[Y_{1M_0} \mid C, M_0] \mid C\}).$$

Putting aside the outer expectation, we work with the inside,

$$\begin{aligned}
\mathrm{E}\{\mathrm{E}[Y_{1M_0} \mid C, M_0] \mid C\} &= \int \mathrm{P}(M_0 = m \mid C)\mathrm{E}[Y_{1M_0} \mid C, M_0 = m]dm &&\text{(writing expectation as integral)}\\
&= \int \mathrm{P}(M_0 = m \mid C)\mathrm{E}[Y_{1m} \mid C, M_0 = m]dm &&\text{(consistency of } M_0\text{)}\\
&= \int \mathrm{P}(M_0 = m \mid C)\mathrm{E}[Y_{1m} \mid C]dm &&(M_0 \perp\!\!\!\perp Y_{1m} \mid C)\\
&= \int \mathrm{P}(M_0 = m \mid C)\mathrm{E}[Y_{1m} \mid C, A = 1]dm &&(A \perp\!\!\!\perp Y_{1m} \mid C)\\
&= \int \mathrm{P}(M_0 = m \mid C)\mathrm{E}[Y_{1m} \mid C, M = m, A = 1]dm &&(M \perp\!\!\!\perp Y_{1m} \mid C, A = 1)
\end{aligned}$$

2

$$= \int P(M_0 = m \mid C) E[Y \mid C, M = m, A = 1] dm \qquad \text{(consistency of } Y_{1m})$$

$$= \int P(M_0 = m \mid C, A = 0) E[Y \mid C, M = m, A = 1] dm \qquad (A \perp\!\!\!\perp M_0 \mid C)$$

$$= \int P(M = m \mid C, A = 0) E[Y \mid C, M = m, A = 1] dm \qquad \text{(consistency of } M_0)$$

$$= E_{M \mid C, A=0} \{ E[Y \mid C, M, A = 1] \}. \qquad \text{(writing integral as expectation)}$$

Combining this with the outer expectation obtains the identification result listed above.

## A.2 First expression

We can manipulate the identification result above to bring out the first formula of the cross-world weights. To simplify presentation, we again put aside the outer expectation and consider the inside,

$$E_{M \mid C, A=0} \{ E[Y \mid C, M, A = 1] \} = \int E[Y \mid C, M = m, A = 1] P(M = m \mid C, A = 0) dm$$

$$= \int E[Y \mid C, M = m, A = 1] \frac{P(M = m \mid C, A = 0)}{P(M = m \mid C, A = 1)} P(M = m \mid C = c, A = 1)$$

$$= \int E\left[ \frac{P(M = m \mid C, A = 0)}{P(M = m \mid C, A = 1)} Y \mid C, M = m, A = 1 \right] P(M = m \mid C, A = 1) dm$$

$$= E\left[ \frac{P(M \mid C, A = 0)}{P(M \mid C, A = 1)} Y \mid C, A = 1 \right]$$

$$= \frac{E\left[ \frac{P(M \mid C, A=0)}{P(M \mid C, A=1)} Y \cdot I(A = 1) \mid C \right]}{P(A = 1 \mid C)}$$

$$= E\left[ \frac{A}{P(A = 1 \mid C)} \frac{P(M \mid C, A = 0)}{P(M \mid C, A = 1)} Y \mid C \right].$$

Combining this with the outer expectation, we have

$$E[Y_{1M_0}] = E\left[ \frac{A}{P(A = 1 \mid C)} \frac{P(M \mid C, A = 0)}{P(M \mid C, A = 1)} Y \right].$$

As $E\left[ \frac{A}{P(A=1|C)} \frac{P(M|C,A=0)}{P(M|C,A=1)} \right] = 1$, it follows that

$$E[Y_{1M_0}] = \frac{E\left[ \frac{A}{P(A=1|C)} \frac{P(M|C,A=0)}{P(M|C,A=1)} Y \right]}{E\left[ \frac{A}{P(A=1|C)} \frac{P(M|C,A=0)}{P(M|C,A=1)} \right]},$$

which is a weighted mean of the outcome in treated units. This gives the first formula of the cross-world weights for treated units,

$$\omega_{\text{x}}(C, M) = \frac{1}{P(A = 1 \mid C)} \frac{P(M \mid C, A = 0)}{P(M \mid C, A = 1)}. \tag{A1}$$

## A.3 Second expression

A similar but slightly different manipulation of the identification result in A.1 will obtain the second expression for $\omega_{\text{x}}(C, M)$. As we already have the first expression though, it is easier to start from there and use Bayes' rule.

$$\omega_{\text{x}}(C, M) = \frac{1}{P(A = 1 \mid C)} \frac{P(M \mid C, A = 0)}{P(M \mid C, A = 1)} \tag{A1}$$

3

$$= \frac{1}{\mathrm{P}(A=1\mid C)} \frac{\mathrm{P}(M,A=0\mid C)/\mathrm{P}(A=0\mid C)}{\mathrm{P}(M,A=1\mid C)/\mathrm{P}(A=1\mid C)}$$

$$= \frac{\mathrm{P}(M,A=0\mid C)}{\mathrm{P}(M,A=1\mid C)} \frac{1}{\mathrm{P}(A=0\mid C)}$$

$$= \frac{\mathrm{P}(M\mid C)\mathrm{P}(A=0\mid C,M)}{\mathrm{P}(M\mid C)\mathrm{P}(A=1\mid C,M)} \frac{1}{\mathrm{P}(A=0\mid C)}$$

$$= \frac{\mathrm{P}(A=0\mid C,M)}{\mathrm{P}(A=1\mid C,M)} \frac{1}{\mathrm{P}(A=0\mid C)}. \tag{A2}$$

## A.4 Third expression

The third expression presented in the main text is an expression of the stabilized version of the cross-world weights. That is, of $\omega_{\mathrm{x}}^{\text{stabilized}}(C,M) = \omega_{\mathrm{x}}(C,M)\mathrm{P}(A=1)$.

We can anchor the stabilized version on either (A1) or (A2) and get the same result. Anchoring on (A1),

$$\omega_{\mathrm{x}}^{\text{stabilized}}(C,M) = \frac{\mathrm{P}(A=1)}{\mathrm{P}(A=1\mid C)} \frac{\mathrm{P}(M\mid C, A=0)}{\mathrm{P}(M\mid C, A=1)}.$$

Again, using Bayes' rule,

$$\omega_{\mathrm{x}}^{\text{stabilized}}(C,M) = \frac{\mathrm{P}(A=1)}{\mathrm{P}(A=1\mid C)} \frac{\mathrm{P}(M,C\mid A=0)/\mathrm{P}(C\mid A=0)}{\mathrm{P}(M,C\mid A=1)/\mathrm{P}(C\mid A=1)}$$

$$= \frac{\mathrm{P}(A=1)}{\mathrm{P}(A=1\mid C)} \frac{\mathrm{P}(M,C\mid A=0)}{\mathrm{P}(M,C\mid A=1)} \frac{\mathrm{P}(C\mid A=1)}{\mathrm{P}(C\mid A=0)}$$

$$= \frac{\mathrm{P}(A=1)}{\mathrm{P}(A=1\mid C)} \frac{\mathrm{P}(M,C\mid A=0)}{\mathrm{P}(M,C\mid A=1)} \frac{\mathrm{P}(C)\mathrm{P}(A=1\mid C)/\mathrm{P}(A=1)}{\mathrm{P}(C)\mathrm{P}(A=0\mid C)/\mathrm{P}(A=0)}$$

$$= \frac{\mathrm{P}(M,C\mid A=0)\frac{\mathrm{P}(A=0)}{\mathrm{P}(A=0\mid C)}}{\mathrm{P}(M,C\mid A=1)}. \tag{A3}$$

# B Robustness of the *more robust* estimators

## B.1 Notation

We denote true conditional mean or density functions, true weighting functions, true potential outcome means, and true average causal effects, etc. with a $^0$ superscript. Specifically, the true weighting functions are denoted by

$$\omega_0^0(C) := \frac{1}{\mathrm{P}(A=0\mid C)},$$

$$\omega_1^0(C) := \frac{1}{\mathrm{P}(A=1\mid C)},$$

$$\omega_{\mathrm{x}}^0(C,M) := \frac{1}{\mathrm{P}(A=1\mid C)} \frac{\mathrm{P}(M\mid C, A=0)}{\mathrm{P}(M\mid C, A=1)} = \frac{\mathrm{P}(A=0\mid C,M)}{\mathrm{P}(A=1\mid C,M)} \frac{1}{\mathrm{P}(A=0\mid C)};$$

the true conditional mean outcome or effect functions are denoted by

$$\mu^0(C,M) := \mathrm{E}[Y\mid C,M,A=1],$$

$$\nu_1^0(C) := \mathrm{E}[Y_1\mid C] = \mathrm{E}[Y\mid C, A=1],$$

$$\nu_0^0(C) := \mathrm{E}[Y_0 \mid C] = \mathrm{E}[Y \mid C, A = 0],$$
$$\nu_x^0(C) := \mathrm{E}[Y_{1M_0} \mid C] = \mathrm{E}\{\mathrm{E}[Y \mid C, M, A = 1] \mid C, A = 0\} = \mathrm{E}[\mu^0(C, M) \mid C, A = 0],$$
$$\xi^0(C) := \mathrm{E}[NDE_0 \mid C] = \nu_x^0(C) - \nu_0^0(C);$$

the true conditional exposure probabily and mediator density functions are denoted by

$$\delta^0(C, A) := \mathrm{P}(A \mid C),$$
$$\gamma^0(C, M, A) := \mathrm{P}(A \mid C, M),$$
$$\lambda_0^0(C, M) := \mathrm{P}(M \mid C, A = 0),$$
$$\lambda_1^0(C, M) := \mathrm{P}(M \mid C, A = 1);$$

the true marginal means of potential outcomes are denoted by

$$\tau_1^0 := \mathrm{E}[Y_1] = \mathrm{E}\{\mathrm{E}[Y \mid C, A = 1]\} = \mathrm{E}[\nu_1^0(C)],$$
$$\tau_0^0 := \mathrm{E}[Y_0] = \mathrm{E}\{\mathrm{E}[Y \mid C, A = 0]\} = \mathrm{E}[\nu_0^0(C)],$$
$$\tau_x^0 := \mathrm{E}[Y_{1M_0}] = \mathrm{E}(\mathrm{E}\{\mathrm{E}[Y \mid C, M, A = 1] \mid C, A = 0\}) = \mathrm{E}\{\mathrm{E}[\mu^0(C, M) \mid C, A = 0]\} = \mathrm{E}[\nu_x^0(C)];$$

and the true effects are denoted by

$$\beta_0^0 := \mathrm{NDE}_0 = \tau_x^0 - \tau_0^0 = \mathrm{E}[\xi^0(C)],$$
$$\beta_1^0 := \mathrm{NIE}_1 = \tau_1^0 - \tau_x^0,$$
$$\beta_2^0 := \mathrm{TE} = \tau_1^0 - \tau_0^0.$$

Conditional mean and density models are indexed with parameters, all called $\theta$ with some subscript to tell them apart if needed. For example, $\mu(C, M; \theta_\mu)$ is the model for $\mu^0(C, M)$. The parameter estimate $\hat{\theta}_\mu$ gives the estimated function $\mu(C, M; \hat{\theta}_\mu)$, which we refer to by the shorthand $\hat{\mu}(C, M)$. For the most part we will refer to the estimators of these functions by this ˆ notation: $\hat{\omega}_0(C), \hat{\omega}_1(C), \hat{\omega}_x(C)$, $\hat{\mu}(C, M)$, $\hat{\nu}_1(C), \hat{\nu}_0(C), \hat{\nu}_x(C)$, $\hat{\xi}(C)$, $\hat{\delta}(C, A), \hat{\gamma}(C, M, A)$, $\hat{\lambda}_0(C, M), \hat{\lambda}_1(C, M)$. We also use ˆ notation to refer to the final estimators of the potential outcome means and the effects: $\hat{\tau}_1, \hat{\tau}_0, \hat{\tau}_x, \hat{\beta}_0, \hat{\beta}_1, \hat{\beta}_2$.

Under regularity conditions (section B.3), estimators converge to their probability limits. These are denoted with a $*$ superscript, e.g., $\omega_0^*(C), \omega_1^*(C), \omega_x^*(C, M)$, $\mu^*(C, M)$, $\nu_1^*(C), \nu_0^*(C), \nu_x^*(C)$, $\xi^*(C)$, $\delta^*(C, A)$, $\gamma^*(C, M, A), \lambda_0^*(C, M), \lambda_1^*(C, M), \tau_1^*, \tau_0^*, \tau_x^*, \beta_0^*, \beta_1^*, \beta_2^*$.

An additional shorthand notation is the use of $\nabla$ (and $\nabla^T$) to denote the partial derivative of a function with a parameter argument with respect to the parameter as a column (row) vector, and $\nabla \nabla^T$ to denote the second partial derivative. For example,

$$\nabla \alpha(Z, \theta_j) := \frac{\partial}{\partial \theta_j} \alpha(Z, \theta_j), \ \nabla^T \alpha(Z, \theta_j) := \frac{\partial}{\partial \theta_j^T} \alpha(Z, \theta_j), \ \nabla \nabla^T \alpha(Z; \theta_j) := \frac{\partial^2}{\partial \theta_j \partial \theta_j^T} \alpha(Z; \theta_j).$$

When we evaluate these derivatives at the probability limit $\theta_j^*$ of $\hat{\theta}_j$, we write the resulting functions using the shorthand $\nabla \alpha^*(Z)$, $\nabla^T \alpha^*(Z)$ and $\nabla \nabla^T \alpha^*(Z)$. (This is an abuse of notation, as the function being evaluated is the derivative function, not the $\alpha()$ function.)

## B.2 Road map

We restrict our attention to the case where weighting and density functions are estimated based on parametric models, and conditional mean models are estimated (semi)parametrically.

5

In this section we show the robustness properties of the *more robust* estimators with the three-step reasoning below, which applies to all the estimators.

1. The first step is to express the estimator as the solution to an estimating equation

$$P_n \psi(Z, \theta) = 0, \tag{A4}$$

   where $P_n$ is the notation for the sample average, $\psi(Z, \theta)$ is a known vector-valued function of data (denoted by $Z = (C, A, M, Y)$) and parameter vector $\theta$; the form of $\psi(Z, \theta)$ is specific to the estimator. As the estimator involves multiple steps, $\psi(Z, \theta)$ include multiple components, one for each step. $\theta$ includes "nuisance" parameters (e.g., parameters of propensity score, mediator density or outcome models) and the parameter of interest (a potential outcome mean or an effect, depending on the estimator). That is, $\theta = (\theta_{\text{nuisance}}^T, \theta_{\text{target}})^T$. Denote the solution to (A4) by $\hat{\theta} = (\hat{\theta}_{\text{nuisance}}^T, \hat{\theta}_{\text{target}})^T$.

2. The second step is to state regularity conditions under which $\hat{\theta}$ converges in probability to the solution (assumed to be unique) of the equation

$$E[\psi(Z, \theta)] = 0, \tag{A5}$$

   where the expectation is taken with respect to the true distribution of $Z$. Denote this solution by $\theta^* = (\theta_{\text{nuisance}}^{*T}, \theta_{\text{target}}^*)^T$. Assuming such convergence, $\theta_{\text{target}}^*$ is the probability limit of the estimator.

3. The last step is to show, based on (A5), that (a) $\theta_{\text{target}}^*$ is equal to the true value of the parameter of interest (which implies the estimator is consistent) when certain modeling components are consistent; and that (b) it is not equal to the true value when certain model components are misspecified – as claimed in the paper.

For a generic estimating function $\psi(Z, \theta)$, we list in section B.3 the regularity conditions for $\hat{\theta}$ to converge in probability to $\theta^*$; this suffices for Step 2 for all of our estimators. Steps 1 and 3, which vary estimator to estimator, are handled in section B.4. Before diving into the details, section B.1 introduces some simplifying notation.

## B.3 Regularity conditions for estimator to converge in probability

For concreteness, take the reg|Ypred.R estimator as an example. This estimator is symmetric with respect to the two regular potential outcome means, $\tau_1^0 = E[Y_1]$ and $\tau_0^0 = E[Y_0]$, so we need to consider only one of the two, say, $\tau_1^0$. As an estimator of $\tau_1^0$, reg|Ypred.R involves: (1) fitting a propensity score model $\delta(C, A, \theta_1)$; (2) using it to compute weights that form the pseudo treated sample and then fitting a model $\nu_1(C, \theta_2)$ regressing $Y$ on $C$ in this pseudo sample; and (3) using this fitted model to predict $Y_1$ for all units in the full sample and averaging these predicted values. These three steps correspond to solving three estimating equations. Equivalently, the three steps combined solve a stacked estimating equation

$$P_n \psi(Z, \theta) = 0 \tag{A4}$$

for $\theta$. In this case

$$\psi(Z, \theta) = \begin{pmatrix} \nabla \log \delta(C, A, \theta_1) \\ \frac{A}{\delta(C, A, \theta_1)} g(C)[Y - \nu_1(C, \theta_2)] \\ \nu_1(C, \theta_2) - \tau_1 \end{pmatrix}, \tag{A6}$$

and $\theta = (\theta_1^T, \theta_2^T, \tau_1)$.

Like reg|Ypred.R, each of the other estimators in the paper involves multiple steps, with functions estimated from earlier steps plugged into subsequent steps, resulting in an estimate of a potential outcome mean (or

of an effect, depending on the estimator) at the last step. This means each estimator may be expressed as the solution of estimating equation (A4) for some estimation function $\psi(Z,\theta)$, where the last element of $\theta$ is the parameter of interest – a potential outcome mean or an effect, depending on the specific estimator. The form of $\psi(Z,\theta)$ will be made explicit for each estimator in section B.4.

We now leave the specific example, and instead consider $\psi(Z,\theta)$ as a generic estimating function corresponding to one of the estimators of interest.

At the start of this section B, we mentioned that one step in explaining the robustness properties of the estimator is to lay out conditions under which $\hat{\theta}$ that solves $P_n\psi(Z,\theta) = 0$ (A4) converges in probability to $\theta^*$ that solves

$$E[\psi(Z,\theta)] = 0. \tag{A5}$$

The following regularity conditions (based on theorem 7.1 in Boos and Stefanski, 2013) provide for such convergence (the convergence is wp1):

- $\theta \in \Theta$ where $\Theta$ is a compact subset of $R^d$, $d$ being the dimension of $\theta$.
- For all $\theta \in \Theta$, $E[\psi(Z,\theta)] < \infty$, and $E[\psi(Z,\theta)]$ has a unique zero, $\theta^*$.
- Every element of vector $\psi(Z,\theta)$ is continuous in $\theta$ and is dominated by a function $h(Z)$, $E[h(Z)] < \infty$.

## B.4 Robustness properties of the estimators

Throughout this section we assume that the regularity conditions listed above hold and $\hat{\theta}$ converges in probability to $\theta^* = (\theta^{*T}_{\text{nuisance}}, \theta^*_{\text{target}})^T$, the unique solution to (A5).

Note that depending on the estimator $\theta^*_{\text{target}}$ may be $\tau_1^*, \tau_0^*, \tau_x^*$ or $\beta_0^*, \beta_1^*, \beta_2^*$.

We will show, for each of the *more robust* estimators, that $\theta^*_{\text{target}}$ (a) equals the true parameter when certain subsets of modeling components are consistent; and (b) generally does not equal the true parameter when certain modeling components are misspecified. Combined with the assumption that $\hat{\theta}$ converges in probability to $\theta^*$, these (a) provide the different chances the estimator has to be consistent, and (b) show the estimator's dependence on correct specification of specific components.

For each estimator, our starting point now is

$$E[\psi(Z,\theta^*)] = 0. \tag{A7}$$

This is an equality, not an equation, as we have plugged in the parameter value $\theta^*$ that, by definition, makes it true. Using the connections among elements of $\psi(Z,\theta^*)$, we will equate (or fail to equate) $\theta^*_{\text{target}}$ with the true parameter in different cases of correct/misspecified model components. The arguments used are similar for all the estimators, so we will spell things out clearly below for the first estimator, but will be more concise with the subsequent ones (skipping steps that are easily reconstructed).

### B.4.1 Estimator of the regular potential outcome mean $E[Y_1]$

#### B.4.1.1 reg|Ypred.R

Again, consider reg|Ypred.R as the estimator of $\tau_1^0 = E[Y_1]$. This is a well-known doubly robust estimator of the mean of a variable in the presence of missing data. As it is the simplest one among the *more robust* estimators covered in the paper, we go through the steps of showing its robustness here, to clarify the type of reasoning which will also apply to the other estimators.

As mentioned above, this estimator has an estimating function of the form,

$$\psi(Z, \theta) = \begin{pmatrix} \nabla \log \delta(C, A, \theta_1) \\ \frac{A}{\delta(C, A, \theta_1)} g(C)[Y - \nu_1(C, \theta_2)] \\ \nu_1(C, \theta_2) - \tau_1 \end{pmatrix}. \quad (A6)$$

Plugging in $\theta^*$, the solution to $\mathrm{E}[\psi(Z, \theta)] = 0$ (A5), we have the function

$$\psi(Z, \theta^*) = \begin{pmatrix} \nabla \log \delta^*(C, A) \\ A \omega_1^*(C) g(C)[Y - \nu_1^*(C)] \\ \nu_1^*(C) - \tau_1^* \end{pmatrix}. \quad (A8)$$

Recall that $\nabla \log \delta^*(C, A)$ and $\delta^*(C, A)$ are shorthand for $\nabla \log \delta(C, A, \theta_1)$ and $\delta(C, A, \theta_1)$ evaluated at $\theta_1^*$, and $\nu_1^*(C)$ is shorthand for $\nu_1(C, \theta_2)$ evaluated at $\theta_2^*$. Also, $\omega_1^*(C) = \frac{1}{\delta^*(C, 1)}$.

Note that the *mean recovery* condition we impose on the outcome model $\nu_1(C, \theta_2)$ fit to the pseudo treated sample means that the vector-valued function $g(C)$ has a constant element (e.g., representing an intercept in a generalized linear model).

The equality

$$\mathrm{E}[\psi(Z, \theta^*)] = 0 \quad (A7)$$

is thus the combination of several equalities. For the current task, we only need to use two of those equalities,

$$\mathrm{E}\{A \omega_1^*(C)[Y - \nu_1^*(C)]\} = 0, \quad (A9)$$
$$\mathrm{E}[\nu_1^*(C) - \tau_1^*] = 0. \quad (A10)$$

**Two chances to be correct**

Case 1: If the weights are consistently estimated, i.e., $\omega_1^*(C) = \omega_1^0(C) := 1/\mathrm{P}(A = 1 \mid C)$, (A9) implies

$$\begin{aligned}
0 &= \mathrm{E}\{A \omega_1^*(C)[Y - \nu_1^*(C)]\} \\
&= \mathrm{E}\left\{A \frac{1}{\mathrm{P}(A = 1 \mid C)}[Y - \nu_1^*(C)]\right\} \\
&= \mathrm{E}\left(\mathrm{E}\left[\mathrm{E}\left\{A \frac{1}{\mathrm{P}(A = 1 \mid C)}[Y - \nu_1^*(C)] \mid C, A\right\} \mid C\right]\right) \\
&= \mathrm{E}[\nu_1^0(C) - \nu_1^*(C)] \\
&= \tau_1^0 - \mathrm{E}[\nu_1^*(C)].
\end{aligned}$$

This, combined with (A10), obtains $\tau_1^* = \tau_1^0$.

Case 2: If the outcome model $\nu_1(C, \theta_2)$ is correctly specified, then $\nu_1^*(C) = \nu_1^0(C)$ regardless of the weighting function. This, combined with (A10), obtains $\tau_1^* = \tau_1^0$.

### B.4.2 Estimators of the cross-world potential outcome mean $\mathrm{E}[Y_{1M_0}]$

We will use the same kind of reasoning as we did with the reg|Ypred.R estimator (see B.4.1.1). Note that we did not need to use the component of $\mathrm{E}[\psi(Z, \theta^*)] = 0$ relating to the model used to estimate the weights (although it will be relevant later for variance derivation). This is true for the rest of this section. This means we can for now ignore the elements of $\psi(Z, \theta)$ that correspond to the weights estimation steps.

### B.4.2.1 crw|psYpred.MR

The estimating function corresponding to this estimator of $\tau_x^0 = \mathrm{E}[Y_{1M_0}]$ is

$$\psi(Z, \theta) = \begin{pmatrix} \text{weights estimation elements with parameters } \theta_1, \theta_2 \\ A\omega_x(C, M, \theta_1, \theta_2)h(C, M)[Y - \mu(C, M, \theta_3)] \\ (1 - A)\omega_0(C, \theta_1)[\mu(C, M, \theta_3) - \tau_x] \end{pmatrix}, \tag{A11}$$

where the vector-valued function $h(C, M)$ has a constant element to satisfy the *mean recovery* condition imposed on the outcome-given-covariates-and-mediators model fit to the pseudo cross-world sample. Based on this, the equality $\mathrm{E}[\psi(Z, \theta^*)] = 0$ includes the following two equalities:

$$\mathrm{E}\left\{A\omega_x^*(C, M)[Y - \mu^*(C, M)]\right\} = 0, \tag{A12}$$
$$\mathrm{E}\left\{(1 - A)\omega_0^*(C)[\mu^*(C, M) - \tau_x^*]\right\} = 0. \tag{A13}$$

#### *Two chances to be correct*

For now, assume that the weights that form the pseudo control sample are consistently estimated, i.e., $\omega_0^*(C) = \omega_0^0(C) = 1/\mathrm{P}(A = 0 \mid C)$. Then (A13) implies

$$\tau_x^* = \mathrm{E}\{\mathrm{E}[\mu^*(C, M) \mid C, A = 0]\}. \tag{A14}$$

Case 1: If in addition to consistent estimation of the control weights, the cross-world weights are also consistently estimated, i.e., $\omega_x^*(C, M) = \omega_x^0(C, M) = \dfrac{1}{\mathrm{P}(A = 1 \mid C)} \dfrac{\mathrm{P}(M \mid C, A = 0)}{\mathrm{P}(M \mid C, A = 1)}$, then (A12) implies

$$\begin{aligned}
0 &= \mathrm{E}\left\{A\omega_x^*(C, M)[Y - \mu^*(C, M)]\right\} \\
&= \mathrm{E}\left[\mathrm{E}\left(\mathrm{E}\left\{\frac{A}{\mathrm{P}(A = 1 \mid C)}\frac{\mathrm{P}(M \mid C, A = 0)}{\mathrm{P}(M \mid C, A = 1)}[Y - \mu^*(C, M)] \mid C, A, M\right\} \mid C, A\right)\right] \\
&= \tau_x^0 - \mathrm{E}\{\mathrm{E}[\mu^*(C, M) \mid C, A = 0]\}.
\end{aligned}$$

This, combined with (A14), obtains $\tau_x^* = \tau_x^0$.

Case 2: If in addition to consistent estimation of the control weights, $\mu(C, M, \theta_3)$ is correctly specified, then $\mu^*(C, M) = \mu^0(C, M)$. This, combined with (A14), obtains $\tau_x^* = \tau_x^0$.

#### *Dependence on consistent $\omega_0(C)$ weights*

If the control weights are not consistent, i.e., $\omega_0^*(C) = \upsilon_0^*(C)\omega_0^0(C)$ where $\upsilon_0^*(C)$ represents an asymptotic bias factor for the weights, which is a function of $C$. Without loss of generality, assume the weights estimation is inconsistent but does not distort the overall scale of the weights, so $\mathrm{E}_C[\upsilon_0^*(C)] = 1$. Then (A13) implies

$$\tau_x^* = \mathrm{E}\{\upsilon_0^*(C)\mathrm{E}[\mu^*(C, M) \mid C, A = 0]\},$$

which is generally not equal to $\tau_x^0 = \mathrm{E}\{\mathrm{E}[\mu^0(C, M) \mid C, A = 0]\}$.

Let us consider several special cases to see if the specific conditions help equate $\tau_x^*$ with $\tau_x^0$.

- If $\mu(C, M, \theta_3)$ is correctly specified, $\mu^*(C, M) = \mu^0(C, M)$. It follows from (A13) that $\tau_x^* = \mathrm{E}[\upsilon_0^*(C)\nu_x^0(C)]$, but this is different from $\tau_x^0 = \mathrm{E}[\nu_x^0(C)]$.

- If the cross-world weights are consistent, (A12) implies $\tau_x^0 = \mathrm{E}\{\mathrm{E}[\mu^*(C, M) \mid C, A = 0]\}$, but this is different from $\tau_x^* = \mathrm{E}\{\upsilon_0^*(C)\mathrm{E}[\mu^*(C, M) \mid C, A = 0]\}$.

- Combining the two conditions above also does not equate $\tau_x^*$ to $\tau_x^0$.

- If the $M$-related component of the cross-world weights is consistent but the other component is not, i.e., $\omega_x^*(C,M) = \upsilon_x^*(C)\omega_x^0(C,M)$, then (A12) implies $\mathrm{E}\{\upsilon_x^*(C)\mathrm{E}[\mu^*(C,M) \mid C, A=0]\} = \mathrm{E}[\upsilon_x^*(C)\nu_x^0(C)]$. If additionally, $\upsilon_x^*(C) = \upsilon_0^*(C)$ (which is the case if we estimate cross-world weights based on the second formula, $\frac{\mathrm{P}(A=0 \mid C,M)}{\mathrm{P}(A=1 \mid C,M)}\frac{1}{\mathrm{P}(A=0 \mid C)}$), then $\tau_x^* = \mathrm{E}[\upsilon_0^*(C)\nu_x^0(C)]$, which is not equal to $\tau_x^0 = \mathrm{E}[\nu_x^0(C)]$.

### B.4.2.2   crw|Ypred.MR

The estimating function for this estimator of $\tau_x^0 = \mathrm{E}[Y_{1M_0}]$ is

$$\psi(Z,\theta) = \begin{pmatrix} \text{weights estimation elements with parameters } \theta_1, \theta_2 \\ A\omega_x(C,M,\theta_1,\theta_2)g(C)[Y - \nu_x(C,\theta_3)] \\ \nu_x(C,\theta_3) - \tau_x \end{pmatrix}, \tag{A15}$$

where $g(C)$ has a constant element to satisfy the *mean recovery* condition imposed on the outcome-given-covariates model fit to the pseudo cross-world sample. Based on this, the equality $\mathrm{E}[\psi(Z,\theta^*)] = 0$ includes the following two equalities:

$$\mathrm{E}\{A\omega_x^*(C,M)[Y - \nu_x^*(C)]\} = 0, \tag{A16}$$
$$\mathrm{E}[\nu_x^*(C) - \tau_x^*] = 0. \tag{A17}$$

***Two chances to be correct***

For now, assume that the $M$-related component of the cross-world weights is consistently estimated, $\omega_x^*(C,M) = \upsilon_x^*(C)\omega_x^0(C,M)$ where $\upsilon_x^*(C)$ represents an asymptotic bias factor function that solely concerns the $C$ distribution. Then (A16) implies

$$\begin{aligned} 0 &= \mathrm{E}\{A\upsilon_x^*(C)\omega_x^0(C,M)[Y - \nu_x^*(C)]\} \\ &= \mathrm{E}[\upsilon_x^*(C)\mathrm{E}\{A\omega_x^0(C,M)[Y - \nu_x^*(C)] \mid C\}] \\ &= \mathrm{E}\{\upsilon_x^*(C)[\nu_x^0(C) - \nu_x^*(C)]\}. \end{aligned} \tag{A18}$$

Case 1: If in addition to consistent estimation of the $M$-related component of the cross-world weights, the $\nu_x(C,\theta_3)$ model is correctly specified, this combined with (A18) gives $\nu_x^*(C) = \nu_x^0$. This, combined with (A17), obtains $\tau_x^* = \tau_x^0$.

Case 2: If instead the cross-world weights are fully consistently estimated, $\upsilon^*(C) = 1$, so (A18) implies $\mathrm{E}[\nu_x^*(C)] = \mathrm{E}[\nu_x^0(C)] = \tau_x^0$. This, combined with (A17), obtains $\tau_x^* = \tau_x^0$.

***Dependence on consistent estimation of the $M$-related component of $\omega_x(C,M)$ weights***

If the $M$-related component of the cross-world weights is not consistent, i.e., $\omega_x^*(C,M) = \chi^*(C,M)\omega_x^0(C,M)$ where $\chi^*(C,M)$ represents the bias factor function, then (A16) implies the following (instead of (A18))

$$\mathrm{E}\{\mathrm{E}[\chi^*(C,M) \mid C, A=0]\nu_x^*(C)\} = \mathrm{E}\{\mathrm{E}[\chi^*(X,M)\mu^0(C,M) \mid C, A=0]\}. \tag{A19}$$

This does not equate $\tau_x^*$ to $\tau_x^0$.

Let us consider the special case where the weights are correct w.r.t. the $C$ distribution (but the $M$-related component of the weights is still inconsistent). Now the LHS of (A19) simplifies to $\mathrm{E}[\nu_x^*(C)]$, which combined with (A17) implies $\tau_x^* = \mathrm{E}\{\mathrm{E}[\chi^*(X,M)\mu^0(C,M) \mid C, A=0]\}$. However, this is still different from $\tau_x^0 = \mathrm{E}\{\mathrm{E}[\mu^0(C,M) \mid C, A=0]\}$. That is, $\tau_x^* \neq \tau_x^0$.

### B.4.2.3 crw|Y2pred.R

The estimating function for this estimator of $\tau_x^0 = \mathrm{E}[Y_{1M_0}]$ is

$$\psi(Z,\theta) = \begin{pmatrix} \text{weights estimation elements with parameters } \theta_1, \theta_2 \\ A\omega_x(C,M,\theta_1,\theta_2)h(C,M)[Y-\mu(C,M,\theta_3)] \\ (1-A)\omega_0(C,\theta_1)g(C)[\mu(C,M,\theta_3)-\nu_x(C,\theta_4)] \\ \nu_x(C,\theta_4)-\tau_x \end{pmatrix}, \quad \text{(A20)}$$

where $h(C,M)$ and $g(C)$ both have constant elements to satisfy the *mean recovery* condition imposed on the two outcome models. Based on this, the equality $\mathrm{E}[\psi(Z,\theta^*)]=0$ includes the following three equalities:

$$\mathrm{E}\{A\omega_x^*(C,M)[Y-\mu^*(C,M)]\} = 0, \quad \text{(A21)}$$
$$\mathrm{E}\{(1-A)\omega_0^*(C)[\mu^*(C,M)-\nu_x^*(C)]\} = 0, \quad \text{(A22)}$$
$$\mathrm{E}[\nu_x^*(C)-\tau_x^*] = 0. \quad \text{(A23)}$$

#### *Three chances to be correct*

Case 1: The cross-world and control weights are both consistently estimated, i.e., $\omega_0^*(C) = \omega_0^0(C)$ and $\omega_x^*(C,M) = \omega_x^0(C,M)$. Then (A21) implies

$$\mathrm{E}\{\mathrm{E}[\mu^*(C,M) \mid C, A=0]\} = \mathrm{E}(\mathrm{E}\{\mathrm{E}[Y \mid C,M,A=1] \mid C, A=0\}) = \tau_x^0,$$

and (A22) implies

$$\mathrm{E}[\nu_x^*(C)] = \mathrm{E}\{\mathrm{E}[\mu^*(C,M) \mid C, A=0]\}.$$

It follows that $\mathrm{E}[\nu_x^*(C)] = \tau_x^0$. This combined with (A23) obtains $\tau_x^* = \tau_x^0$.

Case 2: The cross-world weights are consistent (i.e., $\omega_0^*(C) = \omega_0^0(C)$) and $\mu(C,M,\theta_3)$ is correctly specified (thus $\mu^*(C,M) = \mu^0(C,M)$). It follows from (A22) that

$$\mathrm{E}[\nu_x^*(C)] = \mathrm{E}\{\mathrm{E}[\mu^0(C,M) \mid C, A=0]\} = \tau_x^0.$$

Case 3: $\mu(C,M,\theta_3)$ and $\nu_x(C,\theta_4)$ are correctly specified. Then $\mu^*(C,M) = \mu^0(C,M)$, leading to $\nu_x^*(C) = \nu_x^0(C)$. This combined with (A23) obtains $\tau_x^* = \tau_x^0$.

Strictly speaking, there is another possibility for the estimator to be consistent, which is when cross-world weights are consistent and $\nu_x(C,\theta_4)$ is correctly specified for $\mathrm{E}[\mu^*(C,M) \mid C, A=0]$ (which may be different from $\nu_x^0(C)$). In this Case 4, (A21) implies $\mathrm{E}\{\mathrm{E}[\mu^*(C,M) \mid C, A=0]\} = \tau_x^0$, and (A22) implies $\nu_x^*(C) = \mathrm{E}\{\mathrm{E}[\mu^*(C,M) \mid C, A=0]\}$, which combined means $\tau_x^0 = \mathrm{E}[\nu_x^*(C)] = \tau_x^*$. However, this theoretical possibility is not practically helpful, because given all the weights estimation procedures considered, if the cross-world weights are correct, then the control weights are also correct, which already satisfy the conditions of Case 1.

### B.4.2.4 crw|MsimYpred.MR

The estimating function for this estimator of $\tau_x^0 = \mathrm{E}[Y_{1M_0}]$ is

$$\psi(Z,\theta) = \begin{pmatrix} \text{weights estimation elements} \\ (1-A)\omega_0(C)\nabla \log \lambda_0(C,M,\theta_3) \\ A\omega_x(C,M)h(C,M)[Y-\mu(C,M,\theta_4)] \\ \mathrm{E}^*[\mu^*(C,M) \mid C, A=0] - \tau_x^* \end{pmatrix}, \quad \text{(A24)}$$

where $h(C, M)$ has a constant element to satisfy the *mean recovery* condition imposed on the $\mu(C, M)$ model, and
$$\mathrm{E}^*[\mu^*(C, M) \mid C, A = 0] := \int \mu^*(C, m)\lambda_0^*(C, m)dm.$$

(We are vague about the parameters involved in weights estimation here. Depending on how the cross-world weights are estimated – using expression (A1), (A2) or (A3) – $\omega_x(C, M)$ may or may not depend on $\theta_3$. This does not make a difference for what we do next.)

Based on this, the equality $\mathrm{E}[\psi(Z, \theta^*)] = 0$ includes the following equalities

$$\mathrm{E}[(1 - A)\omega_0^*(C)\nabla \log \lambda_0^*(C, M)] = 0, \tag{A25}$$
$$\mathrm{E}\{A\omega_x^*(C, M)[Y - \mu^*(C, M)]\} = 0, \tag{A26}$$
$$\mathrm{E}\{\mathrm{E}^*[\mu^*(C, M) \mid C, A = 0] - \tau_x^*\} = 0. \tag{A27}$$

**Two chances to be correct**

For now, assume that $\lambda_0(C, M, \theta_3)$ is correctly specified. Then $\lambda_0^*(C, M) = \lambda_0^0(C, M)$. This combined with (A27) implies
$$\tau_x^* = \mathrm{E}\{\mathrm{E}[\mu^*(C, M) \mid C, A = 0]\}. \tag{A28}$$

Case 1: In addition to correct specification of $\lambda_0(C, M, \theta_3)$, if $\mu(C, M, \theta_4)$ is also correctly specified, then $\mu^*(C, M) = \mu^0(C, M)$. This combined with (A28) implies $\tau_x^* = \tau_x^0$.

Case 2: In addition to correct specification of $\lambda_0(C, M, \theta_3)$, if the cross-world weights are consistent, i.e., $\omega_x^*(C, M) = \omega_x^0(C, M)$, then (A26) implies $\mathrm{E}\{\mathrm{E}[\mu^*(C, M) \mid C, A = 0]\} = \tau_x^0$. This combined with (A28) implies $\tau_x^* = \tau_x^0$.

**Dependence on correct specification of the mediator density model $\lambda_0(C, M)$**

We have just shown that when (i) $\lambda_0(C, M, \theta_3)$ is correctly specified, and either (ii-a) $\mu(C, M, \theta_4)$ is correctly specified or (ii-b) the control weights are consistently estimated, the crw|MsimYpred.MR estimator of $\tau_x^0$ is consistent. Now we show that if (i) does not hold, this estimator is generally not consistent, even if (ii-a) and (ii-b) hold.

Assume (ii-a) and (ii-b) (that is, $\mu^*(C, M) = \mu^0(C, M)$, and $\omega_0^*(C, M) = \omega_0^0(C, M)$). We will

- characterize the probability limit $\lambda_0^*(C, M) = \lambda_0(C, M, \theta_3^*)$, when the model $\lambda_0(C, M, \theta_3)$ is misspecified
- having characterized $\lambda_0^*(C, M)$, derive the asymptotic bias of the estimator crw|MsimYpred.MR
- consider this bias in the special case where $M$ is binary to see if it turns out to be zero (unfortunately it does not!)

*Characterization of $\lambda_0^*(C, M)$*

We replace $\omega_0^*(C)$ with $\omega_0^0(C)$ in (A25) and evaluate it using iterated expectation, obtaining
$$\mathrm{E}\{\mathrm{E}[\nabla \log \lambda_0(C, M, \theta_3^*) \mid C, A = 0]\} = 0.$$

This implies that $\theta_3^*$ is the parameter value that maximizes
$$\mathrm{E}\{\mathrm{E}[\log \lambda_0(C, M, \theta_3) \mid C, A = 0]\}. \tag{A29}$$

Note that $\mathrm{E}\{\mathrm{E}[\log \lambda_0^0(C, M) \mid C, A = 0]\}$ is a constant. This means $\theta_3^*$ is the minimizer of

$$\mathrm{E}\{\mathrm{E}[\log \lambda_0^0(C, M) \mid C, A = 0]\} - (\text{A29}) = \mathrm{E}\left\{\mathrm{E}\left[\log \frac{\lambda_0^0(C, M)}{\lambda_0(C, M, \theta_3)} \mid C, A = 0\right]\right\}$$

$$= \mathrm{E}\left[\int \lambda_0^0(C, m) \log \frac{\lambda_0^0(C, m)}{\lambda_0(C, m, \theta_3)} dm\right] \quad (\text{A30})$$

(recall that $\lambda_0^0(C, M) = \mathrm{P}(M \mid C, A = 0)$).

The expression inside the last expectation is a KL-divergence. This means $\lambda_0^*(C, M) = \lambda_0(C, M, \theta_3^*)$ is a KL-projection of the true conditional density $\lambda_0^0(C, M) = \mathrm{P}(M \mid C, A = 0)$ onto the $\lambda_0(C, M, \theta_3)$ model, where the criterion is the expectation of the conditional KL-divergence over the marginal distribution of $C$.

(Had we fit the model to the unweighted control subsample, this expectation would be over the conditional distribution of $C$.)

The only case where this projection $\lambda_0^*(C, M)$ coincides with the true density $\lambda_0^0(C, M)$ is where the model $\lambda_0(C, M, \theta_3)$ includes the true density, i.e., it is correctly specified.

*Bias formula*

This difference between the two densities $\lambda_0^*(C, M)$ and $\lambda_0^0(C, M)$ implies that the estimator has an asymptotic bias of magnitude

$$\begin{aligned}
\tau_\mathrm{x}^* - \tau_\mathrm{x}^0 &= \mathrm{E}\left\{\int \mu^0(C, m) \lambda_0^*(C, m) dm\right\} - \mathrm{E}\{\mathrm{E}[\mu^0(C, M) \mid C, A = 0]\} \\
&= \mathrm{E}\left\{\int \mu^0(C, m) \frac{\lambda_0^*(C, m)}{\lambda_0^0(C, m)} \lambda_0^0(C, m) dm\right\} - \mathrm{E}\{\mathrm{E}[\mu^0(C, M) \mid C, A = 0]\} \\
&= \mathrm{E}\left\{\mathrm{E}\left[\mu^0(C, M) \frac{\lambda_0^*(C, M)}{\lambda_0^0(C, M)} \mid C, A = 0\right]\right\} - \mathrm{E}\{\mathrm{E}[\mu^0(C, M) \mid C, A = 0]\} \\
&= \mathrm{E}\left\{\mathrm{E}\left[\mu^0(C, M) \left(\frac{\lambda_0^*(C, M)}{\lambda_0^0(C, M)} - 1\right) \mid C, A = 0\right]\right\}. \quad (\text{A31})
\end{aligned}$$

This case thus shows that the estimator is not consistent if the model $\lambda_0(C, M)$ for $\mathrm{P}(M \mid C, A = 0)$ is misspecified.

*Special case with binary $M$*

In this case, let $\pi^0(C)$ denote $\mathrm{P}(M = 1 \mid C, A = 0)$ and $\pi^*(C)$ denote $\lambda_0^*(C, 1)$. Then

$$\begin{aligned}
\tau_\mathrm{x}^* - \tau_x^0 &= \mathrm{E}\left\{\mu^0(C, 1)\left(\frac{\pi^*(C)}{\pi^0(C)} - 1\right)\pi^0(C) + \mu^0(C, 0)\left(\frac{1 - \pi^*(C)}{1 - \pi^0(C)} - 1\right)[1 - \pi^0(C)]\right\} \\
&= \mathrm{E}\left\{[\mu^0(C, 1) - \mu^0(C, 0)][\pi^*(C) - \pi^0(C)]\right\}, \quad (\text{A32})
\end{aligned}$$

which is generally nonzero.

### B.4.3   If targeting additive effects

#### B.4.3.1   NDE|YpredEpred.R

The estimating function for this estimator of $\beta_1^0 = \mathrm{E}[Y_{1M_0}] - \mathrm{E}[Y_0]$ is

$$\psi(Z, \theta) = \begin{pmatrix} \text{weights estimation elements with parameters } \theta_1, \theta_2 \\ A\omega_{\mathrm{x}}(C, M, \theta_1, \theta_2) h(C, M)[Y - \mu(C, M, \theta_3)] \\ (1-A)\omega_0(C, \theta_1) g(C)[\mu(C, M, \theta_3) - Y - \xi(C, \theta_4)] \\ \xi(C, \theta_4) - \beta_0 \end{pmatrix}, \quad \text{(A33)}$$

where $h(C, M)$ and $g(C)$ have constant elements to satisfy the *mean recovery* condition imposed on these conditional mean (outcome/effect) models. Based on this, the $\mathrm{E}[\psi(Z, \theta^*)] = 0$ equality includes the following inequalities

$$\mathrm{E}\{A\omega_{\mathrm{x}}^*(C, M)[Y - \mu^*(C, M)]\} = 0, \quad \text{(A34)}$$
$$\mathrm{E}\{(1-A)\omega_0^*(C)[\mu^*(C, M) - Y - \xi^*(C)]\} = 0, \quad \text{(A35)}$$
$$\mathrm{E}[\xi^*(C) - \beta_0^*] = 0. \quad \text{(A36)}$$

**Three chances to be correct**

Case 1: the control and cross-world weights are consistent, i.e., $\omega_0^*(C) = \omega_0^0(C)$ and $\omega_{\mathrm{x}}^*(C, M) = \omega_{\mathrm{x}}^0(C, M)$. Then (A34) implies

$$\mathrm{E}\{\mathrm{E}[\mu^*(C, M) \mid C, A = 0]\} = \tau_{\mathrm{x}}^0,$$

and (A35) implies

$$\mathrm{E}[\xi^*(C)] = \mathrm{E}\{\mathrm{E}[\mu^*(C, M) \mid C, A = 0]\} - \tau_0^0,$$

which together imply $\mathrm{E}[\xi^*(C)] = \tau_{\mathrm{x}}^0 - \tau_0^0 = \beta_0^0$. This combined with (A36) obtains $\beta_0^* = \beta_0^0$.

Case 2: the control weights are consistent and $\mu(C, M, \theta_3)$ is correctly specified, i.e., $\omega_0^*(C) = \omega_0^0(C)$ and $\mu^*(C, M) = \mu^0(C, M)$. It follows from (A35) that

$$\mathrm{E}[\xi^*(C)] = \mathrm{E}\{\mathrm{E}[\mu^0(C, M) - Y \mid C, A = 0]\} = \tau_{\mathrm{x}}^0 - \tau_0^0 = \beta_0^0.$$

This combined with (A36) obtains $\beta_0^* = \beta_0^0$.

Case 3: $\mu(C, M, \theta_3)$ and $\xi(C, \theta_4)$ are correctly specified. Then $\mu^*(C, M) = \mu^0(C, M)$, which entails

$$\xi^*(C) = \mathrm{E}[\mu^0(C, M) - Y \mid C, A = 0] = \nu_{\mathrm{x}}^0(C) - \nu_0^0(C).$$

Combining this with (A36) obtains

$$\beta_0^* = \mathrm{E}[\xi^*(C)] = \mathrm{E}[\nu_{\mathrm{x}}^0(C) - \nu_0^0(C)] = \tau_{\mathrm{x}}^0 - \tau_0^0 = \beta_0^0.$$

# C  Asymptotic variance of the estimators

## C.1  Road map

Again, we restrict our attention to the case where weighting and density functions are estimated based on parametric models, and conditional mean models are estimated (semi)parametrically.

We follow another three-step reasoning, which applies to all estimators

1. The first step is to express the estimator as the solution to an estimating equation. This is exactly the same as step 1 in Section B (page 4), except we will need the full function $\psi(Z, \theta)$ including the elements for weights estimation.

14

2. The second step is to state regularity conditions under which $\hat{\theta} = (\hat{\theta}_{\text{nuisance}}^{*T}, \hat{\theta}_{\text{target}})^T$ is asymptotically normal. These include the regularity conditions for convergence in probability (section B.3) and additional conditions. That $\hat{\theta}$ is asymptotically normal implies that $\hat{\theta}_{\text{target}}$ is asymptotically normal with asymptotic variance equal to the variance of its influence function.

3. The third step is to derive the estimator's influence function. For an effect estimator that is the difference between two potential outcome mean estimators, its influence function is the difference between the influence functions of the component estimators. Therefore we simply derive the influence functions of the component estimators.

    This derivation does not rely on the assumption that any estimation components are consistent for the true functions/parameters, but takes the estimator as a consistent estimator of its probability limit. We point out the form of the influence function in special cases where certain estimation components are consistent. For the *more robust* estimators, the simplification of the influence function when all estimation components are consistent shows the estimator's local efficiency.

For a generic estimating function $\psi(Z, \theta)$, we list the regularity conditions for $\hat{\theta}$ to be asymptotically normal in section C.2; this suffices for Step 2 for all of our estimators. Steps 1 and 3, which vary estimator to estimator, are handled in sections C.3 (for the *nonrobust* estimators) and C.4 (for the *more robust* estimators).

## C.2 Regularity conditions for estimator to be asymptotically normal

The following regularity conditions (based on theorem 7.2 in Boos and Stefanski (2013)) provide for the solution $\hat{\theta}$ to $P_n \psi(Z, \theta) = 0$ to be asymptotically normal:

- $\hat{\theta}$ converges in probability to a unique $\theta^*$ that solves $E[\psi(Z, \theta)] = 0$. (This holds under conditions laid out in B.3.)

- $\psi(z, \theta)$ and its first two derivatives with respect to $\theta$ exist for all $z$ in the support of $Z$ and for all $\theta$ in a neighborhood of $\theta^*$.

- Every element of the second derivative array of $\psi(Z, \theta)$ w.r.t. $\theta$ is dominated by a function $g(Z)$, $E[g(Z)] < \infty$. That is, $\left| \frac{\partial^2}{\partial \theta_k \partial \theta_l} \psi_j(Z, \theta) \right| \leq g(Z)$, for $j, k, l \in \{1, \ldots, \text{length}(\theta)\}$.

- $\Omega = E\left[ \frac{\partial \psi(Z, \theta)}{\partial \theta^T} \Big|_{\theta = \theta^*} \right]$ exists and is nonsingular.

- $V = E\left[ \psi(Z, \theta^*) \psi(Z, \theta^*)^T \right]$ exists and is finite.

Under these conditions,
$$\sqrt{n}(\hat{\theta} - \theta^*) \xrightarrow{d} N\left(0, \ (-\Omega)^{-1} V (-\Omega)^{-1}\right) \quad \text{as } n \to \infty. \tag{A37}$$

Let A denote the row of the matrix $(-\Omega)^{-1}$ that corresponds to element $\theta_{\text{target}}$ of $\theta$. That is, with $\theta_{\text{target}}$ being the last element in vector $\theta$, A is the last row of $(-\Omega)^{-1}$. Then (A37) implies
$$\sqrt{n}(\hat{\theta}_{\text{target}} - \theta_{\text{target}}^*) \xrightarrow{d} N(0, A V A^T) \quad \text{as } n \to \infty. \tag{A38}$$

Note that $A V A^T$ is the variance of $A \psi(Z, \theta^*)$, which is the influence function of the estimator $\hat{\theta}_{\text{target}}$.

## C.3 Influence functions of the *nonrobust* estimators

We now assume that the regularity conditions listed above hold for $\hat{\theta}_{\text{target}}$ to be asymptotically normal, with asymptotic variance equal to the variance of the influence function. This section and section C.4 are

15

concerned with deriving the influence function $A\psi^*(Z)$ for each estimator, where $\psi^*(Z) = \psi(Z, \theta^*)$, using the calculus of M-estimation of Stefanski and Boos (2002).

### C.3.1  Estimators of the regular potential outcome mean E[$Y_1$]

With all reg| estimators, we will just derive the influence function of the estimator of E[$Y_1$]; the result for E[$Y_0$] is simply a mirror image.

#### C.3.1.1  reg|wtd

For this estimator of $\tau_1^0 = \mathrm{E}[Y_1]$, the estimating function is

$$\psi(Z, \theta) = \begin{pmatrix} \nabla \log \delta(C, A, \theta_1) \\ \frac{A}{\delta(C, 1, \theta_1)}(Y - \tau_1) \end{pmatrix}. \tag{A39}$$

The expectation of the derivative of this estimating function w.r.t. $\theta^T$ evaluated at $\theta^*$ is

$$\Omega = \begin{pmatrix} \Omega_{11} & 0 \\ \Omega_{21} & \Omega_{22} \end{pmatrix} = \begin{pmatrix} \mathrm{E}\left[\nabla\nabla^T \log \delta^*(C, A)\right] & 0 \\ \mathrm{E}\left[-\frac{A(Y-\tau_1)}{\delta^*(C,1)}\nabla^T \log \delta^*(C, A)\right] & \mathrm{E}\left[-\frac{A}{\delta^*(C,1)}\right] \end{pmatrix},$$

which is a block triangle matrix, relatively simple to invert. The bottom row of $(-\Omega)^{-1}$ is

$$\mathrm{A} = (-\Omega_{22})^{-1}\begin{bmatrix} \Omega_{21}(-\Omega_{11})^{-1} & 1 \end{bmatrix}.$$

To simplify notation (which is not important here but will help with the more complex estimators that come later), we use the shorthand:

$$\mathrm{A}_{22} = (-\Omega_{22})^{-1},$$
$$\mathrm{A}_{21} = \Omega_{21}(-\Omega_{11})^{-1}.$$

The influence function of the estimator is $A\psi^*(Z)$, where $\psi^*(Z) = \psi(Z, \theta^*, \tau^*)$,

$$A\psi^*(Z) = \mathrm{A}_{22}\begin{pmatrix} \mathrm{A}_{21} & 1 \end{pmatrix}\psi^*(Z), \tag{A40}$$

where

$$\mathrm{A}_{22} = \mathrm{E}\left[\frac{A}{\delta^*(C,1)}\right]^{-1},$$
$$\mathrm{A}_{21} = \mathrm{E}\left[-\frac{A(Y-\tau_1^*)}{\delta^*(C,1)}\nabla^T \log \delta^*(C, 1)\right]\mathrm{E}\left[-\nabla\nabla^T \log \delta^*(C, A)\right]^{-1}.$$
$$\psi^*(Z) = \begin{pmatrix} \nabla \log \delta^*(C, A) \\ \frac{A}{\delta^*(C,1)}(Y - \tau_1^*) \end{pmatrix}.$$

Special case: If $\delta$ is correctly specified,

$$\mathrm{A}_{22} = 1,$$
$$\mathrm{A}_{21} = \underbrace{\mathrm{E}\left[-\frac{A(Y-\tau_1^0)}{\delta^0(C,1)}\nabla^T \log \delta^0(C, 1)\right]}_{\mathrm{E}\left[-\{\nu_1^0(C)-\tau_1^0\}\nabla^T \log \delta^0(C,1)\right]}\mathrm{E}\left[-\nabla\nabla^T \log \delta^0(C, A)\right]^{-1},$$
$$\psi^*(Z) = \begin{pmatrix} \nabla \log \delta^0(C, A) \\ \frac{A}{\delta^0(C,1)}(Y - \boxed{\tau_1^0}) \end{pmatrix}.$$

16

### C.3.1.2 reg|Ypred

$$\psi(Z, \theta) = \begin{pmatrix} Ag(C, \theta_1)[Y - \nu_1(C; \theta_1)] \\ \nu_1(C; \theta_1) - \tau_1 \end{pmatrix}. \tag{A41}$$

$$\Omega = \begin{pmatrix} \Omega_{11} & 0 \\ \Omega_{21} & -1 \end{pmatrix} = \begin{pmatrix} \mathrm{E}\left[-Ag^*(C)\nabla^T\nu_1^*(C) + A\nabla^T g^*(C)[Y - \nu_1^*(C)]\right] & 0 \\ \mathrm{E}[\nabla^T \nu_1^*(C)] & -1 \end{pmatrix}.$$

$$\mathrm{A} = \begin{pmatrix} \Omega_{21}(-\Omega_{11})^{-1} & 1 \end{pmatrix} = \begin{pmatrix} \mathrm{A}_{21} & 1 \end{pmatrix}.$$

The influence function of the estimator is

$$\begin{pmatrix} \mathrm{A}_{21} & 1 \end{pmatrix} \psi^*(Z), \tag{A42}$$

where

$$\mathrm{A}_{21} = \mathrm{E}[\nabla^T \nu_1^*(C)] \mathrm{E}\left[A\left\{g^*(C)\nabla^T \nu_1^*(C) - \nabla^T g^*(C)[Y - \nu_1^*(C)]\right\}\right]^{-1},$$

$$\psi^*(Z) = \begin{pmatrix} Ag^*(C)[Y - \nu_1^*(C)] \\ \nu_1^*(C) - \tau_1^* \end{pmatrix}.$$

Special case: If the $\nu_1$ model is correctly specified,

$$\mathrm{A}_{21} = \mathrm{E}\left[\nabla^T \nu_1^0(C)\right] \mathrm{E}\left[Ag^*(C)\nabla^T \nu_1^0(C)\right]^{-1},$$

$$\psi^*(Z) = \begin{pmatrix} Ag^*(C)[Y - \nu_1^0(C)] \\ \nu_1^0(C) - \boxed{\tau_1^0} \end{pmatrix}.$$

### C.3.2 Estimators of the cross-world potential outcome mean $\mathrm{E}[Y_{1M_0}]$

#### C.3.2.1 crw|wtd

**Using weights estimated by method 1 – via mediator density models**

$$\psi(Z, \theta) = \begin{pmatrix} \nabla \log \delta(C, A; \theta_1) \\ A \nabla \log \lambda_1(C, M; \theta_2) \\ (1 - A) \nabla \log \lambda_0(C, M; \theta_3) \\ \frac{A}{\delta(C, 1; \theta_1)} \frac{\lambda_0(C, M; \theta_3)}{\lambda_1(C, M; \theta_2)} (Y - \tau_x) \end{pmatrix}. \tag{A43}$$

$$\Omega = \begin{pmatrix} \Omega_{11} & 0 & 0 & 0 \\ 0 & \Omega_{22} & 0 & 0 \\ 0 & 0 & \Omega_{33} & 0 \\ \Omega_{41} & \Omega_{42} & \Omega_{43} & \Omega_{44} \end{pmatrix}$$

where

$\Omega_{11} = \mathrm{E}\left[\nabla \nabla^T \log \delta^*(C, A)\right],$ $\qquad \Omega_{41} = \mathrm{E}\left[-\frac{A}{\delta^*(C,1)} \frac{\lambda_0^*(C,M)}{\lambda_1^*(C,M)} (Y - \tau_x) \nabla^T \log \delta^*(C, 1)\right],$

$\Omega_{22} = \mathrm{E}[A \nabla \nabla^T \log \lambda_1^*(C, M)],$ $\qquad \Omega_{42} = \mathrm{E}\left[-\frac{A}{\delta^*(C,1)} \frac{\lambda_0^*(C,M)}{\lambda_1^*(C,M)} (Y - \tau_x) \nabla^T \log \lambda_1^*(C, M)\right],$

$\Omega_{33} = \mathrm{E}[(1 - A) \nabla \nabla^T \log \lambda_0^*(C, M)],$ $\qquad \Omega_{43} = \mathrm{E}\left[\frac{A}{\delta^*(C,1)} \frac{\lambda_0^*(C,M)}{\lambda_1^*(C,M)} (Y - \tau_x) \nabla^T \log \lambda_0^*(C, M)\right],$

$\Omega_{44} = \mathrm{E}\left[-\frac{A}{\delta^*(C,1)} \frac{\lambda_0^*(C,M)}{\lambda_1^*(C,M)}\right].$

17

$$A = (-\Omega_{44})^{-1} \begin{pmatrix} \Omega_{41}(-\Omega_{11})^{-1} & \Omega_{42}(-\Omega_{22})^{-1} & \Omega_{43}(-\Omega_{33})^{-1} & 1 \end{pmatrix}$$
$$= A_{44} \begin{pmatrix} A_{41} & A_{42} & A_{43} & 1 \end{pmatrix}.$$

The influence function of the estimator is

$$A_{44} \begin{pmatrix} A_{41} & A_{42} & A_{43} & 1 \end{pmatrix} \psi^*(Z), \tag{A44}$$

where

$$A_{44} = E\left[\frac{A}{\delta^*(C,1)} \frac{\lambda_0^*(C,M)}{\lambda_1^*(C,M)}\right]^{-1},$$

$$A_{41} = E\left[-\frac{A}{\delta^*(C,1)} \frac{\lambda_0^*(C,M)}{\lambda_1^*(C,M)}(Y-\tau_x^*)\nabla^T \log \delta^*(C,1)\right] E[-\nabla\nabla^T \log \delta^*(C,A)]^{-1},$$

$$A_{42} = E\left[-\frac{A}{\delta^*(C,1)} \frac{\lambda_0^*(C,M)}{\lambda_1^*(C,M)}(Y-\tau_x^*)\nabla^T \log \lambda_1^*(C,M)\right] E[-A\nabla\nabla^T \log \lambda_1^*(C,M)]^{-1}$$

$$A_{43} = E\left[\frac{A}{\delta^*(C,1)} \frac{\lambda_0^*(C,M)}{\lambda_1^*(C,M)}(Y-\tau_x^*)\nabla^T \log \lambda_0^*(C,M)\right] E[-(1-A)\nabla\nabla^T \log \lambda_0^*(C,M)]^{-1}$$

$$\psi^*(Z) = \begin{pmatrix} \nabla \log \delta^*(C,A) \\ A\nabla \log \lambda_1^*(C,M) \\ (1-A)\nabla \log \lambda_0^*(C,M) \\ \frac{A}{\delta^*(C,1)} \frac{\lambda_0^*(C,M)}{\lambda_1^*(C,M)}(Y-\tau_x^*) \end{pmatrix}.$$

Special case: If $\delta, \lambda_1, \lambda_0$ are correctly specified,

$$A_{44} = 1$$

$$A_{41} = \underbrace{E\left[-\frac{A}{\delta^0(C,1)} \frac{\lambda_0^0(C,M)}{\lambda_1^0(C,M)}(Y-\tau_x^0)\nabla^T \log \delta^0(C,1)\right]}_{E[-\{\nu_x^0(C)-\tau_x^0\}\nabla^T \log \delta^0(C,1)]} E[-\nabla\nabla^T \log \delta^0(C,A)]^{-1},$$

$$A_{42} = \underbrace{E\left[-\frac{A}{\delta^0(C,1)} \frac{\lambda_0^0(C,M)}{\lambda_1^0(C,M)}(Y-\tau_x^0)\nabla^T \log \lambda_1^0(C,M)\right]}_{E\left[-E\left\{[\mu_x^0(C,M)-\tau_x^0]\nabla^T \log \lambda_1^0(C,M)|C,A=0\right\}\right]} E[-A\nabla\nabla^T \log \lambda_1^0(C,M)]^{-1},$$

$$A_{43} = \underbrace{E\left[\frac{A}{\delta^0(C,1)} \frac{\lambda_0^0(C,M)}{\lambda_1^0(C,M)}(Y-\tau_x^0)\nabla^T \log \lambda_0^0(C,M)\right]}_{E\left[E\left\{[\mu_x^0(C,M)-\tau_x^0]\nabla^T \log \lambda_0^0(C,M)|C,A=0\right\}\right]} E[-(1-A)\nabla\nabla^T \log \lambda_0^0(C,M)]^{-1},$$

$$\psi^*(Z) = \begin{pmatrix} \nabla \log \delta^0(C,A) \\ A\nabla \log \lambda_1^0(C,M) \\ (1-A)\nabla \log \lambda_0^0(C,M) \\ \frac{A}{\delta^0(C,1)} \frac{\lambda_0^0(C,M)}{\lambda_1^0(C,M)}(Y-\boxed{\tau_x^0}) \end{pmatrix}.$$

**Using weights estimated by method 2 – via conditional exposure models**

$$\psi(Z,\theta) = \begin{pmatrix} \nabla \log \delta(C,A;\theta_1) \\ \nabla \log \gamma(C,M,A;\theta_2) \\ \frac{A\gamma(C,M,0;\theta_2)}{\delta(C,0;\theta_1)\gamma(C,M,1;\theta_2)}(Y-\tau_x) \end{pmatrix}. \tag{A45}$$

18

$$\Omega = \begin{pmatrix} \Omega_{11} & 0 & 0 \\ 0 & \Omega_{22} & 0 \\ \Omega_{31} & \Omega_{32} & \Omega_{33} \end{pmatrix}$$

where

$$\Omega_{11} = \mathrm{E}\left[\nabla\nabla^T \log \delta^*(C,A)\right], \qquad \Omega_{31} = \mathrm{E}\left[-\frac{A\gamma^*(C,M,0)}{\delta^*(C,0)\gamma^*(C,M,1)}(Y-\tau_\mathrm{x}^*)\nabla^T \log \delta^*(C,0)\right],$$
$$\Omega_{22} = \mathrm{E}\left[\nabla\nabla^T \log \gamma^*(C,M,A)\right], \qquad \Omega_{32} = \mathrm{E}\left[-\frac{A}{\delta^*(C,0)\gamma^*(C,M,1)}(Y-\tau_\mathrm{x}^*)\nabla^T \log \gamma^*(C,M,1)\right],$$
$$\Omega_{33} = \mathrm{E}\left[-\frac{A\gamma^*(C,M,0)}{\delta^*(C,0)\gamma^*(C,M,1)}\right].$$

$$\mathrm{A} = (-\Omega_{33})^{-1}\begin{pmatrix}\Omega_{31}(-\Omega_{11})^{-1} & \Omega_{32}(-\Omega_{22})^{-1} & 1\end{pmatrix}$$
$$= \mathrm{A}_{33}\begin{pmatrix}\mathrm{A}_{31} & \mathrm{A}_{32} & 1\end{pmatrix}.$$

The influence function of the estimator is

$$\mathrm{A}_{33}\begin{pmatrix}\mathrm{A}_{31} & \mathrm{A}_{32} & 1\end{pmatrix}\psi^*(Z), \tag{A46}$$

where

$$\mathrm{A}_{33} = \mathrm{E}\left[\frac{A\gamma^*(C,M,0)}{\delta^*(C,0)\gamma^*(C,M,1)}\right]^{-1},$$
$$\mathrm{A}_{31} = \mathrm{E}\left[-\frac{A\gamma^*(C,M,0)}{\delta^*(C,0)\gamma^*(C,M,1)}(Y-\tau_\mathrm{x}^*)\nabla^T \log \delta^*(C,0)\right]\mathrm{E}\left[-\nabla\nabla^T \log \delta^*(C,A)\right]^{-1},$$
$$\mathrm{A}_{32} = \mathrm{E}\left[-\frac{A}{\delta^*(C,0)\gamma^*(C,M,1)}(Y-\tau_\mathrm{x}^*)\nabla^T \log \gamma^*(C,M,1)\right]\mathrm{E}\left[-\nabla\nabla^T \log \gamma^*(C,M,A)\right]^{-1},$$
$$\psi^*(Z) = \begin{pmatrix}\nabla \log \delta^*(C,A) \\ \nabla \log \gamma^*(C,M,A) \\ A\frac{\gamma^*(C,M,0)}{\delta^*(C,0)\gamma^*(C,M,1)}(Y-\tau_\mathrm{x}^*)\end{pmatrix}.$$

Special case: If both $\delta$ and $\gamma$ are correctly specified,

$$\mathrm{A}_{33} = 1,$$
$$\mathrm{A}_{31} = \underbrace{\mathrm{E}\left[-\frac{A\gamma^0(C,M,0)}{\delta^0(C,0)\gamma^0(C,M,1)}(Y-\tau_\mathrm{x}^0)\nabla^T \log \delta^0(C,0)\right]}_{\mathrm{E}[-\{\nu_\mathrm{x}^0(C)-\tau_\mathrm{x}^0\}\nabla^T \log \delta^0(C,0)]}\mathrm{E}\left[-\nabla\nabla^T \log \delta^0(C,A)\right]^{-1},$$
$$\mathrm{A}_{32} = \underbrace{\mathrm{E}\left[-\frac{A}{\delta^0(C,0)\gamma^0(C,M,1)}(Y-\tau_\mathrm{x}^0)\nabla^T \log \gamma^0(C,M,1)\right]}_{\mathrm{E}\left[-\mathrm{E}\left\{[\mu^0(C,M)-\tau_\mathrm{x}^0]\frac{\nabla^T \log \gamma^0(C,M,1)}{\gamma^0(C,M,0)}\big|C,A=0\right\}\right]}\mathrm{E}\left[-\nabla\nabla^T \log \gamma^0(C,M,A)\right]^{-1},$$
$$\psi^*(Z) = \begin{pmatrix}\nabla \log \delta^0(C,A) \\ \nabla \log \gamma^0(C,M,A) \\ A\frac{\gamma^0(C,M,0)}{\delta^0(C,0)\gamma^0(C,M,1)}(Y-\boxed{\tau_\mathrm{x}^0})\end{pmatrix}.$$

**Using weights estimated by method 3 – mimicking pseudo control sample**

$$\psi(Z,\theta) = \begin{pmatrix}\nabla \log \delta(C,A;\theta_1) \\ \left[A+\frac{1-A}{\delta(C,0;\theta_1)}\right]\nabla \log \sigma(C,M,A;\theta_2) \\ A\frac{\sigma(C,M,0;\theta_2)}{\sigma(C,M,1;\theta_2)}(Y-\tau_\mathrm{x})\end{pmatrix}. \tag{A47}$$

19

$$\Omega = \begin{pmatrix} \Omega_{11} & 0 & 0 \\ \Omega_{21} & \Omega_{22} & 0 \\ 0 & \Omega_{32} & \Omega_{33} \end{pmatrix}$$

where

$$\Omega_{11} = \mathrm{E}[\nabla\nabla^T \log \delta^*(C,A)], \qquad \Omega_{21} = \mathrm{E}\left[-\tfrac{1-A}{\delta^*(C,0)}\nabla\log\sigma^*(C,M,0)\nabla^T\log\delta^*(C,0)\right],$$
$$\Omega_{22} = \mathrm{E}\left[\left(A + \tfrac{1-A}{\delta^*(C,0)}\right)\nabla\nabla^T\log\sigma^*(C,M,A)\right], \quad \Omega_{32} = \mathrm{E}\left[-\tfrac{A}{\sigma^*(C,M,1)}(Y-\tau_\mathrm{x})\nabla^T\log\sigma^*(C,M,1)\right],$$
$$\Omega_{33} = \mathrm{E}\left[-A\tfrac{\sigma^*(C,M,0)}{\sigma^*(C,M,1)}\right].$$

$$A = (-\Omega_{33})^{-1}\begin{pmatrix}\Omega_{32}(-\Omega_{22})^{-1}\Omega_{21}(-\Omega_{11})^{-1} & \Omega_{32}(-\Omega_{22})^{-1} & 1\end{pmatrix}$$
$$= A_{33}\begin{pmatrix}A_{32}A_{21} & A_{32} & 1\end{pmatrix}.$$

The influence function of the estimator is

$$A_{33}\begin{pmatrix}A_{32}A_{21} & A_{32} & 1\end{pmatrix}\psi^*(Z), \tag{A48}$$

where

$$A_{33} = \mathrm{E}\left[A\frac{\sigma^*(C,M,0)}{\sigma^*(C,M,1)}\right]^{-1},$$

$$A_{32} = \mathrm{E}\left[-\frac{A}{\sigma^*(C,M,1)}(Y-\tau_\mathrm{x}^*)\nabla^T\log\sigma^*(C,M,1)\right]\mathrm{E}\left[-\left(A + \frac{1-A}{\delta^*(C,0)}\right)\nabla\nabla^T\log\sigma^*(C,M,A)\right]^{-1},$$

$$A_{21} = \mathrm{E}\left[-\frac{1-A}{\delta^*(C,0)}\nabla\log\sigma^*(C,M,0)\nabla^T\log\delta^*(C,0)\right]\mathrm{E}[-\nabla\nabla^T\log\delta^*(C,A)]^{-1},$$

$$\psi^*(Z) = \begin{pmatrix}\nabla\log\delta^*(C,A) \\ \left[A + \frac{1-A}{\delta^*(C,0)}\right]\nabla\log\sigma^*(C,M,A) \\ A\frac{\sigma^*(C,M,0)}{\sigma^*(C,M,1)}(Y-\tau_\mathrm{x}^*)\end{pmatrix}.$$

Special case: If both $\delta$ and $\sigma$ are correctly specified,

$A_{33} = 1,$

$$A_{32} = \underbrace{\mathrm{E}\left[-\frac{A}{\sigma^0(C,M,1)}(Y-\tau_\mathrm{x}^0)\nabla^T\log\sigma^0(C,M,1)\right]\mathrm{E}\left[-\left(A + \frac{1-A}{\delta^0(C,0)}\right)\nabla\nabla^T\log\sigma^0(C,M,A)\right]^{-1}}_{\mathrm{E}\left[-\mathrm{E}\left\{[\mu^0(C,M)-\tau_\mathrm{x}^0]\frac{\nabla^T\log\sigma^0(C,M,1)}{\sigma^0(C,M,0)}\mid C,A=0\right\}\right]},$$

$$A_{21} = \underbrace{\mathrm{E}\left[-\frac{1-A}{\delta^0(C,0)}\nabla\log\sigma^0(C,M,0)\nabla^T\log\delta^0(C,0)\right]\mathrm{E}[-\nabla\nabla^T\log\delta^0(C,A)]^{-1}}_{\mathrm{E}[-\mathrm{E}\{\nabla\log\sigma^0(C,M,0)\mid C,A=0\}\nabla^T\log\delta^0(C,0)]},$$

$$\psi^*(Z) = \begin{pmatrix}\nabla\log\delta^0(C,A) \\ \left[A + \frac{1-A}{\delta^0(C,0)}\right]\nabla\log\sigma^0(C,M,A) \\ A\frac{\sigma^0(C,M,0)}{\sigma^0(C,M,1)}(Y-\boxed{\tau_\mathrm{x}^0})\end{pmatrix}.$$

#### C.3.2.2 crw|psYpred

$$\psi(Z,\theta) = \begin{pmatrix}\nabla\log\delta(C,A;\theta_1) \\ Ah(C,M;\theta_2)[Y - \mu(C,M;\theta_2)] \\ \frac{1-A}{\delta(C,0;\theta_1)}[\mu(C,M;\theta_2) - \tau_\mathrm{x}]\end{pmatrix}. \tag{A49}$$

$$\Omega = \begin{pmatrix} \Omega_{11} & 0 & 0 \\ 0 & \Omega_{22} & 0 \\ \Omega_{31} & \Omega_{32} & \Omega_{33} \end{pmatrix},$$

where

$\Omega_{11} = \mathrm{E}[\nabla \nabla^T \log \delta^*(C, A)],$

$\Omega_{22} = \mathrm{E}\left[-A\{h^*(C, M)\nabla^T \mu^*(C, M) - \nabla^T h^*(C, M)[Y - \mu^*(C, M)]\}\right],$

$\Omega_{33} = \mathrm{E}\left[-\dfrac{1-A}{\delta^*(C, 0)}\right].$

$\Omega_{31} = \mathrm{E}\left[-\dfrac{1-A}{\delta^*(C, 0)}[\mu^*(C, M) - \tau_{\mathrm{x}}^*]\nabla^T \log \delta^*(C, 0)\right],$

$\Omega_{32} = \mathrm{E}\left[\dfrac{1-A}{\delta^*(C, 0)}\nabla^T \mu^*(C, M)\right],$

$$\begin{aligned} \mathrm{A} &= (-\Omega_{33})^{-1} \begin{pmatrix} \Omega_{31}(-\Omega_{11})^{-1} & \Omega_{32}(-\Omega_{22})^{-1} & 1 \end{pmatrix} \\ &= \mathrm{A}_{33} \begin{pmatrix} \mathrm{A}_{31} & \mathrm{A}_{32} & 1 \end{pmatrix}. \end{aligned}$$

The influence function of the estimator is

$$\mathrm{A}_{33} \begin{pmatrix} \mathrm{A}_{31} & \mathrm{A}_{32} & 1 \end{pmatrix} \psi^*(Z), \tag{A50}$$

where

$\mathrm{A}_{33} = \mathrm{E}\left[\dfrac{1-A}{\delta^*(C, 0)}\right]^{-1},$

$\mathrm{A}_{31} = \mathrm{E}\left[-\dfrac{1-A}{\delta^*(C, 0)}[\mu^*(C, M) - \tau_{\mathrm{x}}^*]\nabla^T \log \delta^*(C, 0)\right] \mathrm{E}[-\nabla \nabla^T \log \delta^*(C, A)]^{-1},$

$\mathrm{A}_{32} = \mathrm{E}\left[\dfrac{1-A}{\delta^*(C, 0)}\nabla^T \mu^*(C, M)\right] \mathrm{E}\left[A\{h^*(C, M)\nabla^T \mu^*(C, M) - \nabla^T h^*(C, M)[Y - \mu^*(C, M)]\}\right]^{-1},$

$\psi^*(Z) = \begin{pmatrix} \nabla \log \delta^*(C, A) \\ A h^*(C, M)[Y - \mu^*(C, M)] \\ \dfrac{1-A}{\delta^*(C, 0)}[\mu^*(C, M) - \tau_{\mathrm{x}}^*] \end{pmatrix}.$

Special cases:

- If $\mu$ is correctly specified,

$\mathrm{A}_{31} = \mathrm{E}\left[-\dfrac{1-A}{\delta^*(C, 0)}[\mu^0(C, M) - \tau_{\mathrm{x}}^*]\nabla^T \log \delta^*(C, 0)\right] \mathrm{E}[-\nabla \nabla^T \log \delta^*(C, A)]^{-1},$

$\mathrm{A}_{32} = \mathrm{E}\left[\dfrac{1-A}{\delta^*(C, 0)}\nabla^T \mu^0(C, M)\right] \mathrm{E}\left[A h^*(C, M)\nabla^T \mu^0(C, M)\right]^{-1},$

$\psi^*(Z) = \begin{pmatrix} \nabla \log \delta^*(C, A) \\ A h^*(C, M)[Y - \mu^0(C, M)] \\ \dfrac{1-A}{\delta^*(C, 0)}[\mu^0(C, M) - \tau_{\mathrm{x}}^*] \end{pmatrix}.$

- If both $\mu$ and $\delta$ are correctly specified,

$\mathrm{A}_{33} = 1,$

$\mathrm{A}_{31} = \underbrace{\mathrm{E}\left[-\dfrac{1-A}{\delta^0(C, 0)}[\mu^0(C, M) - \tau_{\mathrm{x}}^0]\nabla^T \log \delta^0(C, 0)\right]}_{\mathrm{E}[-\{\nu_{\mathrm{x}}^0(C) - \tau_{\mathrm{x}}^0\}\nabla^T \log \delta^0(C, 0)]} \mathrm{E}[-\nabla \nabla^T \log \delta^0(C, A)]^{-1},$

21

$$A_{32} = E\left[\frac{1-A}{\delta^0(C,0)}\nabla^T\mu^0(C,M)\right] E\left[Ah^*(C,M)\nabla^T\mu^0(C,M)\right]^{-1},$$

$$\psi^*(Z) = \begin{pmatrix} \nabla \log \delta^0(C,A) \\ Ah^*(C,M)[Y - \mu^0(C,M)] \\ \frac{1-A}{\delta^0(C,0)}[\mu^0(C,M) - \tau_x^0] \end{pmatrix}.$$

### C.3.2.3 crw|Ypred

$$\psi(Z,\theta) = \begin{pmatrix} \nabla \log \gamma(C,M,A;\theta_1) \\ A\frac{\gamma(C,M,0;\theta_1)}{\gamma(C,M,1;\theta_1)}g(C;\theta_2)[Y - \nu_x(C;\theta_2)] \\ \nu_x(C;\theta_2) - \tau_x \end{pmatrix}. \tag{A51}$$

$$\Omega = \begin{pmatrix} \Omega_{11} & 0 & 0 \\ \Omega_{21} & \Omega_{22} & 0 \\ 0 & \Omega_{32} & -1 \end{pmatrix},$$

where

$$\Omega_{11} = E[\nabla\nabla^T \log \gamma^*(C,M,A)], \qquad \Omega_{21} = E\left[-\frac{A}{\gamma^*(C,M,1)}g^*(C)[Y - \nu_x^*(C)]\nabla^T \log \gamma^*(C,M,1)\right],$$

$$\Omega_{22} = E\left[-\frac{A\gamma^*(C,M,0)}{\gamma^*(C,M,1)}\left(\{g^*(C)\nabla^T\nu_x^*(C) - [Y - \nu_x^*(C)]\nabla^T g^*(C)\}\right)\right], \quad \Omega_{32} = E[\nabla^T\nu_x^*(C)].$$

$$A = \begin{pmatrix} \Omega_{32}(-\Omega_{22})^{-1}\Omega_{21}(-\Omega_{11})^{-1} & \Omega_{32}(-\Omega_{22})^{-1} & 1 \end{pmatrix}$$
$$= \begin{pmatrix} A_{32}A_{21} & A_{32} & 1 \end{pmatrix}.$$

The influence function of the estimator is

$$\begin{pmatrix} A_{32}A_{21} & A_{32} & 1 \end{pmatrix} \psi^*(Z), \tag{A52}$$

where

$$A_{32} = E[\nabla^T\nu_x^*(C)] E\left[\frac{A\gamma^*(C,M,0)}{\gamma^*(C,M,1)}\left(g^*(C)\nabla^T\nu_x^*(C) - [Y - \nu_x^*(C)]\nabla^T g^*(C)\right)\right]^{-1},$$

$$A_{21} = E\left[-\frac{A}{\gamma^*(C,M,1)}g^*(C)[Y - \nu_x^*(C)]\nabla^T \log \gamma^*(C,M,1)\right] E[-\nabla\nabla^T \log \gamma^*(C,M,A)]^{-1},$$

$$\psi^*(Z) = \begin{pmatrix} \nabla \log \gamma^*(C,M,A) \\ A\frac{\gamma^*(C,M,0)}{\gamma^*(C,M,1)}g^*(C)[Y - \nu_x^*(C)] \\ \nu_x^*(C) - \tau_x^* \end{pmatrix}.$$

Special case: If $\gamma, \nu_x$ are correctly specified,

$$A_{32} = E[\nabla^T\nu_x^0(C)] \underbrace{E\left[\frac{A\gamma^0(C,M,0)}{\gamma^0(C,M,1)}g^*(C)\nabla^T\nu_x^0(C)\right]^{-1}}_{\frac{1}{P(A=0)}E[g^*(C)\nabla^T\nu_x^0(C)|A=0]^{-1}},$$

$$A_{21} = \underbrace{E\left[-\frac{A}{\gamma^0(C,M,1)}g^*(C)[Y - \nu_x^0(C)]\nabla^T \log \gamma^0(C,M,1)\right]}_{P(A=0)E\left[-g^*(C)[\mu^0(C,M) - \nu_x^0(C)]\frac{\nabla^T \log \gamma^0(C,M,1)}{\gamma^0(C,M,0)}|A=0\right]} E[-\nabla\nabla^T \log \gamma^0(C,M,A)]^{-1},$$

22

$$\psi^*(Z) = \begin{pmatrix} \nabla \log \gamma^0(C, M, A) \\ A \frac{\gamma^0(C,M,0)}{\gamma^0(C,M,1)} g^*(C)[Y - \nu_x^0(C)] \\ \nu_x^0(C) - \tau_x^0 \end{pmatrix}.$$

### C.3.2.4 crw|Y2pred

$$\psi(Z, \theta) = \begin{pmatrix} Ah(C, M; \theta_1)[Y - \mu(C, M; \theta_1)] \\ (1 - A)g(C; \theta_2)[\mu(C, M; \theta_1) - \nu_x(C; \theta_2)] \\ \nu_x(C; \theta_2) - \tau_x \end{pmatrix}. \tag{A53}$$

$$\Omega = \begin{pmatrix} \Omega_{11} & 0 & 0 \\ \Omega_{21} & \Omega_{22} & 0 \\ 0 & \Omega_{32} & -1 \end{pmatrix},$$

where

$$\Omega_{11} = \mathrm{E}\big[ - A \{h^*(C, M)\nabla^T\mu^*(C, M) - [Y - \mu^*(C, M)]\nabla^T h^*(C, M)\}\big], \quad \Omega_{21} = \mathrm{E}[(1 - A)g^*(C)\nabla^T\mu^*(C, M)],$$
$$\Omega_{22} = \mathrm{E}\big[ - (1 - A) \{g^*(C)\nabla^T\nu_x^*(C) - [\mu^*(C, M) - \nu_x^*(C)]\nabla^T g^*(C)\}\big], \quad \Omega_{32} = \mathrm{E}[\nabla^T\nu_x^*(C)].$$

$$\mathrm{A} = \begin{pmatrix} \Omega_{32}(-\Omega_{22})^{-1}\Omega_{21}(-\Omega_{11})^{-1} & \Omega_{32}(-\Omega_{22})^{-1} & 1 \end{pmatrix}$$
$$= \begin{pmatrix} \mathrm{A}_{32}\mathrm{A}_{21} & \mathrm{A}_{32} & 1 \end{pmatrix},$$

and the influence function is

$$\begin{pmatrix} \mathrm{A}_{32}\mathrm{A}_{21} & \mathrm{A}_{32} & 1 \end{pmatrix} \psi^*(Z), \tag{A54}$$

where

$$\mathrm{A}_{32} = \mathrm{E}[\nabla^T\nu_x^*(C)]\mathrm{E}\big[(1 - A) \{g^*(C)\nabla^T\nu_x^*(C) - [\mu^*(C, M) - \nu_x^*(C)]\nabla^T g^*(C)\}\big]^{-1},$$
$$\mathrm{A}_{21} = \mathrm{E}[(1 - A)g^*(C)\nabla^T\mu^*(C, M)]\mathrm{E}\big[A \{h^*(C, M)\nabla^T\mu^*(C, M) - [Y - \mu^*(C, M)]\nabla^T h^*(C, M)\}\big]^{-1},$$
$$\psi^*(Z) = \begin{pmatrix} Ah^*(C, M)[Y - \mu^*(C, M)] \\ (1 - A)g^*(C)[\mu^*(C, M) - \nu_x^*(C)] \\ \nu_x^*(C) - \tau_x^* \end{pmatrix}.$$

Special case: If $\mu$ and $\nu_x$ are correctly specified,

$$\mathrm{A}_{32} = \mathrm{E}[\nabla^T\nu_x^0(C)]\mathrm{E}[(1 - A)g^*(C)\nabla^T\nu_x^0(C)]^{-1},$$
$$\mathrm{A}_{21} = \mathrm{E}[(1 - A)g^*(C)\nabla^T\mu^0(C, M)]\mathrm{E}[Ah^*(C, M)\nabla^T\mu^0(C, M)]^{-1},$$
$$\psi^*(Z) = \begin{pmatrix} Ah^*(C, M)[Y - \mu^0(C, M)] \\ (1 - A)g^*(C)[\mu^0(C, M) - \nu_x^0(C)] \\ \nu_x^0(C) - \tau_x^0 \end{pmatrix}.$$

### C.3.2.5 crw|MsimYpred

$$\psi(Z, \theta) = \begin{pmatrix} Ah(C, M; \theta_1)[Y - \mu(C, M; \theta_1)] \\ (1 - A)\nabla \log \lambda_0(C, M; \theta_2) \\ \int \mu(C, m; \theta_1)\lambda_0(C, m; \theta_2)dm - \tau_x \end{pmatrix}. \tag{A55}$$

23

$$\Omega = \begin{pmatrix} \Omega_{11} & 0 & 0 \\ 0 & \Omega_{22} & 0 \\ \Omega_{31} & \Omega_{32} & -1 \end{pmatrix},$$

where

$$\Omega_{11} = \mathrm{E}\left[-A\{h(C,M)\nabla^T\mu^*(C,M) - [Y-\mu^*(C,M)]\nabla^T h^*(C,M)\}\right], \quad \Omega_{31} = \mathrm{E}\left[\int [\nabla^T\mu^*(C,m)]\lambda_0^*(C,m)dm\right],$$

$$\Omega_{22} = \mathrm{E}\left[(1-A)\nabla\nabla^T\log\lambda_0^*(C,M)\right], \quad \Omega_{32} = \mathrm{E}\left[\int \mu^*(C,m)[\nabla^T\log\lambda_0^*(C,M)]\lambda_0^*(C,M)dm\right].$$

$$A = \begin{pmatrix} \Omega_{31}(-\Omega_{11})^{-1} & \Omega_{32}(-\Omega_{22})^{-1} & 1 \end{pmatrix} = \begin{pmatrix} A_{31} & A_{32} & 1 \end{pmatrix}.$$

The influence function of the estimator is

$$\begin{pmatrix} A_{31} & A_{32} & 1 \end{pmatrix} \psi^*(Z), \tag{A56}$$

where

$$A_{31} = \mathrm{E}\left[\int [\nabla^T\mu^*(C,m)]\lambda_0^*(C,m)dm\right] \mathrm{E}\left[A\{h^*(C,M)\nabla^T\mu^*(C,M) - [Y-\mu^*(C,M)]\nabla^T h^*(C,M)\}\right]^{-1},$$

$$A_{32} = \mathrm{E}\left[\int \mu^*(C,m)[\nabla^T\log\lambda_0^*(C,m)]\lambda_0^*(C,m)dm\right] \mathrm{E}\left[-(1-A)\nabla\nabla^T\log\lambda_0^*(C,M)\right]^{-1},$$

$$\psi^*(Z) = \begin{pmatrix} Ah^*(C,M)[Y-\mu^*(C,M)] \\ (1-A)\nabla\log\lambda_0^*(C,M) \\ \int \mu^*(C,m)\lambda_0^*(C,m)dm - \tau_x^* \end{pmatrix}.$$

Special case: If $\mu$ and $\lambda_0$ are correctly specified,

$$A_{31} = \underbrace{\mathrm{E}\left[\int [\nabla^T\mu^0(C,m)]\lambda_0^0(C,m)dm\right]}_{\mathrm{E}[\mathrm{E}\{\nabla^T\mu^0(C,M)|C,A=0\}]} \mathrm{E}[Ah^*(C,M)\nabla^T\mu^0(C,M)]^{-1},$$

$$A_{32} = \underbrace{\mathrm{E}\left[\int \mu^0(C,m)[\nabla^T\log\lambda_0^0(C,m)]\lambda_0^0(C,m)dm\right]}_{\mathrm{E}[\mathrm{E}\{\mu^0(C,M)\nabla^T\log\lambda_0^0(C,M)|C,A=0\}]} \mathrm{E}\left[-(1-A)\nabla\nabla^T\log\lambda_0^0(C,M)\right]^{-1},$$

$$\psi^*(Z) = \begin{pmatrix} Ah^*(C,M)[Y-\mu^0(C,M)] \\ (1-A)\nabla\log\lambda_0^0(C,M) \\ \underbrace{\int \mu^0(C,m)\lambda_0^0(C,m)dm}_{\nu_x^0(C)} - \boxed{\tau_x^0} \end{pmatrix}.$$

### C.3.3 If targeting additive effects

#### C.3.3.1 NDE|YpredEpred

$$\psi(Z,\theta) = \begin{pmatrix} Ah(C,M;\theta_1)[Y-\mu(C,M;\theta_1)] \\ (1-A)g(C;\theta_2)[\mu(C,M;\theta_1) - Y - \xi(C;\theta_2)] \\ \xi(C;\theta_2) - \beta_0 \end{pmatrix}. \tag{A57}$$

24

$$\Omega = \begin{pmatrix} \Omega_{11} & 0 & 0 \\ \Omega_{21} & \Omega_{22} & 0 \\ 0 & \Omega_{32} & -1 \end{pmatrix},$$

where

$$\Omega_{11} = \mathrm{E}\big[-A\{h^*(C,M)\nabla^T\mu^*(C,M) - [Y - \mu^*(C,M)]\nabla^T h^*(C,M)\}\big], \quad \Omega_{21} = \mathrm{E}[(1-A)g^*(C)\nabla^T\mu^*(C,M)],$$
$$\Omega_{22} = \mathrm{E}\big[-(1-A)\{g^*(C)\nabla^T\xi^*(C) - [\mu^*(C,M) - Y - \xi^*(C)]\nabla^T g^*(C)\}\big], \quad \Omega_{32} = \mathrm{E}[\nabla^T\xi^*(C)].$$

$$\begin{aligned}
\mathrm{A} &= \begin{pmatrix} \Omega_{32}(-\Omega_{22})^{-1}\Omega_{21}(-\Omega_{11})^{-1} & \Omega_{32}(-\Omega_{22})^{-1} & 1 \end{pmatrix} \\
&= \begin{pmatrix} \mathrm{A}_{32}\mathrm{A}_{21} & \mathrm{A}_{32} & 1 \end{pmatrix},
\end{aligned}$$

and the influence function is

$$\begin{pmatrix} \mathrm{A}_{32}\mathrm{A}_{21} & \mathrm{A}_{32} & 1 \end{pmatrix} \psi^*(Z), \tag{A58}$$

where

$$\mathrm{A}_{32} = \mathrm{E}[\nabla^T\xi^*(C)]\mathrm{E}\big[(1-A)\{g^*(C)\nabla^T\xi^*(C) - [\mu^*(C,M) - Y - \xi^*(C)]\nabla^T g^*(C)\}\big]^{-1},$$
$$\mathrm{A}_{21} = \mathrm{E}[(1-A)g^*(C)\nabla^T\mu^*(C,M)]\mathrm{E}\big[A\{h^*(C,M)\nabla^T\mu^*(C,M) - [Y - \mu^*(C,M)]\nabla^T h^*(C,M)\}\big]^{-1},$$
$$\psi^*(Z) = \begin{pmatrix} Ah^*(C,M)[Y - \mu^*(C,M)] \\ (1-A)g^*(C)[\mu^*(C,M) - Y - \xi^*(C)] \\ \xi^*(C) - \beta_0^* \end{pmatrix}.$$

Special case: If $\mu$ and $\xi$ are correctly specified,

$$\mathrm{A}_{32} = \mathrm{E}[\nabla^T\xi^0(C)]\mathrm{E}[(1-A)g^*(C)\nabla^T\xi^0(C)]^{-1},$$
$$\mathrm{A}_{21} = \mathrm{E}[(1-A)g^*(C)\nabla^T\mu^0(C,M)]\mathrm{E}[Ah^*(C,M)\nabla^T\mu^0(C,M)]^{-1},$$
$$\psi^*(Z) = \begin{pmatrix} Ah^*(C,M)[Y - \mu^0(C,M)] \\ (1-A)g^*(C)[\mu^0(C,M) - Y - \xi^0(C)] \\ \xi^0(C) - \beta_0^0 \end{pmatrix}.$$

## C.4 Influence functions of the *more robust* estimators

The paper states that for the more robust estimators, the conditional mean outcome/effect models are required to satisfy the (group) mean recovery condition. This condition may be satisfied in different ways. In this section, we presume that this condition is satisfied by using either a generalized linear model with canonical link and an intercept, or semiparametric model with a certain form of the estimating function.

For concreteness, consider a model for $\mathrm{E}[Y \mid C, A = 1]$ fit to the pseudo treated sample. We suppose that the estimating function corresponding with this model is

$$\frac{A}{\delta(C,1)} g(C) \left[Y - \nu_1\left(\theta_m^T g(C)\right)\right]. \tag{A59}$$

Here the vector $g(C)$ includes a constant element (for the intercept) plus the model's predictor terms. If $\nu_1(\cdot)$ is the inverse of the canonical link function, this estimating function is the weighted score function.

**Lemma.** With $0 < v(C) < \infty$, $M = \mathrm{E}\left[v(C)\begin{pmatrix} 1 & C^T \end{pmatrix}\right] \mathrm{E}\left[v(C)\begin{pmatrix} 1 \\ C \end{pmatrix}\begin{pmatrix} 1 & C^T \end{pmatrix}\right]^{-1} = \begin{pmatrix} 1 & 0 & \ldots & 0 \end{pmatrix}.$

*Proof of Lemma.*

First consider the special case with $v(C) = 1$.

$$M = \begin{pmatrix} 1 & \mathrm{E}C^T \end{pmatrix} \begin{pmatrix} 1 & \mathrm{E}C^T \\ \mathrm{E}C & \mathrm{E}[CC^T] \end{pmatrix}^{-1}.$$

Let $a = \mathrm{E}C$, $b = \mathrm{E}[CC^T]$.

$$M = \begin{pmatrix} 1 & a^T \end{pmatrix} \begin{pmatrix} 1 & a^T \\ a & b \end{pmatrix}^{-1} = \begin{pmatrix} 1 & a^T \end{pmatrix} \begin{pmatrix} 1 + a^T(b-aa^T)^{-1}a & -a^T(b-aa^T)^{-1} \\ -(b-aa^T)^{-1}a & (b-aa^T)^{-1} \end{pmatrix} = \begin{pmatrix} 1 & 0^T \end{pmatrix}.$$

To consider the general case, let $h(C)$ be an arbitrary function of $C$, and let $f(C)$ be the density of $C$.

$$\mathrm{E}[v(C)h(C)] = \int v(c)h(c)f(c)d(c) = \mathrm{E}[v(C)] \int h(c) \underbrace{\frac{v(c)f(c)}{\mathrm{E}[v(C)]}}_{:=f^*(c)} dc$$

Note that $f^*(C) \geq 0$ because $v(C) > 0$, and $\int f^*(c)dc = 1$, so $f^*(C)$ is a density function. Thus $\int h(c)f^*(c)dc$ is the expectation of $h(C)$ w.r.t. this density function. Denote it by $\mathrm{E}^*[h(C)]$. Then

$$M = \begin{pmatrix} 1 & \mathrm{E}^*C^T \end{pmatrix} \begin{pmatrix} 1 & \mathrm{E}^*C^T \\ \mathrm{E}^*C & \mathrm{E}^*[CC^T] \end{pmatrix}^{-1}.$$

As our reasoning for the special case is agnostic to the specific density of $C$, it follows that the result also holds for this density, i.e.,

$$M = \begin{pmatrix} 1 & 0^T \end{pmatrix}.$$

□

### C.4.1 Estimator of the regular potential outcome mean $\mathrm{E}[Y_1]$

#### C.4.1.1 reg|Ypred.R

$$\psi(Z, \theta) = \begin{pmatrix} \nabla \log \delta(C, A; \theta_1) \\ \frac{A}{\delta(C, 1; \theta_1)} g(C)[Y - \nu_1(\theta_2^T g(C))] \\ \nu_1(\theta_2^T g(C)) - \tau_1 \end{pmatrix}. \tag{A60}$$

$$\Omega = \begin{pmatrix} \Omega_{11} & 0 & 0 \\ \Omega_{21} & \Omega_{22} & 0 \\ 0 & \Omega_{32} & -1 \end{pmatrix},$$

where

$\Omega_{11} = \mathrm{E}\left[\nabla\nabla^T \log \delta^*(C, A)\right],$ $\qquad \Omega_{21} = \mathrm{E}\left[-\frac{A}{\delta^*(C,1)} g(C)[Y - \nu_1^*(C)]\nabla^T \log \delta^*(C, 1)\right],$

$\Omega_{22} = \mathrm{E}\left[-\frac{A}{\delta^*(C,1)} \nu_1'^*(C) g(C) g^T(C)\right],$ $\quad \Omega_{32} = \mathrm{E}[\nu_1'^*(C) g^T(C)],$

where $\nu_1'(\cdot)$ is the first derivative of function $\nu_1(\cdot)$, and $\nu_1'^*(C)$ is shorthand for $\nu_1'(\theta_2^{*T} g(C))$.

$$\mathrm{A} = \begin{pmatrix} \Omega_{32}(-\Omega_{22})^{-1}\Omega_{21}(-\Omega_{11})^{-1} & \Omega_{32}(-\Omega_{22})^{-1} & 1 \end{pmatrix} = \begin{pmatrix} \mathrm{A}_{32}\mathrm{A}_{21} & \mathrm{A}_{32} & 1 \end{pmatrix}.$$

26

The influence function of the estimator is

$$\begin{pmatrix} A_{32}A_{21} & A_{32} & 1 \end{pmatrix} \psi^*(Z), \tag{A61}$$

where

$$A_{32} = E[\nu_1'^*(C)g^T(C)] E\left[\frac{A}{\delta^*(C,1)}\nu_1'^*(C)g(C)g^T(C)\right]^{-1},$$

$$A_{21} = E\left[-\frac{A}{\delta^*(C,1)}g(C)[Y-\nu_1^*(C)]\nabla^T \log \delta^*(C,1)\right] E\left[-\nabla\nabla^T \log \delta^*(C,A)\right]^{-1},$$

$$\psi^*(Z) = \begin{pmatrix} \nabla \log \delta^*(C,A) \\ \frac{A}{\delta^*(C,1)} g(C)[Y-\nu_1^*(C)] \\ \nu_1^*(C) - \tau_1^* \end{pmatrix}.$$

Special cases:

- If $\nu_1$ is correctly specified,

$$A_{32} = E[\nu_1'^0(C)g^T(C)] E\left[\frac{A}{\delta^*(C,1)}\nu_1'^0(C)g(C)g^T(C)\right]^{-1},$$

$$A_{21} = 0,$$

and the two relevant elements of $\psi^*(Z)$ are

$$\psi_{2,3}^*(Z) = \begin{pmatrix} \frac{A}{\delta^*(C,1)} g(C)[Y-\nu_1^0(C)] \\ \nu_1^0(C) - \tau_1^0 \end{pmatrix}.$$

- If $\delta$ is correctly specified, $A_{32} = E[\nu_1'^*(C)g^T(C)] E[\nu_1'^*(C)g(C)g^T(C)]^{-1} = \begin{pmatrix} 1 & 0 & \ldots & 0 \end{pmatrix}$ by the Lemma and the fact that $\nu_1'(\cdot) > 0$ (because $\nu_1(\cdot)$ as an inverse link function is monotone increasing). This implies

$$A_{32}A_{21} = \underbrace{E\left[-\frac{A}{\delta^0(C,1)}[Y-\nu_1^*(C)]\nabla^T \log \delta^0(C,1)\right]}_{E\left[[\nu_1^0(C)-\nu_1^*(C)]\nabla^T \log \delta^0(C,1)\right]} E\left[-\nabla\nabla^T \log \delta^0(C,A)\right]^{-1},$$

$$A_{32}\psi_2^*(Z) = \frac{A}{\delta^0(C,1)}[Y-\nu_1^*(C)],$$

and the other two elements of $\psi^*(Z)$ (other than the second element dealt with above) are

$$\psi_{1,3}^*(Z) = \begin{pmatrix} \nabla \log \delta^0(C,A) \\ \nu_1^*(C) - \tau_1^0 \end{pmatrix}.$$

- If both $\nu_1$ and $\delta$ models are correctly specified,

$$A_{21} = 0,$$

$$A_{32}\psi_2^*(Z) = \frac{A}{\delta^0(C,1)}[Y-\nu_1^0(C)],$$

and the IF reduces to the nonparametric IF:

$$\frac{A}{\delta^0(C,1)}[Y-\nu_1^0(C)] + \{\nu_1^0(C) - \tau_1^0\}.$$

27

### C.4.2 Estimators of the cross-world potential outcome mean $\mathrm{E}[Y_{1M_0}]$

#### C.4.2.1 crw|psYpred.MR

**Using weights estimated by method 1 – via mediator density models**

$$\psi(Z,\theta) = \begin{pmatrix} \nabla \log \delta(C, A; \theta_1) \\ A\nabla \log \lambda_1(C, M; \theta_2) \\ (1-A)\nabla \log \lambda_0(C, M; \theta_3) \\ A\dfrac{\lambda_0(C, M; \theta_3)}{\delta(C, 1; \theta_1)\lambda_1(C, M; \theta_2)}h(C, M)[Y - \mu(\theta_4^T h(C, M))] \\ \dfrac{1-A}{\delta(C, 0; \theta_1)}[\mu(\theta_4^T h(C, M)) - \tau_{\mathrm{x}}] \end{pmatrix}. \qquad (A62)$$

$$\Omega = \begin{pmatrix} \Omega_{11} & 0 & 0 & 0 & 0 \\ 0 & \Omega_{22} & 0 & 0 & 0 \\ 0 & 0 & \Omega_{33} & 0 & 0 \\ \Omega_{41} & \Omega_{42} & \Omega_{43} & \Omega_{44} & 0 \\ \Omega_{51} & 0 & 0 & \Omega_{54} & \Omega_{55} \end{pmatrix},$$

where

$\Omega_{11} = \mathrm{E}[\nabla\nabla^T \log \delta^*(C, A)],$

$\Omega_{22} = \mathrm{E}[A\nabla\nabla^T \log \lambda_1^*(C, M)],$

$\Omega_{33} = \mathrm{E}[(1-A)\nabla\nabla^T \log \lambda_0^*(C, M)],$

$\Omega_{44} = \mathrm{E}\left[-\dfrac{A\lambda_0^*(C, M)}{\delta^*(C, 1)\lambda_1^*(C, M)}h(C, M)\mu'^*(C, M)h^T(C, M)\right],$

$\Omega_{55} = \mathrm{E}\left[-\dfrac{1-A}{\delta^*(C, 0)}\right],$

$\Omega_{41} = \mathrm{E}\left[-\dfrac{A\lambda_0^*(C, M)}{\delta^*(C, 1)\lambda_1^*(C, M)}h(C, M)[Y - \mu^*(C, M)]\nabla^T \log \delta^*(C, 1)\right],$

$\Omega_{42} = \mathrm{E}\left[-\dfrac{A\lambda_0^*(C, M)}{\delta^*(C, 1)\lambda_1^*(C, M)}h(C, M)[Y - \mu^*(C, M)]\nabla^T \log \lambda_1^*(C, M)\right],$

$\Omega_{43} = \mathrm{E}\left[\dfrac{A\lambda_0^*(C, M)}{\delta^*(C, 1)\lambda_1^*(C, M)}h(C, M)[Y - \mu^*(C, M)]\nabla^T \log \lambda_0^*(C, M)\right],$

$\Omega_{51} = \mathrm{E}\left[-\dfrac{1-A}{\delta^*(C, 0)}[\mu^*(C, M) - \tau_{\mathrm{x}}^*]\nabla^T \log \delta^*(C, 0)\right],$

$\Omega_{54} = \mathrm{E}\left[\dfrac{1-A}{\delta^*(C, 0)}\mu'^*(C, M)h^T(C, M)\right]$

$$\mathrm{A} = \mathrm{A}_{55}\begin{pmatrix} \mathrm{A}_{54}\mathrm{A}_{41} + \mathrm{A}_{51} & \mathrm{A}_{54}\mathrm{A}_{42} & \mathrm{A}_{54}\mathrm{A}_{43} & \mathrm{A}_{54} & 1 \end{pmatrix},$$

where

$\mathrm{A}_{55} = (-\Omega_{55})^{-1} = \mathrm{E}\left[\dfrac{1-A}{\delta^*(C, 0)}\right]^{-1},$

$\mathrm{A}_{54} = \Omega_{54}(-\Omega_{44})^{-1} = \mathrm{E}\left[\dfrac{1-A}{\delta^*(C, 0)}\mu'^*(C, M)h^T(C, M)\right] \mathrm{E}\left[\dfrac{A\lambda_0^*(C, M)}{\delta^*(C, 1)\lambda_1^*(C, M)}\mu'^*(C, M)h(C, M)h^T(C, M)\right]^{-1},$

$\mathrm{A}_{41} = \Omega_{41}(-\Omega_{11})^{-1} = \mathrm{E}\left[-\dfrac{A\lambda_0^*(C, M)}{\delta^*(C, 1)\lambda_1^*(C, M)}h(C, M)[Y - \mu^*(C, M)]\nabla^T \log \delta^*(C, 1)\right] \mathrm{E}[-\nabla\nabla^T \log \delta^*(C, A)]^{-1},$

$\mathrm{A}_{51} = \Omega_{51}(-\Omega_{11})^{-1} = \mathrm{E}\left[-\dfrac{1-A}{\delta^*(C, 0)}[\mu^*(C, M) - \tau_{\mathrm{x}}^*]\nabla^T \log \delta^*(C, 0)\right] \mathrm{E}[-\nabla\nabla^T \log \delta^*(C, A)]^{-1}$

$\mathrm{A}_{42} = \Omega_{42}(-\Omega_{22})^{-1} = \mathrm{E}\left[-\dfrac{A\lambda_0^*(C, M)}{\delta^*(C, 1)\lambda_1^*(C, M)}h(C, M)[Y - \mu^*(C, M)]\nabla^T \log \lambda_1^*(C, M)\right] \mathrm{E}[-A\nabla\nabla^T \log \lambda_1^*(C, M)]^{-1},$

$\mathrm{A}_{43} = \Omega_{43}(-\Omega_{33})^{-1} = \mathrm{E}\left[\dfrac{A\lambda_0^*(C, M)}{\delta^*(C, 1)\lambda_1^*(C, M)}h(C, M)[Y - \mu^*(C, M)]\nabla^T \log \lambda_0^*(C, M)\right] \mathrm{E}[-(1-A)\nabla\nabla^T \log \lambda_0^*(C, M)]^{-1}.$

The influence function of the estimator is

$$\mathrm{A}_{55}\begin{pmatrix} \mathrm{A}_{54}\mathrm{A}_{41} + \mathrm{A}_{51} & \mathrm{A}_{54}\mathrm{A}_{42} & \mathrm{A}_{54}\mathrm{A}_{43} & \mathrm{A}_{54} & 1 \end{pmatrix}\psi^*(Z), \qquad (A63)$$

where

$$\psi^*(Z) = \begin{pmatrix} \nabla \log \delta^*(C, A) \\ A\nabla \log \lambda_1^*(C, M) \\ (1-A)\nabla \log \lambda_0^*(C, M) \\ A\frac{\lambda_0^*(C,M)}{\delta^*(C,1)\lambda_1^*(C,M)} h(C, M)[Y - \mu^*(C, M)] \\ \frac{1-A}{\delta^*(C,0)}[\mu^*(C, M) - \tau_x^*] \end{pmatrix}.$$

Special cases:

- If $\mu$ is correctly specified,

  $A_{41} = 0,$

  $A_{42} = 0,$

  $A_{43} = 0,$

  $A_{51} = \mathrm{E}\left[-\frac{1-A}{\delta^*(C,0)}[\mu^0(C, M) - \tau_x^*]\nabla^T \log \delta^*(C, 0)\right] \mathrm{E}[-\nabla\nabla^T \log \delta^*(C, A)]^{-1},$

  $A_{54} = \mathrm{E}\left[\frac{1-A}{\delta^*(C,0)}\mu'^0(C, M)h^T(C, M)\right] \mathrm{E}\left[\frac{A\lambda_0^*(C,M)}{\delta^*(C,1)\lambda_1^*(C,M)}\mu'^0(C, M)h(C, M)h^T(C, M)\right]^{-1},$

  and the relevant elements of $\psi^*(Z)$ are

  $$\psi_{1,4,5}^*(Z) = \begin{pmatrix} \nabla \log \delta^*(C, A) \\ \frac{A\lambda_0^*(C,M)}{\delta^*(C,1)\lambda_1^*(C,M)} h(C, M)[Y - \mu^0(C, M)] \\ \frac{1-A}{\delta^*(C,0)}[\mu^0(C, M) - \tau_x^*] \end{pmatrix}.$$

- If in addition to $\mu$, $\delta$ is also correctly specified,

  $A_{55} = 1,$

  $A_{41} = 0,$

  $A_{42} = 0,$

  $A_{43} = 0,$

  $A_{51} = \underbrace{\mathrm{E}\left[-\frac{1-A}{\delta^0(C,0)}[\mu^0(C, M) - \tau_x^0]\nabla^T \log \delta^0(C, 0)\right]}_{\mathrm{E}[-\{\nu_x^0(C) - \tau_x^0\}\nabla^T \log \delta^0(C,0)]} \mathrm{E}[-\nabla\nabla^T \log \delta^0(C, A)]^{-1},$

  $A_{54} = \underbrace{\mathrm{E}\left[\frac{1-A}{\delta^0(C,0)}\mu'^*(C, M)h^T(C, M)\right]}_{\mathrm{E}\{\mathrm{E}[\mu'^*(C,M)h^T(C,M)|C,A=0]\}} \mathrm{E}\left[\frac{A\lambda_0^*(C,M)}{\delta^0(C,1)\lambda_1^*(C,M)}\mu'^*(C, M)h(C, M)h^T(C, M)\right]^{-1},$

  and the relevant elements of $\psi^*(Z)$ are

  $$\psi_{1,4,5}^*(Z) = \begin{pmatrix} \nabla \log \delta^0(C, A) \\ \frac{A\lambda_0^*(C,M)}{\delta^0(C,1)\lambda_1^*(C,M)} h(C, M)[Y - \mu^0(C, M)] \\ \frac{1-A}{\delta^0(C,0)}[\mu^0(C, M) - \boxed{\tau_x^0}] \end{pmatrix}.$$

- If $\delta$, $\lambda_1$ and $\lambda_0$ are correctly specified,

  $A_{55} = 1,$

$$A_{51} = \underbrace{E\left[-\frac{1-A}{\delta^0(C,0)}[\mu^*(C,M) - \tau_x^0]\nabla^T \log \delta^0(C,0)\right]}_{E[-\{E[\mu^*(C,M)|C,A=0]-\tau_x^0\}\nabla^T \log \delta^0(C,0)]} E[-\nabla\nabla^T \log \delta^0(C,A)]^{-1},$$

$$A_{54} = E[E\{\mu'^*(C,M)h^T(C,M) \mid C, A=0\}]E[E\{\mu'^*(C,M)h(C,M)h^T(C,M) \mid C, A=0\}]^{-1} = \begin{pmatrix}1 & 0 & \cdots & 0\end{pmatrix},$$

by the Lemma and the fact that $\mu'(\cdot) > 0$ (because $\mu(\cdot)$ as an inverse link function is monotone increasing), which implies

$$A_{54}A_{41} = \underbrace{E\left[-\frac{A\lambda_0^0(C,M)}{\delta^0(C,1)\lambda_1^0(C,M)}[Y - \mu^*(C,M)]\nabla^T \log \delta^0(C,1)\right]}_{E[-\{\nu_x^0(C)-E[\mu^*(C,M)|C,A=0]\}\nabla^T \log \delta^0(C,1)]} E[-\nabla\nabla^T \log \delta^0(C,A)]^{-1},$$

$$A_{54}A_{42} = \underbrace{E\left[-\frac{A\lambda_0^0(C,M)}{\delta^0(C,1)\lambda_1^0(C,M)}[Y - \mu^*(C,M)]\nabla^T \log \lambda_1^0(C,M)\right]}_{E[-E\{[\mu^0(C,M)-\mu^*(C,M)]\nabla^T \log \lambda_1^0(C,M)|C,A=0\}]} E[-A\nabla\nabla^T \log \lambda_1^0(C,M)]^{-1},$$

$$A_{54}A_{43} = \underbrace{E\left[\frac{A\lambda_0^0(C,M)}{\delta^0(C,1)\lambda_1^0(C,M)}[Y - \mu^*(C,M)]\nabla^T \log \lambda_0^0(C,M)\right]}_{E[E\{[\mu^0(C,M)-\mu^*(C,M)]\nabla^T \log \lambda_0^0(C,M)|C,A=0\}]} E[-(1-A)\nabla\nabla^T \log \lambda_0^0(C,M)]^{-1},$$

$$A_{54}\psi_4^*(Z) = A\frac{\lambda_0^0(C,M)}{\delta^0(C,1)\lambda_1^0(C,M)}[Y - \mu^*(C,M)],$$

and the other elements of $\psi^*(Z)$ (other than the fourth element dealt with above) are

$$\psi_{1,2,3,5}^*(Z) = \begin{pmatrix} \nabla \log \delta^0(C,A) \\ A\nabla \log \lambda_1^0(C,M) \\ (1-A)\nabla \log \lambda_0^0(C,M) \\ \frac{1-A}{\delta^0(C,0)}[\mu^*(C,M) - \tau_x^0] \end{pmatrix}.$$

- If $\delta, \lambda_1, \lambda_0, \mu$ are all correctly specified, the influence function reduces to

$$A\frac{\lambda_0^0(C,M)}{\delta^0(C,1)\lambda_1^0(C,M)}[Y - \mu^0(C,M)] + \frac{1-A}{\delta^0(C,0)}[\mu^0(C,M) - \tau_x^0].$$

**Using weights estimated by method 2 – via conditional exposure models**

$$\psi(Z,\theta) = \begin{pmatrix} \nabla \log \delta(C,A;\theta_1) \\ \nabla \log \gamma(C,M,A;\theta_2) \\ A\frac{\gamma(C,M,0;\theta_2)}{\delta(C,0;\theta_1)\gamma(C,M,1;\theta_2)}h(C,M)[Y - \mu(\theta_3^T h(C,M))] \\ \frac{1-A}{\delta(C,0;\theta_1)}[\mu(\theta_3^T h(C,M)) - \tau_x] \end{pmatrix}. \quad (A64)$$

$$\Omega = \begin{pmatrix} \Omega_{11} & 0 & 0 & 0 \\ 0 & \Omega_{22} & 0 & 0 \\ \Omega_{31} & \Omega_{32} & \Omega_{33} & 0 \\ \Omega_{41} & 0 & \Omega_{43} & \Omega_{44} \end{pmatrix},$$

where

$$\Omega_{11} = \mathrm{E}[\nabla\nabla^T \log \delta^*(C,A)],$$

$$\Omega_{22} = \mathrm{E}[\nabla\nabla^T \log \gamma^*(C,M,A)],$$

$$\Omega_{33} = \mathrm{E}\left[-\frac{A\gamma^*(C,M,0)}{\delta^*(C,0)\gamma^*(C,M,1)}\mu'^*(C,M)h(C,M)h^T(C,M)\right],$$

$$\Omega_{44} = \mathrm{E}\left[-\frac{1-A}{\delta^*(C,0)}\right].$$

$$\Omega_{31} = \mathrm{E}\left[-\frac{A\gamma^*(C,M,0)}{\delta^*(C,0)\gamma^*(C,M,1)}h(C,M)[Y-\mu^*(C,M)]\nabla^T\log\delta^*(C,0)\right],$$

$$\Omega_{41} = \mathrm{E}\left[-\frac{1-A}{\delta^*(C,0)}[\mu^*(C,M)-\tau_\mathrm{x}^*]\nabla^T\log\delta^*(C,0)\right],$$

$$\Omega_{32} = \mathrm{E}\left[-\frac{A}{\delta^*(C,1)\gamma^*(C,M,1)}h(C,M)[Y-\mu^*(C,M)]\nabla^T\log\gamma^*(C,M,1)\right],$$

$$\Omega_{43} = \mathrm{E}\left[\frac{1-A}{\delta^*(C,0)}\mu'^*(C,M)h^T(C,M)\right],$$

$$\mathrm{A} = \mathrm{A}_{44}\begin{pmatrix} \mathrm{A}_{43}\mathrm{A}_{31} + \mathrm{A}_{41} & \mathrm{A}_{43}\mathrm{A}_{32} & \mathrm{A}_{43} & 1 \end{pmatrix},$$

and the influence function of the estimator is

$$\mathrm{A}_{44}\begin{pmatrix} \mathrm{A}_{43}\mathrm{A}_{31} + \mathrm{A}_{41} & \mathrm{A}_{43}\mathrm{A}_{32} & \mathrm{A}_{43} & 1 \end{pmatrix}\psi^*(Z), \tag{A65}$$

where

$$\mathrm{A}_{44} = (-\Omega_{44})^{-1} = \mathrm{E}\left[\frac{1-A}{\delta^*(C,0)}\right]^{-1},$$

$$\mathrm{A}_{43} = \Omega_{43}(-\Omega_{33})^{-1} = \mathrm{E}\left[\frac{1-A}{\delta^*(C,0)}\mu'^*(C,M)h^T(C,M)\right]\mathrm{E}\left[\frac{A\gamma^*(C,M,0)}{\delta^*(C,0)\gamma^*(C,M,1)}\mu'^*(C,M)h(C,M)h^T(C,M)\right]^{-1},$$

$$\mathrm{A}_{31} = \Omega_{31}(-\Omega_{11})^{-1} = \mathrm{E}\left[-\frac{A\gamma^*(C,M,0)}{\delta^*(C,0)\gamma^*(C,M,1)}h(C,M)[Y-\mu^*(C,M)]\nabla^T\log\delta^*(C,0)\right]\mathrm{E}[-\nabla\nabla^T\log\delta^*(C,A)]^{-1},$$

$$\mathrm{A}_{41} = \Omega_{41}(-\Omega_{11})^{-1} = \mathrm{E}\left[-\frac{1-A}{\delta^*(C,0)}[\mu^*(C,M)-\tau_\mathrm{x}^*]\nabla^T\log\delta^*(C,0)\right]\mathrm{E}[-\nabla\nabla^T\log\delta^*(C,A)]^{-1},$$

$$\mathrm{A}_{32} = \Omega_{32}(-\Omega_{22})^{-1} = \mathrm{E}\left[-\frac{A}{\delta^*(C,1)\gamma^*(C,M,1)}h(C,M)[Y-\mu^*(C,M)]\nabla^T\log\gamma^*(C,M,1)\right]\mathrm{E}[-\nabla\nabla^T\log\gamma^*(C,M,A)]^{-1},$$

$$\psi^*(Z) = \begin{pmatrix} \nabla\log\delta^*(C,A) \\ \nabla\log\gamma^*(C,M,A) \\ A\frac{\gamma^*(C,M,0)}{\delta^*(C,0)\gamma^*(C,M,1)}h(C,M)[Y-\mu^*(C,M)] \\ \frac{1-A}{\delta^*(C,0)}[\mu^*(C,M)-\tau_\mathrm{x}^*] \end{pmatrix}.$$

Special cases:

- If $\mu$ is correctly specified,

$$\mathrm{A}_{31} = 0,$$
$$\mathrm{A}_{32} = 0,$$
$$\mathrm{A}_{41} = \mathrm{E}\left[-\frac{1-A}{\delta^*(C,0)}[\mu^0(C,M)-\tau_\mathrm{x}^*]\nabla^T\log\delta^*(C,0)\right]\mathrm{E}[-\nabla\nabla^T\log\delta^*(C,A)]^{-1},$$
$$\mathrm{A}_{43} = \mathrm{E}\left[\frac{1-A}{\delta^*(C,0)}\mu'^*(C,M)h^T(C,M)\right]\mathrm{E}\left[\frac{A\gamma^*(C,M,0)}{\delta^*(C,0)\gamma^*(C,M,1)}\mu'^*(C,M)h(C,M)h^T(C,M)\right]^{-1},$$

and the relevant elements of $\psi^*(Z)$ are

$$\psi^*_{1,2,4}(Z) = \begin{pmatrix} \nabla\log\delta^*(C,A) \\ A\frac{\gamma^*(C,M,0)}{\delta^*(C,0)\gamma^*(C,M,1)}h(C,M)[Y-\mu^0(C,M)] \\ \frac{1-A}{\delta^*(C,0)}[\mu^0(C,M)-\tau_\mathrm{x}^*] \end{pmatrix}.$$

31

- If in addition to $\mu$, $\delta$ is also correctly specified,

$$A_{44} = 1,$$

$$A_{41} = \underbrace{E\left[-\frac{1-A}{\delta^0(C,0)}[\mu^0(C,M) - \tau_x^0]\nabla^T \log \delta^0(C,0)\right]}_{E[-\{\nu_x^0(C) - \tau_x^0\}\nabla^T \log \delta^0(C,0)]} E[-\nabla\nabla^T \log \delta^0(C,A)]^{-1},$$

$$A_{43} = \underbrace{E\left[\frac{1-A}{\delta^0(C,0)}\mu'^*(C,M)h^T(C,M)\right]}_{E\{E[\mu'^*(C,M)h^T(C,M)|C,A=0]\}} E\left[\frac{A\gamma^*(C,M,0)}{\delta^0(C,0)\gamma^*(C,M,1)}\mu'^*(C,M)h(C,M)h^T(C,M)\right]^{-1},$$

and the relevant elements of $\psi^*(Z)$ are

$$\psi_{1,2,4}^*(Z) = \begin{pmatrix} \nabla \log \delta^0(C,A) \\ A\frac{\gamma^*(C,M,0)}{\delta^0(C,0)\gamma^*(C,M,1)}h(C,M)[Y - \mu^0(C,M)] \\ \frac{1-A}{\delta^0(C,0)}[\mu^0(C,M) - \tau_x^*] \end{pmatrix}.$$

- If $\delta$ and $\gamma$ are correctly specified,

$$A_{44} = 1,$$

$$A_{41} = \underbrace{E\left[-\frac{1-A}{\delta^0(C,0)}[\mu^*(C,M) - \tau_x^0]\nabla^T \log \delta^0(C,0)\right]}_{E[-\{E[\mu^*(C,M)|C,A=0] - \tau_x^0\}\nabla^T \log \delta^0(C,0)]} E[-\nabla\nabla^T \log \delta^0(C,A)]^{-1}$$

$$A_{43} = E\left[E\{\mu'^*(C,M)h^T(C,M) \mid C, A = 0\}\right] E\left[\mu'^*(C,M)h(C,M)h^T(C,M)\right]^{-1} = \begin{pmatrix} 1 & 0 & \cdots & 0 \end{pmatrix},$$

by the Lemma and the fact that $\mu'(\cdot) > 0$ (because $\mu(\cdot)$ as an inverse link function is monotone increasing), which implies

$$A_{43}A_{31} = \underbrace{E\left[-\frac{A\gamma^0(C,M,0)}{\delta^0(C,0)\gamma^0(C,M,1)}[Y - \mu^*(C,M)]\nabla^T \log \delta^0(C,0)\right]}_{E[-\{\nu_x^0(C) - E[\mu^*(C,M)|C,A=0]\}\nabla^T \log \delta^0(C,0)]} E[-\nabla\nabla^T \log \delta^0(C,A)]^{-1},$$

$$A_{43}A_{32} = \underbrace{E\left[-\frac{A}{\delta^0(C,1)\gamma^0(C,M,1)}[Y-\mu^*(C,M)]\nabla^T \log \gamma^0(C,M,1)\right]}_{E\left[-E\left\{[\mu^0(C,M) - \mu^*(C,M)]\frac{\nabla^T \log \gamma^0(C,M,1)}{\gamma^0(C,M,0)}\right\}\right]} E[-\nabla\nabla^T \log \gamma^0(C,M,A)]^{-1},$$

$$A_{43}\psi_3^*(Z) = A\frac{\gamma^0(C,M,0)}{\delta^0(C,0)\gamma^0(C,M,1)}[Y - \mu^*(C,M)]$$

and the other element of $\psi^*(Z)$ (other than the third element dealt with above) are

$$\psi_{1,2,4}^*(Z) = \begin{pmatrix} \nabla \log \delta^0(C,A) \\ \nabla \log \gamma^0(C,M,A) \\ \frac{1-A}{\delta^0(C,0)}[\mu^*(C,M) - \tau_x^0] \end{pmatrix}.$$

- If $\delta, \gamma, \mu$ are all correctly specified, the influence function reduces to

$$A\frac{\gamma^0(C,M,0)}{\delta^0(C,0)\gamma^0(C,M,1)}[Y - \mu^0(C,M)] + \frac{1-A}{\delta^0(C,0)}[\mu^0(C,M) - \tau_x^0].$$

32

### C.4.2.2 crw|Ypred.MR

**Using weights estimated by method 1 – via mediator density models**

$$\psi(Z,\theta) = \begin{pmatrix} \nabla \log \delta(C,A;\theta_1) \\ A\nabla \log \lambda_1(C,M;\theta_2) \\ (1-A)\nabla \log \lambda_0(C,M;\theta_3) \\ A\dfrac{\lambda_0(C,M;\theta_3)}{\delta(C,1;\theta_1)\lambda_1(C,M;\theta_2)}g(C)[Y - \nu_{\mathrm{x}}(\theta_4^T g(C))] \\ \nu_{\mathrm{x}}(C;\theta_4) - \tau_{\mathrm{x}} \end{pmatrix}. \quad (A66)$$

$$\Omega = \begin{pmatrix} \Omega_{11} & 0 & 0 & 0 & 0 \\ 0 & \Omega_{22} & 0 & 0 & 0 \\ 0 & 0 & \Omega_{33} & 0 & 0 \\ \Omega_{41} & \Omega_{42} & \Omega_{43} & \Omega_{44} & 0 \\ 0 & 0 & 0 & \Omega_{54} & -1 \end{pmatrix},$$

where

$$\Omega_{11} = \mathrm{E}[\nabla\nabla^T \log \delta^*(C,A)],$$
$$\Omega_{22} = \mathrm{E}[A\nabla\nabla^T \log \lambda_1^*(C,M)],$$
$$\Omega_{33} = \mathrm{E}[(1-A)\nabla\nabla^T \log \lambda_0^*(C,M)],$$
$$\Omega_{44} = \mathrm{E}\left[-\frac{A\lambda_0^*(C,M)}{\delta^*(C,1)\lambda_1^*(C,M)}\nu_{\mathrm{x}}'^*(C)g(C)g^T(C)\right],$$
$$\Omega_{41} = \mathrm{E}\left[-\frac{A\lambda_0^*(C,M)}{\delta^*(C,1)\lambda_1^*(C,M)}g(C)[Y - \nu_{\mathrm{x}}^*(C)]\nabla^T \log \delta^*(C,0)\right]$$
$$\Omega_{42} = \mathrm{E}\left[-\frac{A\lambda_0^*(C,M)}{\delta^*(C,1)\lambda_1^*(C,M)}g(C)[Y - \nu_{\mathrm{x}}^*(C)]\nabla^T \log \lambda_1^*(C,M)\right]$$
$$\Omega_{43} = \mathrm{E}\left[\frac{A\lambda_0^*(C,M)}{\delta^*(C,1)\lambda_1^*(C,M)}g(C)[Y - \nu_{\mathrm{x}}^*(C)]\nabla^T \log \lambda_0^*(C,M)\right]$$
$$\Omega_{54} = \mathrm{E}[\nu_{\mathrm{x}}'^*(C)g^T(C)].$$

$$\mathrm{A} = \begin{pmatrix} \Omega_{54}(-\Omega_{44})^{-1}\Omega_{41}(-\Omega_{11})^{-1} & \Omega_{54}(-\Omega_{44})^{-1}\Omega_{42}(-\Omega_{22})^{-1} & \Omega_{54}(-\Omega_{44})^{-1}\Omega_{43}(-\Omega_{33})^{-1} & \Omega_{54}(-\Omega_{44})^{-1} & 1 \end{pmatrix}$$
$$= \begin{pmatrix} \mathrm{A}_{54}\mathrm{A}_{41} & \mathrm{A}_{54}\mathrm{A}_{42} & \mathrm{A}_{54}\mathrm{A}_{43} & \mathrm{A}_{54} & 1 \end{pmatrix}.$$

The influence function of the estimator is

$$\begin{pmatrix} \mathrm{A}_{54}\mathrm{A}_{41} & \mathrm{A}_{54}\mathrm{A}_{42} & \mathrm{A}_{54}\mathrm{A}_{43} & \mathrm{A}_{54} & 1 \end{pmatrix}\psi^*(Z), \quad (A67)$$

where

$$\mathrm{A}_{54} = \mathrm{E}[\nu_{\mathrm{x}}'^*(C)g^T(C)]\mathrm{E}\left[\frac{A\lambda_0^*(C,M)}{\delta^*(C,1)\lambda_1^*(C,M)}\nu_{\mathrm{x}}'^*(C)g(C)g^T(C)\right]^{-1},$$

$$\mathrm{A}_{41} = \mathrm{E}\left[-\frac{A\lambda_0^*(C,M)}{\delta^*(C,1)\lambda_1^*(C,M)}g(C)[Y - \nu_{\mathrm{x}}^*(C)]\nabla^T \log \delta^*(C,1)\right]\mathrm{E}[-\nabla\nabla^T \log \delta^*(C,A)]^{-1},$$

$$\mathrm{A}_{42} = \mathrm{E}\left[-\frac{A\lambda_0^*(C,M)}{\delta^*(C,1)\lambda_1^*(C,M)}g(C)[Y - \nu_{\mathrm{x}}^*(C)]\nabla^T \log \lambda_1^*(C,M)\right]\mathrm{E}[-A\nabla\nabla^T \log \lambda_1^*(C,M)]^{-1},$$

$$\mathrm{A}_{43} = \mathrm{E}\left[\frac{A\lambda_0^*(C,M)}{\delta^*(C,1)\lambda_1^*(C,M)}g(C)[Y - \nu_{\mathrm{x}}^*(C)]\nabla^T \log \lambda_0^*(C,M)\right]\mathrm{E}[-(1-A)\nabla\nabla^T \log \lambda_0^*(C,M)]^{-1},$$

$$\psi^*(Z) = \begin{pmatrix} \nabla \log \delta^*(C,A) \\ A\nabla \log \lambda_1^*(C,M) \\ (1-A)\nabla \log \lambda_0^*(C,M) \\ A\dfrac{\lambda_0^*(C,M)}{\delta^*(C,1)\lambda_1^*(C,M)}g(C)[Y - \nu_{\mathrm{x}}^*(C)] \\ \nu_{\mathrm{x}}^*(C) - \tau_{\mathrm{x}}^* \end{pmatrix}.$$

Special cases:

- If $\delta, \lambda_1, \lambda_0$ are correctly specified,

$$A_{54} = E[\nu_x'^*(C)g^T(C)]E[\nu_x'^*(C)g(C)g^T(C)]^{-1} = \begin{pmatrix} 1 & 0 & \cdots & 0 \end{pmatrix}$$

by the Lemma and the fact that $\nu_x'(\cdot) > 0$ (because $\nu_x(\cdot)$ as an inverse link function is monotone increasing), which implies

$$A_{54}A_{41} = \underbrace{E\left[-\frac{A\lambda_0^0(C,M)}{\delta^0(C,1)\lambda_1^0(C,M)}[Y - \nu_x^*(C)]\nabla^T \log \delta^0(C,1)\right]}_{E[-\{\nu_x^0(C)-\nu_x^*(C)\}\nabla^T \log \delta^0(C,1)]} E[-\nabla\nabla^T \log \delta^*(C,A)]^{-1},$$

$$A_{54}A_{42} = \underbrace{E\left[-\frac{A\lambda_0^0(C,M)}{\delta^0(C,1)\lambda_1^0(C,M)}[Y - \nu_x^*(C)]\nabla^T \log \lambda_1^0(C,M)\right]}_{E[-E\{[\mu^0(C,M)-\nu_x^*(C)]\nabla^T \log \lambda_1^0(C,M)|C,A=0\}]} E[-A\nabla\nabla^T \log \lambda_1^*(C,M)]^{-1},$$

$$A_{54}A_{43} = \underbrace{E\left[\frac{A\lambda_0^0(C,M)}{\delta^0(C,1)\lambda_1^0(C,M)}[Y - \nu_x^*(C)]\nabla^T \log \lambda_0^0(C,M)\right]}_{E[E\{[\mu^0(C,M)-\nu_x^*(C)]\nabla^T \log \lambda_0^0(C,M)|C,A=0\}]} E[-(1-A)\nabla\nabla^T \log \lambda_0^*(C,M)]^{-1},$$

$$A_{54}\psi_4^*(Z) = A\frac{\lambda_0^0(C,M)}{\delta^0(C,1)\lambda_1^0(C,M)}[Y - \nu_x^*(C)],$$

and the other elements of $\psi^*(Z)$ (other than the fourth element dealt with above) are

$$\psi_{1,2,3,5}^*(Z) = \begin{pmatrix} \nabla \log \delta^0(C,A) \\ A\nabla \log \lambda_1^0(C,M) \\ (1-A)\nabla \log \lambda_0^0(C,M) \\ \nu_x^*(C) - \tau_x^0 \end{pmatrix}.$$

- If $\lambda_1, \lambda_0$ and $\nu_x$ are correctly specified,

$$A_{54} = E[\nu_x'^0(C)g^T(C)]\underbrace{E\left[\frac{A\lambda_0^0(C,M)}{\delta^*(C,1)\lambda_1^0(C,M)}\nu_x'^0(C)g(C)g^T(C)\right]^{-1}}_{E\left[\frac{\delta^0(C,1)}{\delta^*(C,1)}\nu_x'^0(C)g(C)g^T(C)\right]^{-1}},$$

$A_{41} = 0,$

$$A_{42} = \underbrace{E\left[-\frac{A\lambda_0^0(C,M)}{\delta^*(C,1)\lambda_1^0(C,M)}g(C)[Y - \nu_x^0(C)]\nabla^T \log \lambda_1^0(C,M)\right]}_{E\left[-\frac{\delta^0(C,1)}{\delta^*(C,1)}g(c)E\{[\mu^0(C,M)-\nu_x^0(C)]\nabla^T \log \lambda_1^0(C,M)|C,A=0\}\right]} E[-A\nabla\nabla^T \log \lambda_1^0(C,M)]^{-1},$$

$$A_{43} = \underbrace{E\left[\frac{A\lambda_0^0(C,M)}{\delta^*(C,1)\lambda_1^0(C,M)}g(C)[Y - \nu_x^0(C)]\nabla^T \log \lambda_0^0(C,M)\right]}_{E\left[\frac{\delta^0(C,1)}{\delta^*(C,1)}g(c)E\{[\mu^0(C,M)-\nu_x^0(C)]\nabla^T \log \lambda_0^0(C,M)|C,A=0\}\right]} E[-(1-A)\nabla\nabla^T \log \lambda_0^*(C,M)]^{-1},$$

and the relevant elements of $\psi^*(Z)$ are

$$\psi_{2,3,4,5}^*(Z) = \begin{pmatrix} A\nabla \log \lambda_1^0(C,M) \\ (1-A)\nabla \log \lambda_0^0(C,M) \\ A\frac{\lambda_0^0(C,M)}{\delta^*(C,1)\lambda_1^0(C,M)}g(C)[Y - \nu_x^0(C)] \\ \nu_x^0(C) - \tau_x^0 \end{pmatrix}.$$

34

- If $\delta, \lambda_1, \lambda_0, \nu_x$ are all correctly specified,

$$A_{41} = 0,$$

$$A_{54}A_{42} = \underbrace{E\left[-\frac{A\lambda_0^0(C,M)}{\delta^0(C,1)\lambda_1^0(C,M)}[Y-\nu_x^0(C)]\nabla^T\log\lambda_1^0(C,M)\right]}_{E\left[-E\{[\mu^0(C,M)-\nu_x^0(C)]\nabla^T\log\lambda_1^0(C,M)|C,A=0\}\right]}E[-A\nabla\nabla^T\log\lambda_1^0(C,M)]^{-1},$$

$$A_{54}A_{43} = \underbrace{E\left[\frac{A\lambda_0^0(C,M)}{\delta^0(C,1)\lambda_1^0(C,M)}[Y-\nu_x^0(C)]\nabla^T\log\lambda_0^0(C,M)\right]}_{E\left[E\{[\mu^0(C,M)-\nu_x^0(C)]\nabla^T\log\lambda_0^0(C,M)|C,A=0\}\right]}E[-(1-A)\nabla\nabla^T\log\lambda_0^0(C,M)]^{-1},$$

$$A_{54}\psi_4^*(Z) = A\frac{\lambda_0^0(C,M)}{\delta^0(C,1)\lambda_1^0(C,M)}[Y-\nu_x^0(C)],$$

and the other relevant elements of $\psi^*(Z)$ (other than the fourth element dealt with above) are

$$\psi_{2,3,5}^*(Z) = \begin{pmatrix} A\nabla\log\lambda_1^0(C,M) \\ (1-A)\nabla\log\lambda_0^0(C,M) \\ \nu_x^0(C) - \tau_x^0 \end{pmatrix}.$$

**Using weights estimated by method 2 – via conditional exposure models**

$$\psi(Z,\theta) = \begin{pmatrix} \nabla\log\delta(C,A;\theta_1) \\ \nabla\log\gamma(C,M,A;\theta_2) \\ A\frac{\gamma(C,M,0;\theta_2)}{\delta(C,0;\theta_1)\gamma(C,M,1;\theta_2)}g(C)[Y-\nu_x(\theta_3^T g(C))] \\ \nu_x(C;\theta_3) - \tau_x \end{pmatrix}. \quad (A68)$$

$$\Omega = \begin{pmatrix} \Omega_{11} & 0 & 0 & 0 \\ 0 & \Omega_{22} & 0 & 0 \\ \Omega_{31} & \Omega_{32} & \Omega_{33} & 0 \\ 0 & 0 & \Omega_{43} & -1 \end{pmatrix},$$

where

$\Omega_{11} = E[\nabla\nabla^T\log\delta^*(C,A)],$

$\Omega_{22} = E[\nabla\nabla^T\log\gamma^*(C,M,A)],$

$\Omega_{33} = E\left[-\frac{A\gamma^*(C,M,0)}{\delta^*(C,0)\gamma^*(C,M,1)}\nu_x'^*(C)g(C)g^T(C)\right],$

$\Omega_{31} = E\left[-\frac{A\gamma^*(C,M,0)}{\delta^*(C,0)\gamma^*(C,M,1)}g(C)[Y-\nu_x^*(C)]\nabla^T\log\delta^*(C,0)\right]$

$\Omega_{32} = E\left[-\frac{A}{\delta^*(C,0)\gamma^*(C,M,1)}g(C)[Y-\nu_x^*(C)]\nabla^T\log\gamma^*(C,M,1)\right],$

$\Omega_{43} = E[\nu_x'^*(C)g^T(C)].$

$$A = \begin{pmatrix} \Omega_{43}(-\Omega_{33})^{-1}\Omega_{31}(-\Omega_{11})^{-1} & \Omega_{43}(-\Omega_{33})^{-1}\Omega_{32}(-\Omega_{22})^{-1} & \Omega_{43}(-\Omega_{33})^{-1} & 1 \end{pmatrix}$$
$$= \begin{pmatrix} A_{43}A_{31} & A_{43}A_{32} & A_{43} & 1 \end{pmatrix}.$$

The influence function of the estimator is

$$\begin{pmatrix} A_{43}A_{31} & A_{43}A_{32} & A_{43} & 1 \end{pmatrix}\psi^*(Z), \quad (A69)$$

where

$$A_{43} = E[\nu_x'^*(C)g^T(C)]E\left[\frac{A\gamma^*(C,M,0)}{\delta^*(C,0)\gamma^*(C,M,1)}\nu_x'^*(C)g(C)g^T(C)\right]^{-1},$$

$$A_{31} = E\left[-\frac{A\gamma^*(C,M,0)}{\delta^*(C,0)\gamma^*(C,M,1)}g(C)[Y-\nu_x^*(C)]\nabla^T\log\delta^*(C,0)\right]E[-\nabla\nabla^T\log\delta^*(C,A)]^{-1},$$

$$A_{32} = E\left[-\frac{A}{\delta^*(C,0)\gamma^*(C,M,1)}g(C)[Y-\nu_x^*(C)]\nabla^T\log\gamma^*(C,M,1)\right]E[-\nabla\nabla^T\log\gamma^*(C,M,A)]^{-1},$$

$$\psi^*(Z) = \begin{pmatrix} \nabla\log\delta^*(C,A) \\ \nabla^T\log\gamma^*(C,M,A) \\ \frac{A\gamma^*(C,M,0)}{\delta^*(C,0)\gamma^*(C,M,1)}g(C)[Y-\nu_x^*(C)] \\ \nu_x^*(C) - \tau_x^* \end{pmatrix}.$$

Special cases:

- If $\delta$ and $\gamma$ are correctly specified,

$$A_{43} = E[\nu_x'^*(C)g^T(C)]E[\nu_x'^*(C)g(C)g^T(C)]^{-1} = \begin{pmatrix} 1 & 0 & \cdots & 0 \end{pmatrix},$$

by the Lemma and the fact that $\nu_x'(\cdot) > 0$ (because $\nu_x(\cdot)$ as an inverse link function is monotone increasing), which implies

$$A_{43}A_{31} = \underbrace{E\left[-\frac{A\gamma^0(C,M,0)}{\delta^0(C,0)\gamma^0(C,M,1)}[Y-\nu_x^*(C)]\nabla^T\log\delta^0(C,0)\right]}_{E[-[\nu_x^0(C)-\nu_x^*(C)]\nabla^T\log\delta^0(C,0)]}E[-\nabla\nabla^T\log\delta^0(C,A)]^{-1},$$

$$A_{43}A_{32} = \underbrace{E\left[-\frac{A}{\delta^0(C,0)\gamma^0(C,M,1)}[Y-\nu_x^*(C)]\nabla^T\log\gamma^0(C,M,1)\right]}_{E\left[-E\left\{[\mu^0(C,M)-\nu_x^*(C)]\frac{\nabla^T\log\gamma^0(C,M,1)}{\gamma^0(C,M,0)}|C,A=0\right\}\right]}E[-\nabla\nabla^T\log\gamma^0(C,M,A)]^{-1},$$

$$A_{43}\psi_3^*(Z) = \frac{A\gamma^0(C,M,0)}{\delta^0(C,0)\gamma^0(C,M,1)}[Y-\nu_x^*(C)],$$

and the other elements of $\psi^*(Z)$ (other than the third element dealt with above) are

$$\psi_{1,2,4}^*(Z) = \begin{pmatrix} \nabla\log\delta^0(C,A) \\ \nabla^T\log\gamma^0(C,M,A) \\ \nu_x^*(C) - \tau_x^0 \end{pmatrix}.$$

- If $\gamma$ and $\nu_x$ are correctly specified,

$$A_{31} = 0,$$

$$A_{43} = E[\nu_x'^0(C)g^T(C)]\underbrace{E\left[\frac{A\gamma^0(C,M,0)}{\delta^*(C,0)\gamma^0(C,M,1)}\nu_x'^0(C)g(C)g^T(C)\right]^{-1}}_{E\left[\frac{\delta^0(C,0)}{\delta^*(C,0)}\nu_x'^0(C)g(C)g^T(C)\right]^{-1}},$$

$$A_{32} = \underbrace{E\left[-\frac{A}{\delta^*(C,0)\gamma^0(C,M,1)}g(C)[Y-\nu_x^0(C)]\nabla^T\log\gamma^0(C,M,1)\right]}_{E\left\{-\frac{\delta^0(C,0)}{\delta^*(C,0)}g(C)E\left[[\mu^0(C,M)-\nu_x^0(C)]\frac{\nabla^T\log\gamma^0(C,M,1)}{\gamma^0(C,M,0)}|C,A=0\right]\right\}}E[-\nabla\nabla^T\log\gamma^0(C,M,A)]^{-1},$$

36

and the relevant elements of $\psi^*(Z)$ are

$$\psi^*_{2,3,4}(Z) = \begin{pmatrix} \nabla^T \log \gamma^0(C,M,A) \\ \frac{A\gamma^0(C,M,0)}{\delta^*(C,0)\gamma^0(C,M,1)} g(C)[Y - \nu^0_{\mathrm{x}}(C)] \\ \nu^0_{\mathrm{x}}(C) - \tau^0_{\mathrm{x}} \end{pmatrix}.$$

- If $\delta, \gamma, \nu_{\mathrm{x}}$ are all correctly specified,

$$A_{31} = 0,$$

$$A_{43}A_{32} = \underbrace{\mathrm{E}\left[-\frac{A}{\delta^0(C,0)\gamma^0(C,M,1)}[Y - \nu^0_{\mathrm{x}}(C)]\nabla^T \log \gamma^0(C,M,1)\right]}_{\mathrm{E}\left\{-\mathrm{E}\left[[\mu^0(C,M) - \nu^0_{\mathrm{x}}(C)]\frac{\nabla^T \log \gamma^0(C,M,1)}{\gamma^0(C,M,0)}|C,A=0\right]\right\}} \mathrm{E}[-\nabla\nabla^T \log \gamma^0(C,M,A)]^{-1},$$

$$A_{43}\psi^*_3(Z) = \frac{A\gamma^0(C,M,0)}{\delta^0(C,0)\gamma^0(C,M,1)}[Y - \nu^0_{\mathrm{x}}(C)],$$

and the other relevant elements of $\psi^*(Z)$ are

$$\psi^*_{2,4}(Z) = \begin{pmatrix} \nabla^T \log \gamma^0(C,M,A) \\ \nu^0_{\mathrm{x}}(C) - \tau^0_{\mathrm{x}} \end{pmatrix}.$$

### C.4.2.3 crw|Y2pred.R

**Using weights estimated by method 1 – via mediator density models**

$$\psi(Z,\theta) = \begin{pmatrix} \nabla \log \delta(C,A;\theta_1) \\ A\nabla \log \lambda_1(C,M;\theta_2) \\ (1-A)\nabla \log \lambda_0(C,M;\theta_3) \\ A\frac{\lambda_0(C,M;\theta_3)}{\delta(C,1;\theta_1)\lambda_1(C,M;\theta_2)}h(C,M)[Y - \mu(\theta_4^T h(C,M))] \\ \frac{1-A}{\delta(C,0;\theta_1)}g(C)[\mu(\theta_4^T h(C,M)) - \nu_{\mathrm{x}}(\theta_5^T g(C))] \\ \nu_{\mathrm{x}}(\theta_5^T g(C)) - \tau_{\mathrm{x}} \end{pmatrix} \tag{A70}$$

$$\Omega = \begin{pmatrix} \Omega_{11} & 0 & 0 & 0 & 0 & 0 \\ 0 & \Omega_{22} & 0 & 0 & 0 & 0 \\ 0 & 0 & \Omega_{33} & 0 & 0 & 0 \\ \Omega_{41} & \Omega_{42} & \Omega_{43} & \Omega_{43} & 0 & 0 \\ \Omega_{51} & 0 & 0 & \Omega_{54} & \Omega_{55} & 0 \\ 0 & 0 & 0 & 0 & \Omega_{65} & -1 \end{pmatrix},$$

where

$\Omega_{11} = \mathrm{E}[\nabla\nabla^T \log \delta^*(C,A)],$

$\Omega_{22} = \mathrm{E}[A\nabla\nabla^T \log \lambda_1^*(C,M)],$

$\Omega_{33} = \mathrm{E}[(1-A)\nabla\nabla^T \log \lambda_0^*(C,M)],$

$\Omega_{44} = \mathrm{E}\left[-\frac{A\lambda_0^*(C,M)}{\delta^*(C,0)\lambda_1^*(C,M)}\mu'^*(C,M)h(C,M)h^T(C,M)\right],$

$\Omega_{55} = \mathrm{E}\left[-\frac{1-A}{\delta^*(C,0)}\nu'^*_{\mathrm{x}}(C)g(C)g^T(C)\right],$

$\Omega_{41} = \mathrm{E}\left[-\frac{A\lambda_0^*(C,M)}{\delta^*(C,1)\lambda_1^*(C,M)}h(C,M)[Y - \mu^*(C,M)]\nabla^T \log \delta^*(C,1)\right],$

$\Omega_{51} = \mathrm{E}\left[-\frac{1-A}{\delta^*(C,0)}g(C)[\mu^*(C,M) - \nu^*_{\mathrm{x}}(C)]\nabla^T \log \delta^*(C,0)\right],$

$\Omega_{42} = \mathrm{E}\left[-\frac{A\lambda_0^*(C,M)}{\delta^*(C,1)\lambda_1^*(C,M)}h(C,M)[Y - \mu^*(C,M)]\nabla^T \log \lambda_1^*(C,M)\right],$

$\Omega_{43} = \mathrm{E}\left[\frac{A\lambda_0^*(C,M)}{\delta^*(C,1)\lambda_1^*(C,M)}h(C,M)[Y - \mu^*(C,M)]\nabla^T \log \lambda_0^*(C,M)\right],$

$\Omega_{54} = \mathrm{E}\left[\frac{1-A}{\delta^*(C,0)}g(C)\mu'^*(C,M)h^T(C,M)\right],$

$\Omega_{65} = \mathrm{E}[\nu'^*_{\mathrm{x}}(C)g^T(C)].$

$$A = \begin{pmatrix} A_{65}(A_{51} + A_{54}A_{41}) & A_{65}A_{54}A_{42} & A_{65}A_{54}A_{43} & A_{65}A_{54} & A_{65} & 1 \end{pmatrix}.$$

The influence function of the estimator is

$$\begin{pmatrix} A_{65}(A_{51} + A_{54}A_{41}) & A_{65}A_{54}A_{42} & A_{65}A_{54}A_{43} & A_{65}A_{54} & A_{65} & 1 \end{pmatrix} \psi^*(Z), \quad (A71)$$

where

$$A_{65} = \Omega_{65}(-\Omega_{55})^{-1} = E[\nu_x'^*(C)g^T(C)] E\left[\frac{1-A}{\delta^*(C,0)}\nu_x'^*(C)g(C)g^T(C)\right]^{-1},$$

$$A_{51} = \Omega_{51}(-\Omega_{11})^{-1} = E\left[-\frac{1-A}{\delta^*(C,0)}g(C)[\mu^*(C,M) - \nu_x^*(C)]\nabla^T \log \delta^*(C,0)\right] E[-\nabla\nabla^T \log \delta^*(C,A)]^{-1},$$

$$A_{54} = \Omega_{54}(-\Omega_{44})^{-1} = E\left[\frac{1-A}{\delta^*(C,0)}g(C)\mu'^*(C,M)h^T(C,M)\right] E\left[\frac{A\lambda_0^*(C,M)}{\delta^*(C,1)\lambda_1^*(C,M)}\mu'^*(C,M)h(C,M)h^T(C,M)\right]^{-1},$$

$$A_{41} = \Omega_{41}(-\Omega_{11})^{-1} = E\left[-\frac{A\lambda_0^*(C,M)}{\delta^*(C,1)\lambda_1^*(C,M)}h(C,M)[Y-\mu^*(C,M)]\nabla^T \log \delta^*(C,1)\right] E[-\nabla\nabla^T \log \delta^*(C,A)]^{-1},$$

$$A_{42} = \Omega_{42}(-\Omega_{22})^{-1} = E\left[-\frac{A\lambda_0^*(C,M)}{\delta^*(C,1)\lambda_1^*(C,M)}h(C,M)[Y-\mu^*(C,M)]\nabla^T \log \lambda_1^*(C,M)\right] E[-A\nabla\nabla^T \log \lambda_1^*(C,M)]^{-1},$$

$$A_{43} = \Omega_{43}(-\Omega_{33})^{-1} = E\left[\frac{A\lambda_0^*(C,M)}{\delta^*(C,1)\lambda_1^*(C,M)}h(C,M)[Y-\mu^*(C,M)]\nabla^T \log \lambda_0^*(C,M)\right] E[-(1-A)\nabla\nabla^T \log \lambda_0^*(C,M)]^{-1},$$

$$\psi^*(Z) = \begin{pmatrix} \nabla \log \delta^*(C,A) \\ A\nabla \log \lambda_1^*(C,M) \\ (1-A)\nabla \log \lambda_0^*(C,M) \\ A\frac{\lambda_0^*(C,M)}{\delta^*(C,1)\lambda_1^*(C,M)}h(C,M)[Y-\mu^*(C,M)] \\ \frac{1-A}{\delta^*(C,0)}g(C)[\mu^*(C,M) - \nu_x^*(C)] \\ \nu_x^*(C) - \tau_x^* \end{pmatrix}.$$

Special cases:

- If $\mu$ and $\nu_x$ are correctly specified,

  $A_{51} = 0,$
  $A_{41} = 0,$
  $A_{42} = 0,$
  $A_{43} = 0,$
  $$A_{65} = E[\nu_x'^0(C)g^T(C)] E\left[\frac{1-A}{\delta^*(C,0)}\nu_x'^0(C)g(C)g^T(C)\right]^{-1},$$
  $$A_{54} = E\left[\frac{1-A}{\delta^*(C,0)}g(C)\mu'^0(C,M)h^T(C,M)\right] E\left[\frac{A\lambda_0^*(C,M)}{\delta^*(C,1)\lambda_1^*(C,M)}\mu'^0(C,M)h(C,M)h^T(C,M)\right]^{-1},$$

  and the relevant elements of $\psi^*(Z)$ are

  $$\psi_{4,5,6}^*(Z) = \begin{pmatrix} A\frac{\lambda_0^*(C,M)}{\delta^*(C,1)\lambda_1^*(C,M)}h(C,M)[Y-\mu^0(C,M)] \\ \frac{1-A}{\delta^*(C,0)}g(C)[\mu^0(C,M) - \nu_x^0(C)] \\ \nu_x^0(C) - \tau_x^0 \end{pmatrix}.$$

38

- If $\delta$ and $\mu$ are correctly specified,

$$A_{41} = 0,$$
$$A_{42} = 0,$$
$$A_{43} = 0,$$
$$A_{65} = E[\nu_x'^*(C) g^T(C)] E[\nu_x'^*(C) g(C) g^T(C)]^{-1} = \begin{pmatrix} 1 & 0 & \cdots & 0 \end{pmatrix}$$

by the Lemma and the fact that $\nu_x'(\cdot) > 0$ (because $\nu_x(\cdot)$ as an inverse link function is monotone increasing), which implies

$$A_{65} A_{51} = E\left[-\frac{1-A}{\delta^0(C,0)}[\mu^0(C,M) - \nu_x^*(C)]\nabla^T \log \delta^0(C,0)\right] E[-\nabla\nabla^T \log \delta^0(C,A)]^{-1},$$

$$A_{65} A_{54} = E\left[\frac{1-A}{\delta^*(C,0)}\mu'^0(C,M) h^T(C,M)\right] E\left[\frac{A\lambda_0^*(C,M)}{\delta^0(C,1)\lambda_1^*(C,M)}\mu'^0(C,M) h(C,M) h^T(C,M)\right]^{-1},$$

$$A_{65} \psi_5^*(Z) = \frac{1-A}{\delta^0(C,0)}[\mu^0(C,M) - \nu_x^*(C)],$$

and the other relevant elements of $\psi^*(Z)$ (other than the fifth element dealt with above) are

$$\psi_{1,4,6} = \begin{pmatrix} \nabla \log \delta^0(C,A) \\ A\frac{\lambda_0^*(C,M)}{\delta^0(C,1)\lambda_1^*(C,M)} h(C,M)[Y - \mu^0(C,M)] \\ \nu_x^*(C) - \boxed{\tau_x^0} \end{pmatrix}.$$

- If in addition to $\delta$ and $\mu$, $\nu_x$ is also correctly specified,

$$A_{51} = 0,$$
$$A_{41} = 0,$$
$$A_{42} = 0,$$
$$A_{43} = 0,$$
$$A_{65} A_{54} = E\left[\frac{1-A}{\delta^*(C,0)}\mu'^0(C,M) h^T(C,M)\right] E\left[\frac{A\lambda_0^*(C,M)}{\delta^0(C,1)\lambda_1^*(C,M)}\mu'^0(C,M) h(C,M) h^T(C,M)\right]^{-1},$$
$$A_{65} \psi_5^*(Z) = \frac{1-A}{\delta^0(C,0)}[\mu^0(C,M) - \nu_x^0(C)],$$

and the other relevant elements of $\psi^*(Z)$ (other than the fifth element dealt with above) are

$$\psi_{1,4,6} = \begin{pmatrix} A\frac{\lambda_0^*(C,M)}{\delta^0(C,1)\lambda_1^*(C,M)} h(C,M)[Y - \mu^0(C,M)] \\ \nu_x^0(C) - \boxed{\tau_x^0} \end{pmatrix}.$$

- If $\delta, \lambda_1, \lambda_0$ are correctly specified

$$A_{65} = \begin{pmatrix} 1 & 0 & \cdots & 0 \end{pmatrix},$$

which implies

$$A_{65} A_{51} = \underbrace{E\left[-\frac{1-A}{\delta^0(C,0)}[\mu^*(C,M) - \nu_x^*(C)]\nabla^T \log \delta^0(C,0)\right]}_{E[-\{E[\mu^*(C,M)|C,A=0] - \nu_x^*(C)\}\nabla^T \log \delta^0(C,0)]} E[-\nabla\nabla^T \log \delta^0(C,A)]^{-1},$$

39

$$A_{65}\psi_5^*(Z) = \frac{1-A}{\delta^0(C,0)}[\mu^*(C,M) - \nu_x^*(C)],$$

$$A_{65}A_{54} = E\left[E\{\mu'^*(C,M)h^T(C,M) \mid C, A=0\}\right] E\left[E\{\mu'^*(C,M)h(C,M)h^T(C,M) \mid C, A=0\}\right]^{-1} = \begin{pmatrix} 1 & 0 & \cdots & 0 \end{pmatrix}$$

by the Lemma and the fact that $\mu'(\cdot) > 0$ (because $\mu(\cdot)$ as an inverse link function is monotone increasing), which in turns implies

$$A_{65}A_{54}A_{41} = \underbrace{E\left[-\frac{A\lambda_0^0(C,M)}{\delta^0(C,1)\lambda_1^0(C,M)}[Y-\mu^*(C,M)]\nabla^T \log \delta^0(C,1)\right]}_{E[-\{\nu_x^0(C) - E[\mu^*(C,M)|C,A=0]\}\nabla^T \log \delta^0(C,1)]} E[-\nabla\nabla^T \log \delta^0(C,A)]^{-1},$$

$$A_{65}A_{54}A_{42} = \underbrace{E\left[-\frac{A\lambda_0^0(C,M)}{\delta^0(C,1)\lambda_1^0(C,M)}[Y-\mu^*(C,M)]\nabla^T \log \lambda_1^0(C,M)\right]}_{E\left[-E\{[\mu^0(C,M)-\mu^*(C,M)]\nabla^T \log \lambda_1^0(C,M)|C,A=0\}\right]} E[-A\nabla\nabla^T \log \lambda_1^0(C,M)]^{-1},$$

$$A_{65}A_{54}A_{43} = \underbrace{E\left[\frac{A\lambda_0^0(C,M)}{\delta^0(C,1)\lambda_1^0(C,M)}[Y-\mu^*(C,M)]\nabla^T \log \lambda_0^0(C,M)\right]}_{E\left[E\{[\mu^0(C,M)-\mu^*(C,M)]\nabla^T \log \lambda_0^0(C,M)|C,A=0\}\right]} E[-(1-A)\nabla\nabla^T \log \lambda_0^0(C,M)]^{-1},$$

$$A_{65}A_{54}\psi_4^*(Z) = \frac{A\lambda_0^0(C,M)}{\delta^0(C,1)\lambda_1^0(C,M)}[Y-\mu^*(C,M)],$$

and the other elements of $\psi^*(Z)$ (other than the fourth and fifth elements dealt with above) are

$$\psi_{1,2,3,6}^*(Z) = \begin{pmatrix} \nabla \log \delta^0(C,A) \\ A\nabla \log \lambda_1^0(C,M) \\ (1-A)\nabla \log \lambda_0^0(C,M) \\ \nu_x^*(C) - \tau_x^0 \end{pmatrix}.$$

- If in addition to $\delta, \lambda_1$ and $\lambda_0$, $\mu$ is also correctly specified,

$$A_{41} = 0,$$
$$A_{42} = 0,$$
$$A_{43} = 0,$$
$$A_{65}A_{51} = \underbrace{E\left[-\frac{1-A}{\delta^0(C,0)}[\mu^0(C,M) - \nu_x^*(C)]\nabla^T \log \delta^0(C,0)\right]}_{E[-[\nu_x^0(C) - \nu_x^*(C)]\nabla^T \log \delta^0(C,0)]} E[-\nabla\nabla^T \log \delta^0(C,A)]^{-1},$$

$$A_{65}\psi_5^*(Z) = \frac{1-A}{\delta^0(C,0)}[\mu^0(C,M) - \nu_x^*(C)],$$

$$A_{65}A_{54}\psi_4^*(Z) = \frac{A\lambda_0^0(C,M)}{\delta^0(C,1)\lambda_1^0(C,M)}[Y-\mu^0(C,M)],$$

and the other elements of $\psi^*(Z)$ (other than the fourth and fifth elements dealt with above) are

$$\psi_{1,6}^*(Z) = \begin{pmatrix} \nabla \log \delta^0(C,A) \\ \nu_x^*(C) - \tau_x^0 \end{pmatrix}.$$

- If $\delta, \gamma, \mu, \nu_x$ are all correctly specified, the IF reduces to the nonparametric IF:

$$\frac{A\lambda_0^0(C,M)}{\delta^0(C,1)\lambda_1^0(C,M)}[Y-\mu^0(C,M)] \;+\; \frac{1-A}{\delta^0(C,0)}[\mu^0(C,M) - \nu_x^0(C)] \;+\; \nu_x^0(C) - \tau_x^0.$$

**Using weights estimated by method 2 – via conditional exposure models**

$$\Psi(Z,\theta) = \begin{pmatrix} \nabla \log \delta(C, A; \theta_1) \\ \nabla \log \gamma(C, M; \theta_2) \\ A \frac{\gamma(C, M, 0; \theta_2)}{\delta(C, 0; \theta_1) \gamma(C, M, 1; \theta_2)} h(C, M)[Y - \mu(\theta_3^T h(C, M))] \\ \frac{1 - A}{\delta(C, 0; \theta_1)} g(C)[\mu(\theta_3^T h(C, M)) - \nu_x(\theta_4^T g(C))] \\ \nu_x(\theta_4^T g(C)) - \tau_x \end{pmatrix} \quad (A72)$$

$$\Omega = \begin{pmatrix} \Omega_{11} & 0 & 0 & 0 & 0 \\ 0 & \Omega_{22} & 0 & 0 & 0 \\ \Omega_{31} & \Omega_{32} & \Omega_{33} & 0 & 0 \\ \Omega_{41} & 0 & \Omega_{43} & \Omega_{44} & 0 \\ 0 & 0 & 0 & \Omega_{54} & -1 \end{pmatrix},$$

where

$\Omega_{11} = \mathrm{E}[\nabla \nabla^T \log \delta^*(C, A)],$

$\Omega_{31} = \mathrm{E}\left[-\frac{A \gamma^*(C, M, 0)}{\delta^*(C, 0) \gamma^*(C, M, 1)} h(C, M)[Y - \mu^*(C, M)] \nabla^T \log \delta^*(C, 0)\right],$

$\Omega_{41} = \mathrm{E}\left[-\frac{1 - A}{\delta^*(C, 0)} g(C)[\mu^*(C, M) - \nu_x^*(C)] \nabla^T \log \delta^*(C, 0)\right],$

$\Omega_{22} = \mathrm{E}[\nabla \nabla^T \log \gamma^*(C, M)],$

$\Omega_{32} = \mathrm{E}\left[-\frac{A}{\delta^*(C, 0) \gamma^*(C, M, 1)} h(C, M)[Y - \mu^*(C, M)] \nabla^T \log \gamma^*(C, M, 1)\right],$

$\Omega_{33} = \mathrm{E}\left[-\frac{A \gamma^*(C, M, 0)}{\delta^*(C, 0) \gamma^*(C, M, 1)} \mu'^*(C, M) h(C, M) h^T(C, M)\right],$

$\Omega_{43} = \mathrm{E}\left[\frac{1 - A}{\delta^*(C, 0)} g(C) \mu'^*(C, M) h^T(C, M)\right],$

$\Omega_{44} = \mathrm{E}\left[-\frac{1 - A}{\delta^*(C, 0)} \nu_x'^*(C) g(C) g^T(C)\right],$

$\Omega_{54} = \mathrm{E}[\nu_x'^*(C) g^T(C)].$

$$\mathrm{A} = \begin{pmatrix} \mathrm{A}_{54}(\mathrm{A}_{41} + \mathrm{A}_{43}\mathrm{A}_{31}) & \mathrm{A}_{54}\mathrm{A}_{43}\mathrm{A}_{32} & \mathrm{A}_{54}\mathrm{A}_{43} & \mathrm{A}_{54} & 1 \end{pmatrix}.$$

The influence function of the estimator is

$$\begin{pmatrix} \mathrm{A}_{54}(\mathrm{A}_{41} + \mathrm{A}_{43}\mathrm{A}_{31}) & \mathrm{A}_{54}\mathrm{A}_{43}\mathrm{A}_{32} & \mathrm{A}_{54}\mathrm{A}_{43} & \mathrm{A}_{54} & 1 \end{pmatrix} \psi^*(Z), \quad (A73)$$

where

$$\mathrm{A}_{54} = \Omega_{54}(-\Omega_{44})^{-1} = \mathrm{E}[\nu_x'^*(C) g^T(C)] \mathrm{E}\left[\frac{1 - A}{\delta^*(C, 0)} \nu_x'^*(C) g(C) g^T(C)\right]^{-1},$$

$$\mathrm{A}_{41} = \Omega_{41}(-\Omega_{11})^{-1} = \mathrm{E}\left[-\frac{1 - A}{\delta^*(C, 0)} g(C)[\mu^*(C, M) - \nu_x^*(C)] \nabla^T \log \delta^*(C, 0)\right] \mathrm{E}[-\nabla \nabla^T \log \delta^*(C, A)]^{-1},$$

$$\mathrm{A}_{43} = \Omega_{43}(-\Omega_{33})^{-1} = \mathrm{E}\left[\frac{1 - A}{\delta^*(C, 0)} g(C) \mu'^*(C, M) h^T(C, M)\right] \mathrm{E}\left[\frac{A \gamma^*(C, M, 0)}{\delta^*(C, 0) \gamma^*(C, M, 1)} \mu'^*(C, M) h(C, M) h^T(C, M)\right]^{-1},$$

$$\mathrm{A}_{31} = \Omega_{31}(-\Omega_{11})^{-1} = \mathrm{E}\left[-\frac{A \gamma^*(C, M, 0)}{\delta^*(C, 0) \gamma^*(C, M, 1)} h(C, M)[Y - \mu^*(C, M)] \nabla^T \log \delta^*(C, 0)\right] \mathrm{E}[-\nabla \nabla^T \log \delta^*(C, A)]^{-1},$$

$$\mathrm{A}_{32} = \Omega_{32}(-\Omega_{22})^{-1} = \mathrm{E}\left[-\frac{A}{\delta^*(C, 0) \gamma^*(C, M, 1)} h(C, M)[Y - \mu^*(C, M)] \nabla^T \log \gamma^*(C, M, 1)\right] \mathrm{E}[-\nabla \nabla^T \log \gamma^*(C, M, A)]^{-1},$$

$$\psi^*(Z) = \begin{pmatrix} \nabla \log \delta^*(C, A) \\ \nabla \log \gamma^*(C, M, A) \\ A \frac{\gamma^*(C, M, 0)}{\delta^*(C, 0) \gamma^*(C, M, 1)} h(C, M)[Y - \mu^*(C, M)] \\ \frac{1 - A}{\delta^*(C, 0)} g(C)[\mu^*(C, M) - \nu_x^*(C)] \\ \nu_x^*(C) - \tau_x^* \end{pmatrix}.$$

Special cases:

- If $\mu$ and $\nu_x$ are correctly specified,

$$A_{41} = 0,$$
$$A_{31} = 0,$$
$$A_{32} = 0,$$
$$A_{54} = E[\nu_x'^0(C)g^T(C)]E\left[\frac{1-A}{\delta^*(C,0)}\nu_x'^0(C)g(C)g^T(C)\right]^{-1},$$
$$A_{43} = E\left[\frac{1-A}{\delta^*(C,0)}g(C)\mu'^0(C,M)h^T(C,M)\right] E\left[\frac{A\gamma^*(C,M,0)}{\delta^*(C,0)\gamma^*(C,M,1)}\mu'^0(C,M)h(C,M)h^T(C,M)\right]^{-1},$$

and the relevant elements of $\psi^*(Z)$ are

$$\psi_{3,4,5}^*(Z) = \begin{pmatrix} A\frac{\gamma^*(C,M,0)}{\delta^*(C,0)\gamma^*(C,M,1)}h(C,M)[Y-\mu^0(C,M)] \\ \frac{1-A}{\delta^*(C,0)}g(C)[\mu^0(C,M)-\nu_x^0(C)] \\ \nu_x^0(C) - \tau_x^0 \end{pmatrix}.$$

- If $\delta$ and $\mu$ are correctly specified,

$$A_{31} = 0,$$
$$A_{32} = 0,$$
$$A_{54} = E[\nu_x'^*(C)g^T(C)]E[\nu_x'^*(C)g(C)g^T(C)]^{-1} = \begin{pmatrix} 1 & 0 & \cdots & 0 \end{pmatrix}$$

by the Lemma and the fact that $\nu_x'(\cdot) > 0$ (because $\nu_x(\cdot)$ as an inverse link function is monotone increasing), which implies

$$A_{54}A_{41} = \underbrace{E\left[-\frac{1-A}{\delta^0(C,0)}[\mu^0(C,M) - \nu_x^*(C)]\nabla^T \log \delta^0(C,0)\right]}_{E[-\{\nu_x^0(C)-\nu_x^*(C)\}\nabla^T \log \delta^0(C,0)]} E[-\nabla\nabla^T \log \delta^0(C,A)]^{-1},$$

$$A_{54}A_{43} = E\left[\frac{1-A}{\delta^0(C,0)}\mu'^0(C,M)h^T(C,M)\right] E\left[\frac{A\gamma^*(C,M,0)}{\delta^0(C,0)\gamma^*(C,M,1)}\mu'^0(C,M)h(C,M)h^T(C,M)\right]^{-1},$$

$$A_{54}\psi_4^*(Z) = \frac{1-A}{\delta^0(C,0)}[\mu^0(C,M) - \nu_x^*(C)],$$

and the other relevant elements of $\psi^*(Z)$ (other than the fourth element dealt with above) are

$$\psi_{1,3,5} = \begin{pmatrix} \nabla \log \delta^0(C,A) \\ \frac{A\gamma^*(C,M,0)}{\delta^0(C,0)\gamma^*(C,M,1)}h(C,M)[Y-\mu^0(C,M)] \\ \nu_x^*(C) - \tau_x^0 \end{pmatrix}.$$

- If in addition to $\delta$ and $\mu$, $\nu_x$ is also correctly specified,

$$A_{41} = 0,$$
$$A_{31} = 0,$$
$$A_{32} = 0,$$
$$A_{54}A_{43} = E\left[\frac{1-A}{\delta^0(C,0)}\mu'^0(C,M)h^T(C,M)\right] E\left[\frac{A\gamma^*(C,M,0)}{\delta^0(C,0)\gamma^*(C,M,1)}\mu'^0(C,M)h(C,M)h^T(C,M)\right]^{-1},$$

$$A_{54}\psi_4^*(Z) = \frac{1-A}{\delta^0(C,0)}[\mu^0(C,M) - \nu_x^0(C)],$$

and the other relevant elements of $\psi^*(Z)$ (other than the fourth element dealt with above) are

$$\psi_{3,5} = \begin{pmatrix} \frac{A\gamma^*(C,M,0)}{\delta^0(C,0)\gamma^*(C,M,1)} h(C,M)[Y - \mu^0(C,M)] \\ \nu_x^*(C) - \boxed{\tau_x^0} \end{pmatrix}.$$

- If $\delta$ and $\gamma$ are correctly specified

$$A_{54} = \mathrm{E}[\nu_x'^*(C)g^T(C)]\mathrm{E}[\nu_x'^*(C)g(C)g^T(C)]^{-1} = \begin{pmatrix} 1 & 0 & \cdots & 0 \end{pmatrix}$$

by the Lemma and the fact that $\nu_x'(\cdot) > 0$ (because $\nu_x(\cdot)$ as an inverse link function is monotone increasing), which implies

$$A_{54}A_{41} = \underbrace{\mathrm{E}\left[-\frac{1-A}{\delta^0(C,0)}[\mu^*(C,M) - \nu_x^*(C)]\nabla^T \log \delta^0(C,0)\right]}_{\mathrm{E}[-\{\mathrm{E}[\mu^*(C,M)|C,A=0] - \nu_x^*(C)\}\nabla^T \log \delta^0(C,0)]} \mathrm{E}[-\nabla\nabla^T \log \delta^0(C,A)]^{-1},$$

$$A_{54}\psi_4^*(Z) = \frac{1-A}{\delta^0(C,0)}[\mu^*(C,M) - \nu_x^*(C)],$$

$$A_{54}A_{43} = \mathrm{E}\left[\mathrm{E}\{\mu'^0(C,M)h^T(C,M) \mid C, A=0\}\right] \mathrm{E}\left[\mathrm{E}\{\mu'^0(C,M)h(C,M)h^T(C,M) \mid C, A=0\}\right]^{-1} = \begin{pmatrix} 1 & 0 & \cdots & 0 \end{pmatrix},$$

which in turns implies

$$A_{54}A_{43}A_{31} = \underbrace{\mathrm{E}\left[-\frac{A\gamma^0(C,M,0)}{\delta^0(C,0)\gamma^0(C,M,1)}[Y - \mu^*(C,M)]\nabla^T \log \delta^0(C,0)\right]}_{\mathrm{E}[-\{\nu_x^0(C) - \mathrm{E}[\mu^*(C,M)|C,A=0]\}\nabla^T \log \delta^0(C,0)]} \mathrm{E}[-\nabla\nabla^T \log \delta^0(C,A)]^{-1},$$

$$A_{54}A_{43}A_{32} = \underbrace{\mathrm{E}\left[-\frac{A}{\delta^0(C,0)\gamma^0(C,M,1)}[Y - \mu^*(C,M)]\nabla^T \log \gamma^0(C,M,1)\right]}_{\mathrm{E}\left[-\mathrm{E}\left\{[\mu^0(C,M) - \mu^*(C,M)]\frac{\nabla^T \log \gamma^0(C,M,1)}{\gamma^0(C,M,0)}\right\}\right]} \mathrm{E}[-\nabla\nabla^T \log \gamma^0(C,M,A)]^{-1},$$

$$A_{54}A_{43}\psi_3^*(Z) = \frac{A\gamma^0(C,M,0)}{\delta^0(C,0)\gamma^0(C,M,1)}[Y - \mu^*(C,M)],$$

and the other elements of $\psi^*(Z)$ (other than the third and fourth elements dealt with above) are

$$\psi_{1,2,5}^*(Z) = \begin{pmatrix} \nabla \log \delta^0(C,A) \\ \nabla \log \gamma^0(C,M,A) \\ \nu_x^*(C) - \boxed{\tau_x^0} \end{pmatrix}.$$

- If in addition to $\delta$ and $\gamma$, $\mu$ is also correctly specified,

$$A_{31} = 0,$$
$$A_{32} = 0,$$
$$A_{54}A_{41} = \underbrace{\mathrm{E}\left[-\frac{1-A}{\delta^0(C,0)}[\mu^0(C,M) - \nu_x^*(C)]\nabla^T \log \delta^0(C,0)\right]}_{\mathrm{E}[-\{\nu_x^0(C) - \nu_x^*(C)\}\nabla^T \log \delta^0(C,0)]} \mathrm{E}[-\nabla\nabla^T \log \delta^0(C,A)]^{-1},$$

43

$$A_{54}\psi_4^*(Z) = \frac{1-A}{\delta^0(C,0)}[\mu^0(C,M) - \nu_x^*(C)],$$

$$A_{54}A_{43}\psi_3^*(Z) = \frac{A\gamma^0(C,M,0)}{\delta^0(C,0)\gamma^0(C,M,1)}[Y - \mu^0(C,M)],$$

and the other relevant elements of $\psi^*(Z)$ (other than the third and fourth elements dealt with above) are

$$\psi_{1,5}^*(Z) = \begin{pmatrix} \nabla \log \delta^0(C,A) \\ \nu_x^*(C) - \tau_x^0 \end{pmatrix}.$$

- If $\delta, \gamma, \mu, \nu_x$ are all correctly specified, the IF reduces to the nonparametric IF:

$$\frac{A\gamma^0(C,M,0)}{\delta^0(C,0)\gamma^0(C,M,1)}[Y - \mu^0(C,M)] \;+\; \frac{1-A}{\delta^0(C,0)}[\mu^0(C,M) - \nu_x^0(C)] \;+\; \nu_x^0(C) - \tau_x^0.$$

### C.4.2.4 crw|MsimYpred.MR

**Using weights estimated by method 1 – via mediator density models**

This is the estimator in Section 5 of Tchetgen Tchetgen and Shpitser (2012).

$$\psi(Z,\theta) = \begin{pmatrix} \nabla \log \delta(C,A;\theta_1) \\ \frac{A}{\delta(C,1;\theta_1)} \nabla \log \lambda_1(C,M;\theta_2) \\ \frac{1-A}{\delta(C,0;\theta_1)} \nabla \log \lambda_0(C,M;\theta_3) \\ \frac{A\lambda_0(C,M;\theta_3)}{\delta(C,1;\theta_1)\lambda_1(C,M;\theta_2)} h(C,M)[Y - \mu(\theta_4^T h(C,M))] \\ \int \mu(C,m;\theta_4)\lambda_0(C,m;\theta_3)dm - \tau_x \end{pmatrix}. \quad (A74)$$

$$\Omega = \begin{pmatrix} \Omega_{11} & 0 & 0 & 0 & 0 \\ \Omega_{21} & \Omega_{22} & 0 & 0 & 0 \\ \Omega_{31} & 0 & \Omega_{33} & 0 & 0 \\ \Omega_{41} & \Omega_{42} & \Omega_{43} & \Omega_{44} & 0 \\ 0 & 0 & \Omega_{53} & \Omega_{54} & -1 \end{pmatrix},$$

where

$\Omega_{11} = E[\nabla \nabla^T \log \delta^*(C,A)],$

$\Omega_{22} = E[\frac{A}{\delta^*(C,1)} \nabla \nabla^T \log \lambda_1^*(C,M)],$

$\Omega_{33} = E[\frac{1-A}{\delta^*(C,0)} \nabla \nabla^T \log \lambda_0^*(C,M)],$

$\Omega_{44} = E\left[-\frac{A\lambda_0^*(C,M)}{\delta^*(C,1)\lambda_1^*(C,M)} \mu'^*(C,M)h(C,M)h^T(C,M)\right],$

$\Omega_{41} = E\left[-\frac{A\lambda_0^*(C,M)}{\delta^*(C,1)\lambda_1^*(C,M)} h(C,M)[Y - \mu^*(C,M)]\nabla^T \log \delta^*(C,1)\right],$

$\Omega_{31} = E\left[-\frac{1-A}{\delta^*(C,0)} \nabla \log \lambda_0^*(C,M)\nabla^T \log \delta^*(C,0)\right],$

$\Omega_{21} = E\left[-\frac{A}{\delta^*(C,1)} \nabla \log \lambda_1^*(C,M)\nabla^T \log \delta^*(C,1)\right],$

$\Omega_{42} = E\left[-\frac{A\lambda_0^*(C,M)}{\delta^*(C,1)\lambda_1^*(C,M)} h(C,M)[Y - \mu^*(C,M)]\nabla^T \log \lambda_1^*(C,M)\right],$

$\Omega_{43} = E\left[\frac{A\lambda_0^*(C,M)}{\delta^*(C,1)\lambda_1^*(C,M)} h(C,M)[Y - \mu^*(C,M)]\nabla^T \log \lambda_0^*(C,M)\right],$

$\Omega_{53} = E\left[\int \mu^*(C,m)[\nabla^T \log \lambda_0^*(C,m)]\lambda_0^*(C,m)dm\right],$

$\Omega_{54} = E\left[\int \mu'^*(C,m)h^T(C,m)\lambda_0^*(C,m)dm\right].$

$$A = \begin{pmatrix} (A_{53}A_{31} + A_{54}(A_{41} + A_{42}A_{21} + A_{43}A_{31})) & A_{54}A_{42} & (A_{53} + A_{54}A_{43}) & A_{54} & 1 \end{pmatrix}.$$

The influence function of the estimator is

$$\left((A_{53}A_{31} + A_{54}(A_{41} + A_{42}A_{21} + A_{43}A_{31}))\quad A_{54}A_{42}\quad (A_{53} + A_{54}A_{43})\quad A_{54}\quad 1\right)\psi^*(Z), \tag{A75}$$

where

$$A_{53} = \mathrm{E}\left[\int \mu^*(C,m)[\nabla^T\log\lambda_0^*(C,m)]\lambda_0^*(C,m)dm\right]\mathrm{E}\left[-\frac{1-A}{\delta^*(C,0)}\nabla\nabla^T\log\lambda_0^*(C,M)\right]^{-1},$$

$$A_{31} = \mathrm{E}\left[-\frac{1-A}{\delta^*(C,0)}\nabla\log\lambda_0^*(C,M)\nabla^T\log\delta^*(C,0)\right]\mathrm{E}[-\nabla\nabla^T\log\delta^*(C,A)]^{-1},$$

$$A_{54} = \mathrm{E}\left[\int \mu'^*(C,m)h^T(C,m)\lambda_0^*(C,m)dm\right]\mathrm{E}\left[\frac{A\lambda_0^*(C,M)}{\delta^*(C,1)\lambda_1^*(C,M)}\mu'^*(C,M)h(C,M)h^T(C,M)\right]^{-1},$$

$$A_{41} = \mathrm{E}\left[-\frac{A\lambda_0^*(C,M)}{\delta^*(C,1)\lambda_1^*(C,M)}h(C,M)[Y-\mu^*(C,M)]\nabla^T\log\delta^*(C,1)\right]\mathrm{E}[-\nabla\nabla^T\log\delta^*(C,A)]^{-1},$$

$$A_{42} = \mathrm{E}\left[-\frac{A\lambda_0^*(C,M)}{\delta^*(C,1)\lambda_1^*(C,M)}h(C,M)[Y-\mu^*(C,M)]\nabla^T\log\lambda_1^*(C,M)\right]\mathrm{E}\left[-\frac{A}{\delta^*(C,1)}\nabla\nabla^T\log\lambda_1^*(C,M)\right]^{-1},$$

$$A_{21} = \mathrm{E}\left[-\frac{A}{\delta^*(C,1)}\nabla\log\lambda_1^*(C,M)\nabla^T\log\delta^*(C,1)\right]\mathrm{E}[-\nabla\nabla^T\log\delta^*(C,A)]^{-1},$$

$$A_{43} = \mathrm{E}\left[\frac{A\lambda_0^*(C,M)}{\delta^*(C,1)\lambda_1^*(C,M)}h(C,M)[Y-\mu^*(C,M)]\nabla^T\log\lambda_0^*(C,M)\right]\mathrm{E}\left[-\frac{1-A}{\delta^*(C,0)}\nabla\nabla^T\log\lambda_0^*(C,M)\right]^{-1},$$

$$\psi^*(Z) = \begin{pmatrix}\nabla\log\delta^*(C,A)\\ \frac{A}{\delta^*(C,1)}\nabla\log\lambda_1^*(C,M)\\ \frac{1-A}{\delta^*(C,0)}\nabla\log\lambda_0^*(C,M)\\ \frac{A\lambda_0^*(C,M)}{\delta^*(C,1)\lambda_1^*(C,M)}h(C,M)[Y-\mu^*(C,M)]\\ \int\mu^*(C,m)\lambda_0^*(C,m)dm - \tau_x^*\end{pmatrix}.$$

Special cases:

- If $\mu$ and $\lambda_0$ are correctly specified,

  $A_{41} = 0,$
  
  $A_{42} = 0,$
  
  $A_{43} = 0,$
  
  $$A_{53} = \underbrace{\mathrm{E}\left[\int \mu^0(C,m)[\nabla^T\log\lambda_0^*(C,m)]\lambda_0^0(C,m)dm\right]}_{\mathrm{E}[\mathrm{E}\{\mu^0(C,M)\nabla^T\log\lambda_0^0(C,M)|C,A=0\}]}\mathrm{E}\left[-\frac{1-A}{\delta^*(C,0)}\nabla\nabla^T\log\lambda_0^0(C,M)\right]^{-1},$$
  
  $$A_{31} = \mathrm{E}\left[-\frac{1-A}{\delta^*(C,0)}\nabla\log\lambda_0^0(C,M)\nabla^T\log\delta^*(C,0)\right]\mathrm{E}[-\nabla\nabla^T\log\delta^*(C,A)]^{-1},$$
  
  $$A_{54} = \underbrace{\mathrm{E}\left[\int \mu'^0(C,m)h^T(C,m)\lambda_0^0(C,m)dm\right]}_{\mathrm{E}[\mathrm{E}\{\mu'^0(C,M)h^T(C,M)|C,A=0\}]}\mathrm{E}\left[\frac{A\lambda_0^0(C,M)}{\delta^*(C,1)\lambda_1^*(C,M)}\mu'^0(C,M)h(C,M)h^T(C,M)\right]^{-1},$$

  and the relevant elements of $\psi^*(Z)$ are

45

$$\psi^*_{1,3,4,5}(Z) = \begin{pmatrix} \nabla \log \delta^*(C, A) \\ \dfrac{1-A}{\delta^*(C,0)} \nabla \log \lambda_0^0(C, M) \\ \dfrac{A \lambda_0^0(C, M)}{\delta^*(C,1)\lambda_1^*(C, M)} h(C, M)[Y - \mu^0(C, M)] \\ \underbrace{\int \mu^0(C, m) \lambda_0^0(C, m) dm}_{\nu_x^0(C)} - \boxed{\tau_x^0} \end{pmatrix}.$$

- If $\delta, \lambda_0, \lambda_1$ are correctly specified,

$$A_{21} = E\left[ -\underbrace{E\{\nabla \log \lambda_1^0(C, M) \mid C, A=1\}}_{0} \nabla^T \log \delta^0(C, 1) \right] = 0,$$

$$A_{31} = E\left[ -\underbrace{E\{\nabla \log \lambda_0^0(C, M) \mid C, A=0\}}_{0} \nabla^T \log \delta^0(C, 0) \right] = 0,$$

$$A_{53} = \underbrace{E\left[ \int \mu^*(C, m)[\nabla^T \log \lambda_0^0(C, m)] \lambda_0^0(C, m) dm \right]}_{E[E\{\mu^*(C,M)\nabla^T \log \lambda_0^0(C,M)|C,A=0\}]} \underbrace{E\left[ -\dfrac{1-A}{\delta^0(C,0)} \nabla \nabla^T \log \lambda_0^0(C, M) \right]^{-1}}_{E[-E\{\nabla\nabla^T \log \lambda_0^0(C,M)|C,A=0\}]},$$

$$A_{54} = E\left[ E\{\mu'^*(C, M) h^T(C, M) \mid C, A=0\} \right] E\left[ E\{\mu'^*(C, M) h(C, M) h^T(C, M) \mid C, A=0\} \right]^{-1} = \begin{pmatrix} 1 & 0 & \cdots & 0 \end{pmatrix}$$

by the Lemma and the fact that $\mu'(\cdot) > 0$ (because $\mu(\cdot)$ as an inverse link function is monotone increasing), which implies

$$A_{54}A_{41} = \underbrace{E\left[ -\dfrac{A\lambda_0^0(C, M)}{\delta^0(C,1)\lambda_1^0(C, M)} [Y - \mu^*(C, M)] \nabla^T \log \delta^0(C, 1) \right]}_{E[-\{\nu_x^0(C) - E[\mu^*(C,M)|C,A=0]\}\nabla^T \log \delta^0(C,1)]} E[-\nabla\nabla^T \log \delta^0(C, A)]^{-1},$$

$$A_{54}A_{42} = \underbrace{E\left[ -\dfrac{A\lambda_0^0(C, M)}{\delta^0(C,1)\lambda_1^0(C, M)} [Y - \mu^*(C, M)] \nabla^T \log \lambda_1^0(C, M) \right]}_{E[-E\{[\mu^0(C,M)-\mu^*(C,M)]\nabla^T \log \lambda_1^0(C,M)|C,A=0\}]} \underbrace{E\left[ -\dfrac{A}{\delta^0(C,1)} \nabla \nabla^T \log \lambda_1^0(C, M) \right]^{-1}}_{E[-E\{\nabla\nabla^T \log \lambda_1^0(C,M)|C,A=1\}]^{-1}},$$

$$A_{54}A_{43} = \underbrace{E\left[ \dfrac{A\lambda_0^0(C, M)}{\delta^0(C,1)\lambda_1^0(C, M)} [Y - \mu^*(C, M)] \nabla^T \log \lambda_0^0(C, M) \right]}_{E[E\{[\mu^0(C,M)-\mu^*(C,M)]\nabla^T \log \lambda_0^0(C,M)|C,A=0\}]} \underbrace{E\left[ -\dfrac{1-A}{\delta^0(C,0)} \nabla \nabla^T \log \lambda_0^0(C, M) \right]^{-1}}_{E[-E\{\nabla\nabla^T \log \lambda_0^0(C,M)|C,A=0\}]^{-1}},$$

$$A_{54}\psi_4^*(Z) = \dfrac{A\lambda_0^0(C, M)}{\delta^0(C,1)\lambda_1^0(C, M)}[Y - \mu^*(C, M)],$$

and the elements of $\psi^*(Z)$ (other than the fourth element dealt with above) are

$$\psi^*_{1,2,3,5}(Z) = \begin{pmatrix} \nabla \log \delta^0(C, A) \\ \dfrac{A}{\delta^0(C,1)} \nabla \log \lambda_1^0(C, M) \\ \dfrac{1-A}{\delta^0(C,0)} \nabla \log \lambda_0^0(C, M) \\ \underbrace{\int \mu^*(C, m) \lambda_0^0(C, m) dm}_{E[\mu^*(C,M)|C,A=0]} - \boxed{\tau_x^0} \end{pmatrix}.$$

46

- If $\delta, \lambda_1, \lambda_0, \mu$ are all correctly specified,

$$A_{21} = 0,$$
$$A_{31} = 0,$$
$$A_{41} = 0,$$
$$A_{42} = 0,$$
$$A_{43} = 0,$$
$$A_{53} = \underbrace{E\left[\int \mu^0(C,m)[\nabla^T\log\lambda_0^0(C,m)]\lambda_0^0(C,m)dm\right]}_{E[E\{\mu^0(C,M)\nabla^T\log\lambda_0^0(C,M)|C,A=0\}]} \underbrace{E\left[-\frac{1-A}{\delta^0(C,0)}\nabla\nabla^T\log\lambda_0^0(C,M)\right]^{-1}}_{E[-E\{\nabla\nabla^T\log\lambda_0^0(C,M)|C,A=0\}]},$$
$$A_{54}\psi_4^*(Z) = \frac{A\lambda_0^0(C,M)}{\delta^0(C,1)\lambda_1^0(C,M)}[Y - \mu^0(C,M)],$$

and the elements of $\psi^*(Z)$ (other than the fourth element dealt with above) are

$$\psi_{3,5}^*(Z) = \begin{pmatrix} \frac{1-A}{\delta^0(C,0)}\nabla\log\lambda_0^0(C,M) \\ \underbrace{\int \mu^0(C,m)\lambda_0^0(C,m)dm}_{\nu_x^0(C)} - \boxed{\tau_x^0} \end{pmatrix}.$$

- If $\delta, \mu, \lambda_1$ are correctly specified but only $\lambda_0$ is misspecified,

$$A_{41} = 0,$$
$$A_{42} = 0,$$
$$A_{43} = 0,$$
$$A_{53} = \underbrace{E\left[\int \mu^0(C,m)[\nabla^T\log\lambda_0^*(C,m)]\lambda_0^*(C,m)dm\right]}_{E[E_{\lambda_0^*}\{\mu^0(C,M)\nabla^T\log\lambda_0(C,M)|C,A=0\}]} \underbrace{E\left[-\frac{1-A}{\delta^0(C,0)}\nabla\nabla^T\log\lambda_0^*(C,M)\right]^{-1}}_{E[-E\{\nabla\nabla^T\log\lambda_0^*(C,M)|C,A=0\}]^{-1}},$$
$$A_{31} = \underbrace{E\left[-\frac{1-A}{\delta^0(C,0)}\nabla\log\lambda_0^*(C,M)\nabla^T\log\delta^0(C,0)\right]}_{E[-E\{\nabla\log_0^*(C,M)|C,A=0\}\nabla^T\log\delta^0(C,0)]} E[-\nabla\nabla^T\log\delta^0(C,A)]^{-1},$$
$$A_{54} = E[E_{\lambda_0^*}\{\mu'^0(C,M)h^T(C,M) \mid C,A=0\}]E\left[E_{\lambda_0^*}\{\mu'^0(C,M)h(C,M)h^T(C,M) \mid C,A=0\}\right]^{-1} = \begin{pmatrix} 1 & 0 & \cdots & 0 \end{pmatrix}$$

by the Lemma and the fact that $\mu'(\cdot) > 0$ (because $\mu(\cdot)$ as an inverse link function is monotone increasing), which implies

$$A_{54}\psi_4^*(Z) = \frac{A\lambda_0^0(C,M)}{\delta^0(C,1)\lambda_1^0(C,M)}[Y - \mu^0(C,M)],$$

and the other relevant elements of $\psi^*(Z)$ (other than the fourth element dealt with above) are

$$\psi_{1,3,5}^*(Z) = \begin{pmatrix} \nabla\log\delta^0(C,A) \\ \frac{1-A}{\delta^0(C,0)}\nabla\log\lambda_0^*(C,M) \\ \underbrace{\int \mu^0(C,m)\lambda_0^*(C,m)dm}_{E_{\lambda_0^*}[\mu(C,M)|C,A=0]} - \tau_x^* \end{pmatrix}.$$

47

**Using weights estimated by method 2 – via conditional exposure models**

$$\psi(Z, \theta) = \begin{pmatrix} \nabla \log \delta(C, A; \theta_1) \\ \nabla \log \gamma(C, M, A; \theta_2) \\ \dfrac{A\gamma(C, M, 0; \theta_2)}{\delta(C, 0; \theta_1)\gamma(C, M, 1; \theta_2)} h(C, M)[Y - \mu(\theta_3^T h(C, M))] \\ \dfrac{1-A}{\delta(C, 0; \theta_1)} \nabla \log \lambda_0(C, M; \theta_4) \\ \int \mu(\theta_3^T h(C, m)) \lambda_0(C, m; \theta_4) dm - \tau_{\mathrm{x}} \end{pmatrix}. \tag{A76}$$

$$\Omega = \begin{pmatrix} \Omega_{11} & 0 & 0 & 0 & 0 \\ 0 & \Omega_{22} & 0 & 0 & 0 \\ \Omega_{31} & \Omega_{32} & \Omega_{33} & 0 & 0 \\ \Omega_{41} & 0 & 0 & \Omega_{44} & 0 \\ 0 & 0 & \Omega_{53} & \Omega_{54} & -1 \end{pmatrix},$$

where

$\Omega_{11} = \mathrm{E}[\nabla \nabla^T \log \delta^*(C, A)]$,

$\Omega_{22} = \mathrm{E}[\nabla \nabla^T \log \gamma^*(C, M, A)]$,

$\Omega_{33} = \mathrm{E}\left[ -\dfrac{A\gamma^*(C, M, 0)}{\delta^*(C, 0)\gamma^*(C, M, 1)} \mu'^*(C, M) h(C, M) h^T(C, M) \right]$,

$\Omega_{44} = \mathrm{E}\left[ \dfrac{1-A}{\delta^*(C, 0)} \nabla \nabla^T \log \lambda_0^*(C, M) \right]$,

$\Omega_{31} = \mathrm{E}\left[ -\dfrac{A\gamma^*(C, M, 0)}{\delta^*(C, 0)\gamma^*(C, M, 1)} h(C, M)[Y - \mu^*(C, M)] \nabla^T \log \delta^*(C, 0) \right]$,

$\Omega_{41} = \mathrm{E}\left[ -\dfrac{1-A}{\delta^*(C, 0)} \nabla \log \lambda_0^*(C, M) \nabla^T \log \delta^*(C, 0) \right]$,

$\Omega_{32} = \mathrm{E}\left[ -\dfrac{A}{\delta^*(C, 0)\gamma^*(C, M, 1)} h(C, M)[Y - \mu^*(C, M)] \nabla^T \log \gamma^*(C, M, 1) \right]$,

$\Omega_{53} = \mathrm{E}\left[ \int \mu'^*(C, m) h^T(C, m) \lambda_0^*(C, m) dm \right]$,

$\Omega_{54} = \mathrm{E}\left[ \int \mu^*(C, m) [\nabla^T \log \lambda_0^*(C, m)] \lambda_0^*(C, m) dm \right]$.

$$\mathrm{A} = \begin{pmatrix} (\mathrm{A}_{53}\mathrm{A}_{31} + \mathrm{A}_{54}\mathrm{A}_{41}) & \mathrm{A}_{53}\mathrm{A}_{32} & \mathrm{A}_{53} & \mathrm{A}_{54} & 1 \end{pmatrix}.$$

The influence function of the estimator is

$$\begin{pmatrix} (\mathrm{A}_{53}\mathrm{A}_{31} + \mathrm{A}_{54}\mathrm{A}_{41}) & \mathrm{A}_{53}\mathrm{A}_{32} & \mathrm{A}_{53} & \mathrm{A}_{54} & 1 \end{pmatrix} \psi^*(Z), \tag{A77}$$

where

$\mathrm{A}_{53} = \Omega_{53}(-\Omega_{33})^{-1} = \mathrm{E}\left[ \int \mu'^*(C, m) h^T(C, m) \lambda_0^*(C, m) dm \right] \mathrm{E}\left[ \dfrac{A\gamma^*(C, M, 0)}{\delta^*(C, 0)\gamma^*(C, M, 1)} \mu'^*(C, M) h(C, M) h^T(C, M) \right]^{-1}$,

$\mathrm{A}_{31} = \Omega_{31}(-\Omega_{11})^{-1} = \mathrm{E}\left[ -\dfrac{A\gamma^*(C, M, 0)}{\delta^*(C, 0)\gamma^*(C, M, 1)} h(C, M)[Y - \mu^*(C, M)] \nabla^T \log \delta^*(C, 0) \right] \mathrm{E}[-\nabla \nabla^T \log \delta^*(C, A)]^{-1}$,

$\mathrm{A}_{54} = \Omega_{54}(-\Omega_{44})^{-1} = \mathrm{E}\left[ \int \mu^*(C, m)[\nabla^T \log \lambda_0^*(C, M)] \lambda_0^*(C, m) dm \right] \mathrm{E}\left[ -\dfrac{1-A}{\delta^*(C, 0)} \nabla \nabla^T \log \lambda_0^*(C, M) \right]^{-1}$,

$\mathrm{A}_{41} = \Omega_{41}(-\Omega_{11})^{-1} = \mathrm{E}\left[ -\dfrac{1-A}{\delta^*(C, 0)} \nabla \log \lambda_0^*(C, M) \nabla^T \log \delta^*(C, 0) \right] \mathrm{E}[-\nabla \nabla^T \log \delta^*(C, A)]^{-1}$

$\mathrm{A}_{32} = \Omega_{32}(-\Omega_{22})^{-1} = \mathrm{E}\left[ -\dfrac{A}{\delta^*(C, 0)\gamma^*(C, M, 1)} h(C, M)[Y - \mu^*(C, M)] \nabla^T \log \gamma^*(C, M, 1) \right] \mathrm{E}[-\nabla \nabla^T \log \gamma^*(C, M, A)]^{-1}$,

$$\psi^*(Z) = \begin{pmatrix} \nabla \log \delta^*(C, A) \\ \nabla \log \gamma^*(C, M, A) \\ \dfrac{A\gamma^*(C, M, 0)}{\delta^*(C, 0)\gamma^*(C, M, 1)} h(C, M)[Y - \mu^*(C, M)] \\ \dfrac{1-A}{\delta^*(C, 0)} \nabla \log \lambda_0^*(C, M) \\ \int \mu^*(C, m) \lambda_0^*(C, m) dm - \tau_{\mathrm{x}}^* \end{pmatrix}.$$

Special cases:

- If $\mu$ and $\lambda_0$ are correctly specified,

$$A_{31} = 0,$$
$$A_{32} = 0,$$
$$A_{54} = \underbrace{E\left[\int \mu^0(C,m)[\nabla^T \log \lambda_0^0(C,M)]\lambda_0^0(C,m)dm\right]}_{E[E\{\mu^0(C,M)\nabla^T \log \lambda_0^0(C,M)|C,A=0\}]} E\left[-\frac{1-A}{\delta^*(C,0)}\nabla\nabla^T \log \lambda_0^0(C,M)\right]^{-1},$$
$$A_{41} = E\left[-\frac{1-A}{\delta^*(C,0)}\nabla \log \lambda_0^0(C,M)\nabla^T \log \delta^*(C,0)\right] E[-\nabla\nabla^T \log \delta^*(C,A)]^{-1},$$
$$A_{53} = E\left[E\left\{\mu'^0(C,M)h^T(C,M) \mid C,A=0\right\}\right] E\left[\frac{A\gamma^*(C,M,0)}{\delta^*(C,0)\gamma^*(C,M,1)}\mu'^0(C,M)h(C,M)h^T(C,M)\right]^{-1},$$

and the relevant elements of $\psi^*(Z)$ are

$$\psi^*_{1,3,4,5}(Z) = \begin{pmatrix} \nabla \log \delta^*(C,A) \\ \frac{A\gamma^*(C,M,0)}{\delta^*(C,0)\gamma^*(C,M,1)}h(C,M)[Y-\mu^0(C,M)] \\ \frac{1-A}{\delta^*(C,0)}\nabla \log \lambda_0^0(C,M) \\ \underbrace{\int \mu^0(C,m)\lambda_0^0(C,m)dm}_{\nu_x^0(C)} - \boxed{\tau_x^0} \end{pmatrix}.$$

- If $\delta, \gamma$ and $\lambda_0$ are correctly specified,

$$A_{41} = E\left[-\underbrace{E\left\{\nabla \log \lambda_0^0(C,M) \mid C,A=0\right\}}_{0} \nabla^T \log \delta^0(C,0)\right] = 0,$$
$$A_{54} = \underbrace{E\left[\int \mu^*(C,m)[\nabla^T \log \lambda_0^0(C,M)]\lambda_0^0(C,m)dm\right]}_{E[E\{\mu^*(C,M)\nabla^T \log \lambda_0^0(C,M)|C,A=0\}]} \underbrace{E\left[-\frac{1-A}{\delta^0(C,0)}\nabla\nabla^T \log \lambda_0^0(C,M)\right]^{-1}}_{E[-E\{\nabla\nabla^T \log \lambda_0^0(C,M)|C,A=0\}]^{-1}},$$
$$A_{53} = E\left[E\left\{\mu'^*(C,M)h^T(C,M) \mid C,A=0\right\}\right] E\left[E\left\{\mu'^*(C,M)h(C,M)h^T(C,M) \mid C,A=0\right\}\right] = \begin{pmatrix} 1 & 0 & \cdots & 0 \end{pmatrix}$$

by the Lemma and the fact that $\mu'(\cdot) > 0$ (because $\mu(\cdot)$ as an inverse link function is monotone increasing), which implies

$$A_{53}A_{31} = \underbrace{E\left[-\frac{A\gamma^0(C,M,0)}{\delta^0(C,0)\gamma^0(C,M,1)}[Y-\mu^*(C,M)]\nabla^T \log \delta^0(C,0)\right]}_{E[-\{\nu_x(C)-E[\mu^*(C,M)|C,A=0]\}\nabla^T \log \delta^0(C,0)]} E[-\nabla\nabla^T \log \delta^0(C,A)]^{-1},$$
$$A_{53}A_{32} = \underbrace{E\left[-\frac{A}{\delta^0(C,0)\gamma^0(C,M,1)}[Y-\mu^*(C,M)]\nabla^T \log \gamma^0(C,M,1)\right]}_{E\left[-E\left\{[\mu^0(C,M)-\mu^*(C,M)]\frac{\nabla^T \log \gamma^0(C,M,1)}{\gamma^0(C,M,0)}|C,A=0\right\}\right]} E[-\nabla\nabla^T \log \gamma^0(C,M,A)]^{-1},$$

49

$$A_{53}\psi_3^*(Z) = \frac{A\gamma^0(C,M,0)}{\delta^0(C,0)\gamma^0(C,M,1)}[Y - \mu^*(C,M)],$$

and the other elements of $\psi^*(Z)$ (other than the third element dealt with above) are

$$\psi_{1,2,4,5}^*(Z) = \begin{pmatrix} \nabla \log \delta^0(C,A) \\ \nabla \log \gamma^0(C,M,A) \\ \frac{1-A}{\delta^0(C,0)}\nabla \log \lambda_0^0(C,M) \\ \underbrace{\int \mu^*(C,m)\lambda_0^0(C,m)dm}_{\mathrm{E}[\mu^*(C,M)|C,A=0]} - \boxed{\tau_{\mathrm{x}}^0} \end{pmatrix}.$$

- If $\delta, \gamma, \lambda_0, \mu$ are all correctly specified,

$$A_{41} = 0,$$
$$A_{31} = 0,$$
$$A_{32} = 0,$$
$$A_{54} = \underbrace{\mathrm{E}\left[\int \mu^0(C,m)[\nabla^T \log \lambda_0^0(C,M)]\lambda_0^0(C,m)dm\right]}_{\mathrm{E}[\mathrm{E}\{\mu^0(C,M)\nabla^T \log \lambda_0^0(C,M)|C,A=0\}]} \underbrace{\mathrm{E}\left[-\frac{1-A}{\delta^0(C,0)}\nabla\nabla^T \log \lambda_0^0(C,M)\right]^{-1}}_{\mathrm{E}[-\mathrm{E}\{\nabla\nabla^T \log \lambda_0^0(C,M)|C,A=0\}]^{-1}},$$
$$A_{53}\psi_3^*(Z) = \frac{A\gamma^0(C,M,0)}{\delta^0(C,0)\gamma^0(C,M,1)}[Y - \mu^0(C,M)],$$

and the other elements of $\psi^*(Z)$ (other than the third element dealt with above) are

$$\psi_{4,5}^*(Z) = \begin{pmatrix} \frac{1-A}{\delta^0(C,0)}\nabla \log \lambda_0^0(C,M) \\ \underbrace{\int \mu^0(C,m)\lambda_0^0(C,m)dm}_{\nu_{\mathrm{x}}^0(C)} - \boxed{\tau_{\mathrm{x}}^0} \end{pmatrix}.$$

- If $\delta, \gamma, \mu$ are correctly specified but $\lambda_0$ is misspecified,

$$A_{31} = 1,$$
$$A_{32} = 1,$$
$$A_{54} = \underbrace{\mathrm{E}\left[\int \mu^0(C,m)[\nabla^T \log \lambda_0^*(C,M)]\lambda_0^*(C,m)dm\right]}_{\mathrm{E}\left[\mathrm{E}_{\lambda_0^*}\{\mu^0(C,M)\nabla^T \log \lambda_0^*(C,M)|C,A=0\}\right]} \underbrace{\mathrm{E}\left[-\frac{1-A}{\delta^0(C,0)}\nabla\nabla^T \log \lambda_0^*(C,M)\right]^{-1}}_{\mathrm{E}[-\mathrm{E}\{\nabla\nabla^T \log \lambda_0^*(C,M)|C,A=0\}]^{-1}},$$
$$A_{41} = \underbrace{\mathrm{E}\left[-\frac{1-A}{\delta^0(C,0)}\nabla \log \lambda_0^*(C,M)\nabla^T \log \delta^0(C,0)\right]}_{\mathrm{E}\left[-\mathrm{E}\{\nabla \log \lambda_0^*(C,M)|C,A=0\}\nabla^T \log \delta^0(C,0)\right]} \mathrm{E}[\nabla\nabla^T \log \delta^0(C,A)]^{-1},$$
$$A_{53} = \underbrace{\mathrm{E}\left[\int \mu'^0(C,m)h^T(C,m)\lambda_0^*(C,m)dm\right]}_{\mathrm{E}\left[\mathrm{E}_{\lambda_0^*}\{\mu'^0(C,M)h^T(C,M)|C,A=0\}\right]} \underbrace{\mathrm{E}\left[\frac{A\gamma^0(C,M,0)}{\delta^0(C,0)\gamma^0(C,M,1)}\mu'^0(C,M)h(C,M)h^T(C,M)\right]^{-1}}_{\mathrm{E}[\mathrm{E}\{\mu'^0(C,M)h(C,M)h^T(C,M)|C,A=0\}]^{-1}},$$

50

and the relevant elements of $\psi^*(Z)$ are

$$\psi^*_{1,3,4,5}(Z) = \begin{pmatrix} \nabla \log \delta^0(C, A) \\ \frac{A\gamma^0(C, M, 0)}{\delta^0(C, 0)\gamma^*(C, M, 1)} h(C, M)[Y - \mu^0(C, M)] \\ \frac{1-A}{\delta^0(C, 0)} \nabla \log \lambda^*_0(C, M) \\ \int \mu^0(C, m)\lambda^*_0(C, m)dm - \tau^*_x \end{pmatrix}.$$

### C.4.3 If targeting additive effects

#### C.4.3.1 NDE|YpredEpred.R

**Using weights estimated by method 2 – via conditional exposure models**

$$\Psi(Z, \theta) = \begin{pmatrix} \nabla \log \delta(C, A; \theta_1) \\ \nabla \log \gamma(C, M; \theta_2) \\ A\frac{\gamma(C, M, 0; \theta_2)}{\delta(C, 0; \theta_1)\gamma(C, M, 1; \theta_2)} h(C, M)[Y - \mu(\theta_3^T h(C, M))] \\ \frac{1-A}{\delta(C, 0; \theta_1)} g(C)[\mu(\theta_3^T h(C, M)) - Y - \xi(\theta_4^T g(C))] \\ \xi(\theta_4^T g(C)) - \beta_0 \end{pmatrix} \quad (A78)$$

$$\Omega = \begin{pmatrix} \Omega_{11} & 0 & 0 & 0 & 0 \\ 0 & \Omega_{22} & 0 & 0 & 0 \\ \Omega_{31} & \Omega_{32} & \Omega_{33} & 0 & 0 \\ \Omega_{41} & 0 & \Omega_{43} & \Omega_{44} & 0 \\ 0 & 0 & 0 & \Omega_{54} & -1 \end{pmatrix},$$

where

$\Omega_{11} = \mathrm{E}[\nabla\nabla^T \log \delta^*(C, A)],$

$\Omega_{22} = \mathrm{E}[\nabla\nabla^T \log \gamma^*(C, M)],$

$\Omega_{33} = \mathrm{E}\left[-\frac{A\gamma^*(C, M, 0)}{\delta^*(C, 0)\gamma^*(C, M, 1)} \mu'^*(C, M) h(C, M) h^T(C, M)\right],$

$\Omega_{44} = \mathrm{E}\left[-\frac{1-A}{\delta^*(C, 0)} \xi'^*(C) g(C) g^T(C)\right],$

$\Omega_{31} = \mathrm{E}\left[-\frac{A\gamma^*(C, M, 0)}{\delta^*(C, 0)\gamma^*(C, M, 1)} h(C, M)[Y - \mu^*(C, M)]\nabla^T \log \delta^*(C, 0)\right],$

$\Omega_{41} = \mathrm{E}\left[-\frac{1-A}{\delta^*(C, 0)} g(C)[\mu^*(C, M) - Y - \xi^*(C)]\nabla^T \log \delta^*(C, 0)\right],$

$\Omega_{32} = \mathrm{E}\left[-\frac{A}{\delta^*(C, 0)\gamma^*(C, M, 1)} h(C, M)[Y - \mu^*(C, M)]\nabla^T \log \gamma^*(C, M, 1)\right],$

$\Omega_{43} = \mathrm{E}\left[\frac{1-A}{\delta^*(C, 0)} g(C)\mu'^*(C, M) h^T(C, M)\right],$

$\Omega_{54} = \mathrm{E}[\xi'^*(C) g^T(C)].$

$$\mathrm{A} = \begin{pmatrix} \mathrm{A}_{54}(\mathrm{A}_{41} + \mathrm{A}_{43}\mathrm{A}_{31}) & \mathrm{A}_{54}\mathrm{A}_{43}\mathrm{A}_{32} & \mathrm{A}_{54}\mathrm{A}_{43} & \mathrm{A}_{54} & 1 \end{pmatrix}.$$

The influence function of the estimator is

$$\begin{pmatrix} \mathrm{A}_{54}(\mathrm{A}_{41} + \mathrm{A}_{43}\mathrm{A}_{31}) & \mathrm{A}_{54}\mathrm{A}_{43}\mathrm{A}_{32} & \mathrm{A}_{54}\mathrm{A}_{43} & \mathrm{A}_{54} & 1 \end{pmatrix}\psi^*(Z), \quad (A79)$$

where

$$\mathrm{A}_{54} = \Omega_{54}(-\Omega_{44})^{-1} = \mathrm{E}[\xi'^*(C)g^T(C)]\mathrm{E}\left[\frac{1-A}{\delta^*(C, 0)}\xi'^*(C)g(C)g^T(C)\right]^{-1},$$

$$\mathrm{A}_{41} = \Omega_{41}(-\Omega_{11})^{-1} = \mathrm{E}\left[-\frac{1-A}{\delta^*(C, 0)}g(C)[\mu^*(C, M) - Y - \xi^*(C)]\nabla^T \log \delta^*(C, 0)\right]\mathrm{E}[-\nabla\nabla^T \log \delta^*(C, A)]^{-1},$$

$$A_{43} = \Omega_{43}(-\Omega_{33})^{-1} = E\left[\frac{1-A}{\delta^*(C,0)}g(C)\mu'^*(C,M)h^T(C,M)\right] E\left[\frac{A\gamma^*(C,M,0)}{\delta^*(C,0)\gamma^*(C,M,1)}\mu'^*(C,M)h(C,M)h^T(C,M)\right]^{-1},$$

$$A_{31} = \Omega_{31}(-\Omega_{11})^{-1} = E\left[-\frac{A\gamma^*(C,M,0)}{\delta^*(C,0)\gamma^*(C,M,1)}h(C,M)[Y-\mu^*(C,M)]\nabla^T\log\delta^*(C,0)\right] E[-\nabla\nabla^T\log\delta^*(C,A)]^{-1},$$

$$A_{32} = \Omega_{32}(-\Omega_{22})^{-1} = E\left[-\frac{A}{\delta^*(C,0)\gamma^*(C,M,1)}h(C,M)[Y-\mu^*(C,M)]\nabla^T\log\gamma^*(C,M,1)\right] E[-\nabla\nabla^T\log\gamma^*(C,M,A)]^{-1},$$

$$\psi^*(Z) = \begin{pmatrix} \nabla\log\delta^*(C,A) \\ \nabla\log\gamma^*(C,M,A) \\ A\frac{\gamma^*(C,M,0)}{\delta^*(C,0)\gamma^*(C,M,1)}h(C,M)[Y-\mu^*(C,M)] \\ \frac{1-A}{\delta^*(C,0)}g(C)[\mu^*(C,M)-Y-\xi^*(C)] \\ \xi^*(C) - \beta_0^* \end{pmatrix}.$$

Special cases:

- If $\mu$ and $\xi$ are correctly specified,

  $A_{41} = 0,$

  $A_{31} = 0,$

  $A_{32} = 0,$

  $$A_{54} = E[\xi'^0(C)g^T(C)]E\left[\frac{1-A}{\delta^*(C,0)}\xi'^0(C)g(C)g^T(C)\right]^{-1},$$

  $$A_{43} = E\left[\frac{1-A}{\delta^*(C,0)}g(C)\mu'^0(C,M)h^T(C,M)\right] E\left[\frac{A\gamma^*(C,M,0)}{\delta^*(C,0)\gamma^*(C,M,1)}\mu'^0(C,M)h(C,M)h^T(C,M)\right]^{-1},$$

  and the relevant elements of $\psi^*(Z)$ are

  $$\psi_{3,4,5}^*(Z) = \begin{pmatrix} A\frac{\gamma^*(C,M,0)}{\delta^*(C,0)\gamma^*(C,M,1)}h(C,M)[Y-\mu^0(C,M)] \\ \frac{1-A}{\delta^*(C,0)}g(C)[\mu^0(C,M)-Y-\xi^0(C)] \\ \xi^0(C) - \beta_0^0 \end{pmatrix}.$$

- If $\delta$ and $\mu$ are correctly specified,

  $A_{31} = 0,$

  $A_{32} = 0,$

  $A_{54} = E[\xi'^*(C)g^T(C)]E[\xi'^*(C)g(C)g^T(C)]^{-1} = \begin{pmatrix} 1 & 0 & \cdots & 0 \end{pmatrix}$

  by the Lemma and the fact that $\xi'(\cdot) > 0$ (because $\xi(\cdot)$ as an inverse link function is monotone increasing) which implies

  $$A_{54}A_{41} = \underbrace{E\left[-\frac{1-A}{\delta^0(C,0)}[\mu^0(C,M)-Y-\xi^*(C)]\nabla^T\log\delta^0(C,0)\right]}_{E[-\{\xi^0(C)-\xi^*(C)\}\nabla^T\log\delta^0(C,0)]} E[-\nabla\nabla^T\log\delta^0(C,A)]^{-1},$$

  $$A_{54}A_{43} = E\left[\frac{1-A}{\delta^0(C,0)}\mu'^0(C,M)h^T(C,M)\right] E\left[\frac{A\gamma^*(C,M,0)}{\delta^0(C,0)\gamma^*(C,M,1)}\mu'^0(C,M)h(C,M)h^T(C,M)\right]^{-1},$$

  $$A_{54}\psi_4^*(Z) = \frac{1-A}{\delta^0(C,0)}[\mu^0(C,M)-Y-\xi^*(C)],$$

52

and the other relevant elements of $\psi^*(Z)$ (other than the fourth element dealt with above) are

$$\psi_{1,3,5} = \begin{pmatrix} \nabla \log \delta^0(C, A) \\ \frac{A\gamma^*(C,M,0)}{\delta^0(C,0)\gamma^*(C,M,1)} h(C, M)[Y - \mu^0(C, M)] \\ \xi^*(C) - \beta_0^0 \end{pmatrix}.$$

- If in addition to $\delta$ and $\mu$, $\xi$ is also correctly specified,

$A_{41} = 0,$

$A_{31} = 0,$

$A_{32} = 0,$

$$A_{54}A_{43} = E\left[\frac{1-A}{\delta^0(C,0)}\mu'^0(C,M)h^T(C,M)\right] E\left[\frac{A\gamma^*(C,M,0)}{\delta^0(C,0)\gamma^*(C,M,1)}\mu'^0(C,M)h(C,M)h^T(C,M)\right]^{-1},$$

$$A_{54}\psi_4^*(Z) = \frac{1-A}{\delta^0(C,0)}[\mu^0(C,M) - Y - \xi^0(C)],$$

and the other relevant elements of $\psi^*(Z)$ (other than the fourth element dealt with above) are

$$\psi_{3,5} = \begin{pmatrix} \frac{A\gamma^*(C,M,0)}{\delta^0(C,0)\gamma^*(C,M,1)} h(C, M)[Y - \mu^0(C, M)] \\ \xi^0(C) - \beta_0^0 \end{pmatrix}.$$

- If $\delta$ and $\gamma$ are correctly specified

$$A_{54} = E[\xi'^*(C)g^T(C)]E[\xi'^*(C)g(C)g^T(C)]^{-1} = \begin{pmatrix} 1 & 0 & \cdots & 0 \end{pmatrix},$$

which implies

$$A_{54}A_{41} = \underbrace{E\left[-\frac{1-A}{\delta^0(C,0)}[\mu^*(C,M) - Y - \xi^*(C)]\nabla^T \log \delta^0(C,0)\right]}_{E[-\{E[\mu^*(C,M)|C,A=0]-\nu_0^0(C)-\xi^*(C)\}\nabla^T \log \delta^0(C,0)]} E[-\nabla\nabla^T \log \delta^0(C,A)]^{-1},$$

$$A_{54}\psi_4^*(Z) = \frac{1-A}{\delta^0(C,0)}[\mu^*(C,M) - Y - \xi^*(C)],$$

$$A_{54}A_{43} = E\left[E\{\mu'^0(C,M)h^T(C,M) \mid C, A = 0\}\right] E\left[E\{\mu'^0(C,M)h(C,M)h^T(C,M) \mid C, A = 0\}\right]^{-1} = \begin{pmatrix} 1 & 0 & \cdots & 0 \end{pmatrix},$$

by the Lemma and the fact that $\mu'(\cdot) > 0$ (because $\mu(\cdot)$ as an inverse link function is monotone increasing), which in turns implies

$$A_{54}A_{43}A_{31} = \underbrace{E\left[-\frac{A\gamma^0(C,M,0)}{\delta^0(C,0)\gamma^0(C,M,1)}[Y - \mu^*(C,M)]\nabla^T \log \delta^0(C,0)\right]}_{E[-\{\nu_x^0(C)-E[\mu^*(C,M)|C,A=0]\}\nabla^T \log \delta^0(C,0)]} E[-\nabla\nabla^T \log \delta^0(C,A)]^{-1},$$

$$A_{54}A_{43}A_{32} = \underbrace{E\left[-\frac{A}{\delta^0(C,0)\gamma^0(C,M,1)}[Y - \mu^*(C,M)]\nabla^T \log \gamma^0(C,M,1)\right]}_{E\left[-E\left\{[\mu^0(C,M)-\mu^*(C,M)]\frac{\nabla^T \log \gamma^0(C,M,1)}{\gamma^0(C,M,0)}\right\}\right]} E[-\nabla\nabla^T \log \gamma^0(C,M,A)]^{-1},$$

$$A_{54}A_{43}\psi_3^*(Z) = \frac{A\gamma^0(C,M,0)}{\delta^0(C,0)\gamma^0(C,M,1)}[Y - \mu^*(C,M)],$$

53

and the other elements of $\psi^*(Z)$ (other than the third and fourth elements dealt with above) are

$$\psi^*_{1,2,5}(Z) = \begin{pmatrix} \nabla \log \delta^0(C,A) \\ \nabla \log \gamma^0(C,M,A) \\ \xi^*(C) - \beta^0_0 \end{pmatrix}.$$

- If in addition to $\delta$ and $\gamma$, $\mu$ is also correctly specified,

$$A_{31} = 0,$$
$$A_{32} = 0,$$
$$A_{54}A_{41} = \underbrace{E\left[-\frac{1-A}{\delta^0(C,0)}[\mu^0(C,M) - Y - \xi^*(C)]\nabla^T \log \delta^0(C,0)\right]}_{E[-\{\xi^0(C) - \xi^*(C)\}\nabla^T \log \delta^0(C,0)]} E[-\nabla\nabla^T \log \delta^0(C,A)]^{-1},$$
$$A_{54}\psi^*_4(Z) = \frac{1-A}{\delta^0(C,0)}[\mu^0(C,M) - Y - \xi^*(C)],$$
$$A_{54}A_{43}\psi^*_3(Z) = \frac{A\gamma^0(C,M,0)}{\delta^0(C,0)\gamma^0(C,M,1)}[Y - \mu^0(C,M)],$$

and the other relevant elements of $\psi^*(Z)$ (other than the third and fourth elements dealt with above) are

$$\psi^*_{1,5}(Z) = \begin{pmatrix} \nabla \log \delta^0(C,A) \\ \xi^*(C) - \beta^0_0 \end{pmatrix}.$$

- If $\delta, \gamma, \mu, \xi$ are all correctly specified, the IF reduces to the nonparametric IF:

$$\frac{A\gamma^0(C,M,0)}{\delta^0(C,0)\gamma^0(C,M,1)}[Y - \mu^0(C,M)] + \frac{1-A}{\delta^0(C,0)}[\mu^0(C,M) - Y - \xi^0(C)] + \xi^0(C) - \beta^0_0.$$

## D  Detailed response to a referee's comment about the bridge distribution

In the article (section 4.5), we mention a referee's comment that logit models for both $E[Y \mid C, M, A = 1]$ and $E[Y \mid C, A = 1]$ imply that $M$ given $C, A = 1$ follows a bridge distribution. We state there that this is not true. In this section of the Technical Appendix, we now elaborate on this statement.

Bridge distributions are defined by Wang and Louis (2003) in the context of random intercepts models for clustered data. A bridge distribution (if it exists) is a specific random intercept distribution that connects two models for the same response variable: a random intercepts model (aka *conditional* model) and a model without random effects (aka *marginal* model), where both models use the same link function and the fixed effect coefficients in the conditional model and the coefficients in the marginal model coincide (up to a proportionality constant). In general form,

$$h\big(E[Y_{ij} \mid X_{ij}]\big) = \beta_1^T X_{ij}, \tag{A80}$$
$$h\big(E[Y_{ij} \mid b_i, X_{ij}]\big) = b_i + \beta_2^T X_{ij}, \tag{A81}$$
$$b_i \mid X_i \sim g(b_i), \tag{A82}$$
$$\beta_1 = \phi\beta_2, \tag{A83}$$

54

where $i$ indexes cluster, $j$ indexes unit within cluster; $h$ is the inverse link function, $X_i$ is the combination of all $X_{ij}$ within cluster $i$, $b_i$ is the random intercept, $g$ is the bridge distribution; $\phi$ is a proportionality constant (also known as the rescaling factor or attenuation factor), which depends on the bridge distribution $g$. Here the random intercept $b_i$ is assumed to be independent of $X_i$, but in Wang and Louis (2004) its variance is allowed to depend on cluster-level covariates.

Wang and Louis (2003) posed the question what distribution $g$ satisfies the specified connection between the two models (same link function, proportional coefficients), and solves this problem for a range of link functions for a binary outcome. For the logit link (which is of interest here), where $h$ is the expit function, Wang and Louis (2003) showed that the bridge distribution has density function

$$g(b) = \frac{1}{2\pi} \frac{\sin(\phi\pi)}{\cosh(\phi b) + \cos(\phi\pi)} \quad (0 < \phi < 1, -\infty < b < \infty). \tag{A84}$$

This distribution has mean 0 and variance $\frac{1}{3}\pi^2(\frac{1}{\phi^2} - 1)$. In the Wang and Louis (2004) version, the variance of this distribution depends on cluster-level covariates, but its mean remains 0.

Now we turn to summarize our current mediation analysis setting before relating it to this bridge distribution setting. To make notation more comparable, we use $X$ instead of $C$ for covariates and suppress the conditioning event $A = 1$ from the notation. When using logit regression for the outcome, the structure is

$$h(\mathrm{E}[Y_i \mid X_i]) = \alpha_1^T X_i, \tag{A85}$$

$$h(\mathrm{E}[Y_i \mid M_i, X_i]) = \gamma M_i + \alpha_2^T X_i, \tag{A86}$$

$$M_i \mid X_i \sim g^*(M_i \mid X_i), \tag{A87}$$

where the two outcome models are specified and the distribution $g^*(M_i \mid X_i)$ is implied by the specified outcome models. Our article accommodates generic $M$, but for the current discussion, let us take $M$ to be a univariate continuous variable. To simplify notation, let $h_1(x) := h(\mathrm{E}[Y_i \mid X_i = x])$ and $h_2(x, m) := h(\mathrm{E}[Y_i \mid M_i = m, X_i = x])$. For each value of $X_i$, $g^*$ is the solution to the equation

$$h_1(X_i) = \int h_2(X_i, m) g^*(m \mid X_i) dm. \tag{A88}$$

Intuitively, within levels of $X_i$, $g^*$ is a weight function such that the weighted mean of $h_2(X_i, m)$ equals $h_1(X_i)$, which means $g^*$ is whatever it needs to be depending on what $h_1$ and $h_2$ happen to be.

The referee's comment is that $g^*$ is the same as $g$ in (A84). This is not the case because the two settings are very different. The bridge distribution setting is one with clustered data, while our current setting has one-level data, although it is unclear whether this difference is important. Putting this aside, there are two differences that are crucial. One is between the two pairs of models: in the bridge distribution setting $\beta_1$ and $\beta_2$ in the two outcome models are restricted to be proportional, whereas there is no such restriction on $\alpha_1$ and $\alpha_2$ in our models. The other difference is a key feature of $g$ that is not shared by $g^*$: random intercepts are defined to have mean zero given model covariates, a restriction not imposed on the mediator given covariates distribution in our setting. In fact an assumption of conditional mean zero for the mediator would be grossly wrong because the covariates are causes of the mediator.

In brief, other than the one similarity that the same link function is used for two outcome models, our current case is fundamentally different from the bridge distribution case, so the mediator given covariates distribution implied by the two outcome models is not the bridge distribution.



\end{document}